%% file: 6d5dSUSPv2.tex
\documentclass[11pt,a4paper]{article}
\pdfoutput=1
\usepackage{jheppub}
\usepackage{amsthm,amsbsy,amsfonts,mathrsfs,enumerate,float,wrapfig}

\title{6d SCFTs, 5d Dualities and Tao Web Diagrams}

\author[a,b]{Hirotaka Hayashi,}
\author[c,e]{Sung-Soo Kim,}
\author[e]{Kimyeong Lee,}
\author[d,e]{and Futoshi Yagi}

\affiliation[a]{Department of Physics, School of Science, Tokai University, \\4-1-1 Kitakaname, Hiratsuka-shi, Kanagawa 259-1292, Japan}
\affiliation[b]{Departamento de F\'isica Te\'orica and Instituto de F\'isica Te\'orica UAM/CSIC,\\ Universidad Aut\'onoma de Madrid, Cantoblanco, 28049 Madrid, Spain}
\affiliation[c]{School of Physics, University of Electronic Science and Technology of China, \\
No.4, Section 2, North Jianshe Road, Chengdu, Sichuan 610054, China}
\affiliation[d]{School of Mathematics, Southwest Jiaotong University,\\ West zone, high-tech district, Chengdu, Sichuan 611756, China}
\affiliation[e]{Korea Institute for Advanced Study, 
85 Hoegi-ro Dongdaemun-gu, Seoul, 02455, Korea}

\emailAdd{h.hayashi@tokai.ac.jp}
\emailAdd{sungsoo.kim@uestc.edu.cn}
\emailAdd{klee@kias.re.kr}
\emailAdd{futoshi\_yagi@swjtu.edu.cn}

\abstract{We propose 5d descriptions of 6d ${\cal N}=(1,0)$ superconformal field theories arising from Type IIA brane configurations with an $O8^-$-plane. We T-dualize the brane diagram along a compactification circle and obtain a 5-brane web diagram with two $O7^-$-planes. The gauge theory description of the resulting 5d theory for a given 6d superconformal field theory is not unique, and we argue that the non-uniqueness leads to various dual 5d gauge theories. There are three sources which lead to the 5d dualities. One type comes from either resolving both or one of the two $O7^-$-planes. The two situations give us two different ways to read off a 5d gauge theory from essentially the same web diagram. The second type originates from different distributions of D5 or D7-branes, shifting the gauge group ranks of the 5d quiver theory.  The last one comes from the 90 or 45 degree rotations of the 5-brane web diagram, which is a part of the $SL(2,\mathbb{Z})$ duality of Type IIB string theory, leading to completely different group structure. These lead to a very rich class of dualities between 5d gauge theories whose UV completion is the same 6d superconformal field theory. We also explore  Higgsing of the 6d theories and their 5d counterparts. Decoupling the same flavors from the dual 5d theories gives rise to another dual 5d theories whose UV completion is the same 5d superconformal field theory. Finally we propose the 6d description of 5d theories which is obtained by a generalization of 5d $T_N$ theories with additional flavors, which turns out not to be in the class of Type IIA brane construction generically. 
}

\begin{document}
\preprint{
{\tt 
\begin{flushright}
IFT-UAM/CSIC-15-097\\
KIAS-P15038
\end{flushright}
}
}
\maketitle

\input{intro.tex}

\bigskip
\input{section2.tex}

\bigskip

\input{section3.tex}

\bigskip

\input{section4-1.tex}

\input{section4-2.tex}

\bigskip

\input{section5.tex}

\bigskip

\input{section6.tex}

\bigskip

\input{discussion.tex}

\bigskip

\acknowledgments
We thank Kang-Sin Choi, Hee-Cheol Kim, Seok Kim, Xiao Liu, Soo-Jong Rey, Ashoke Sen and Piljin Yi for useful discussions. We especially thank Masato Taki for early collaboration and discussions. We are thankful to the Workshop on Geometric Correspondences of Gauge Theories at SISSA. S.K. is grateful to ``APCTP focus program on holography and its application'' (2015) where a part of work is done. The work of H.H. is supported by the grant FPA2012-32828 from the MINECO, the REA grant agreement PCIG10-GA-2011-304023 from the People Programme of FP7 (Marie Curie Action), the ERC Advanced Grant SPLE under contract ERC-2012-ADG-20120216-320421 and the grant SEV-2012-0249 of the ``Centro de Excelencia Severo Ochoa'' Programme. The work of HH is also supported in part by Perimeter Institute for Theoretical Physics where a part of the work was done. Research at Perimeter Institute is supported by the Government of Canada through Industry Canada and by the Province of Ontario through Ministry of Economic Development \& Innovation.
The work of K.L. is supported in part by the National Research Foundation of Korea (NRF) Grants No. 2006-0093850.  \bigskip

\input{appendix.tex}

\providecommand{\href}[2]{#2}\begingroup\raggedright\endgroup

\end{document}

%% file: intro.tex
\section{Introduction and Summary}

6d $\mathcal{N}=(1, 0)$ superconformal field theories (SCFTs) possess 
many interesting as well as mysterious features. For instance, they may not have an ultraviolet (UV) Lagrangian description. They may have tensionless self-dual strings as fundamental degrees of freedom. Still, they are expected to be well-defined local quantum field theories. Therefore, a better understanding of the 6d SCFTs will deepen our understandings of quantum field theories. Various aspects of 6d $\mathcal{N} = (1,0)$ SCFTs have been analyzed recently, for example, the classification from F-theory \cite{Heckman:2013pva, Gaiotto:2014lca, DelZotto:2014hpa, Heckman:2014qba, Heckman:2015bfa, Bhardwaj:2015xxa}, the elliptic genus computation of the self-dual strings \cite{Haghighat:2013gba, Haghighat:2013tka, Haghighat:2014pva, Kim:2014dza, Haghighat:2014vxa, Gadde:2015tra, Haghighat:2015ega}, the anomaly polynomials \cite{Ohmori:2014pca, Ohmori:2014kda, Intriligator:2014eaa, Heckman:2015ola, Heckman:2015axa, Cordova:2015fha, Bobev:2015kza} and also their torus compactifications \cite{Ohmori:2015pua, DelZotto:2015rca, Ohmori:2015pia} as well as compactifications on other Riemann surfaces \cite{Gaiotto:2015usa, Franco:2015jna, Hanany:2015pfa}.  

Another important way of analyzing a 6d $\mathcal{N}=(1, 0)$ SCFT is to study 
a 5d $\mathcal{N}=1$ supersymmetric quantum field theory whose UV completion is the 6d SCFT. 
Given a 6d SCFT, we first move to a tensor branch where scalars in tensor multiplets acquire a vacuum expectation value (VEV), and then perform a circle compactification with gauge or flavor Wilson lines along the $S^1$. After the compactification, we obtain a certain 5d $\mathcal{N}=1$ supersymmetric quantum field theory. From this point of view, the Kaluza-Klein (KK) modes become instantons in the 5d theory. In other words, the KK modes are dynamically generated in the 5d theory, and they become massless at the UV fixed point, namely we recover the 6d SCFT at the UV fixed point of the 5d theory. Although this picture is useful for the analysis of 6d SCFTs, it is typically difficult to identify what is a 5d description of a 6d SCFT on $S^1$. 

A new way to see a direct connection between a 6d SCFT and a 5d theory was discovered in \cite{Hayashi:2015fsa}. It starts from a brane configuration in Type IIA string theory developed in \cite{Brunner:1997gf, Hanany:1997gh}, realizing a 6d $Sp(N)$ gauge theory with $N_f = 2N+8$ hypermultiplets in the fundamental representation 
coupled to a tensor multiplet. 
The fixed point of the 6d theory is the $(D_{N+4}, D_{N+4})$ minimal conformal matter theory. 
One then compactifies the 6d theory on a circle with Wilson lines and perform T-duality along the circle.
Then the brane configuration in Type IIA string theory becomes a 5-brane web in Type IIB string theory. Moreover, it possible to see the 5-brane web diagram yields a 5d $SU(N+2)$ gauge theory with $N_f = 2N+8$ hypermultiplets in the fundamental representation. The corresponding web diagram has a particular feature of an infinitely expanding spiral shape. Hence, it is in a class of so-called Tao web diagram introduced in \cite{Kim:2015jba}, which has been conjectured to give a 5d theory whose UV completion is a 6d SCFT. This way gives us a direct connection between the 6d $(D_{N+4}, D_{N+4})$ minimal conformal matter theory and the 5d $SU(N+2)$ gauge theory with $N_f=2N+8$ flavors\footnote{The same claim that the UV completion of the 5d $SU(N+2)$ gauge theory with $N_f=2N+8$ flavors is the 6d $(D_{N+4}, D_{N+4})$ minimal conformal matter was also obtained in \cite{Yonekura:2015ksa} by the analysis of the instanton operator in the 5d theory. The instanton operator analysis is also a useful way to explore 6d SCFTs from 5d theories \cite{Tachikawa:2015mha, Zafrir:2015uaa, Yonekura:2015ksa, Gaiotto:2015una}. The quantization under a one-instanton background can yield instanton states. 
One may construct a Dynkin diagram of an enhanced flavor symmetry from the instanton states. If the Dynkin diagram is affine, then the 5d theory is conjectured to have a 6d UV completion.}. 

The main aim of this paper is to generalize the connection between the 6d SCFT and the 5d theory to a broader class of 6d SCFTs realized by a brane configuration in Type IIA string theory constructed in \cite{Brunner:1997gf, Hanany:1997gh}. The system consists of NS5-branes, D6-branes, D8-branes and an $O8^-$-plane. The D6-branes are divided into pieces by NS5-branes, and D8-branes introduce some hypermultiplets in the fundamental representation. The worldvolume theory on the D6-branes yield a 6d theory on the 
tensor branch of a 6d SCFT. It is possible to apply the new method \cite{Hayashi:2015fsa} to various 6d SCFTs that can be constructed by the brane system with an $O8^{-}$-plane in Type IIA string theory. The 6d theories on the tensor branch of the 6d SCFTs we will consider are in the following two classes\footnote{In our notation, $[k]_{R}$ means $k$ hypermultiplets in the representation $R$. When $R$ is the fundamental representation, we will omit it for simplicity.}:
\begin{equation}
6d  \; Sp(N) - SU(2N+8) - SU(2N+16) - \cdots - SU(2N+8(n-1)) - [2N+8n], \label{6dquiver1}
\end{equation} 
and 
\begin{equation}
6d  \; [1]_{A} - SU(N) - SU(N+8) - SU(N+16) - \cdots - SU(N+8(n-1)) - [N+8n], \label{6dquiver2}
\end{equation}
where $n$ is a positive integer and the subscript $A$ stands for the anti-symmetric representation. With the sufficient number of the flavors at the end node of the quiver theories, 
one may Higgs the theories as in \cite{Gaiotto:2014lca}. The renormalization group (RG) flow triggered by the Higgs VEV yields a different 6d theory at low energies. The corresponding gauge theories may have different ranks, and fundamental hypermultiplets will be attached to various nodes, depending on the Higgsing. We will consider such two large families of 6d SCFTs realized by Type IIA branes with an $O8^-$-plane. 
As in \cite{Hayashi:2015fsa},
we find that a circle compactification of the system 
followed by T-duality along the $S^1$ always yields a Tao web diagram leading to a certain 5d theory which have the 6d UV fixed point. Furthermore, the Tao web diagram enables one to read off a gauge theory description of the 5d theory.

In the process of going down to 5d from 6d, we find that there are several 
different choices which yields 5d theories with different gauge groups and matter. We, however, claim that those 5d theories are dual to each other in the sense that they have the same UV completion as a 6d SCFT.  
We find that there are three different types of the origin for this 5d duality.

\paragraph{(i) $SU-Sp$ duality:} 
The first type of the 5d duality that we argue comes from the choice 
of whether we resolve two $O7^-$-planes or only one $O7^-$-plane after the T-duality. The starting 6d brane setup includes an $O8^-$-plane, and hence the T-duality induces two $O7^-$-planes. An $O7^-$-plane can be split into two 7-branes by the non-perturbative effects in the string coupling \cite{Sen:1996vd}. As we have two $O7^-$-planes, we may split the both $O7^-$-planes or only one of them into a pair of 7-branes 
\footnote{Keeping two $O7^-$-planes unresolved will not give a 5d description due to the explicit existence of the $S^1$.}. The two descriptions, in fact, express two different 5d gauge theories, an $SU$ gauge theory for two resolved $8^-$-planes and an $Sp$ gauge theory for an $O8^-$-plane, and thus yield different 5-brane web diagrams. On the other hand, since these two web diagrams arise essentially from the same 6d brane configuration, the two 5d theories are dual to each other with the same 6d UV completion.  
It is also worthy noting that given these dual 5d theories which have the same 6d UV fixed point, we can decouple flavors of one of the 5d gauge theories, say $SU$ theories, one by one by sending the masses to $\pm \infty$, leading to the 5d theory which has the 5d UV fixed point. In the same way, we decouple the flavors in the dual 5d theories, $Sp$ theories, and then the resulting two 5d theories should be also dual to each and have the same 5d UV fixed point. Hence, we obtain a 5d duality between 5d $SU$ gauge theories and $Sp$ gauge theories with the same flavors in the sense that the dual 5d theories have the same 5d UV fixed point. In particular, the 5d $SU-Sp$ duality proposed in \cite{Gaiotto:2015una} which is an example in the class of dual pairs that we consider in this paper.

\paragraph{(ii) Distribution duality:} 
The second type of the 5d duality 
arises from a choice 
of how we allocate D5-branes after the T-duality. In the Type IIA brane configuration with an $O8^-$-plane, the fundamental region is either the left side or the right side of the $O8^-$-plane since the transverse space to the $O8^-$-plane is a real one-dimensional space. After the T-duality, the $O8^-$-plane becomes two $O7^-$-planes and the transverse space to the $O7^-$-planes is a real two-dimensional space. Each $O7^-$-plane induces a point-symmetric 5-brane web configuration with respect to the point where the $O7^-$-plane is located in the two-dimensional plane. Then, the fundamental region is a region bounded by two parallel infinite mirror lines with each line passing through a different $O7^-$-plane. In this case, it is possible to allocate different numbers of D5-branes between the left side and the right side of a line passing through the two $O7^-$-planes by introducing Wilson lines. In other words, along the circumference of the $S^1$ where two $O7^-$-planes maximally apart, the position of each D5-brane along this circumference can be distributed by proper Wilson lines. After splitting the two $O7^-$-planes, the different distributions give a different 5d quiver theory with different ranks of the gauge groups. Again, the different 5d theories have the same 6d UV fixed point and thus they are dual to each other. 
It is, in fact, possible for one to show the duality explicitly at the level of the 5-brane web diagrams. For instance, these different distributions can be more naturally understood as certain mass deformations in their S-dual pictures. Again, decoupling flavors in the same way as (i) above 
yield another 5d dualities in the sense that the 5d theories dual to each other have the same 5d UV fixed point. 

\paragraph{(iii) $SL(2,\mathbb{Z})$-duality:} 
The third type of the 5d dualities arises from utilizing the $SL(2,\mathbb{Z})$ duality in Type IIB string theory.  
The rotation of a 5-brane web in a real two-dimensional space called the $(p,q)$ plane is a part of the $SL(2,\mathbb{Z})$ duality. In particular, S-duality corresponds to the rotation by $90$ degrees. Therefore, the $90$ degree rotation of any 5-brane webs obtained after the T-duality from the 6d brane configuration leads to another dual picture. We also find that there is another dual frame corresponding to a rotation by $45$ degrees or the $TST$-duality. After the $45$ degree rotation in the $(p,q)$ plane, the 5-brane web may still admit a 5d gauge theory description. In that case, this rotation gives another class of dual 5d theories.\\

From the 6d quiver theories of \eqref{6dquiver1} and \eqref{6dquiver2} and their Higgsed theories, we find that combinations of the above three types of the 5d dualities can give rise to various dual pairs of 5d theories which have the same 6d UV fixed point. Furthermore, it is interesting to note that we will find a sort of ``exotic'' examples after a chain of the dualities. Namely, those exotic 5d theories do not show an enhancement of the flavor symmetry to an affine type when one applies the instanton operator analysis performed in \cite{Yonekura:2015ksa}. However, this is not a contradiction since the analysis in \cite{Yonekura:2015ksa} only includes instanton states whose total instanton number is $1$. Therefore, we argue that the exotic 5d theories should also show the enhancement of the flavor symmetry to an affine type when we include instanton states with the higher instanton number.

Among the 6d theories \eqref{6dquiver1} and \eqref{6dquiver2} with their various Higgsings, we find that it is particularly interesting to focus on a case where the Type IIA brane configuration includes eight D8-branes with an $O8^-$-plane on top of each other after the Higgsing. An important point in this case is that the eight D8-branes with the $O8^-$-plane become two sets of 
four D7-branes with an $O7^-$-plane after the T-duality. The combination of 
four D7-branes with an $O7^-$-plane is special in the sense that the S-dual of this combination gives the same 7-brane configuration. Namely, we can still interpret it as four 
D7-branes and an $O7^-$-plane even after the S-duality. Due to this feature, we find that the resulting S-dualized 5d quiver may have an $SU$ gauge node with an anti-symmetric hypermultiplet or an $Sp$ gauge node at either or both ends of the quiver, depending on the splitting type of an $O7^-$-plane. Namely, these special 6d theories yield yet another class of dual 5d theories.  

As described above, we can relate various brane configurations in Type IIA string theory to Tao web diagrams in Type IIB string theory. We also note that there exists another class of Tao web diagrams that do not arise from the T-duality from a brane system in Type IIA string theory. The 5d theories from that class of the Tao web diagrams can be obtained by adding some flavors at the end nodes of the 5d linear quiver theory realizing the 5d $T_N$ theory \cite{Kim:2015jba}. We will call the 5d theory as 5d ``$T_N$ Tao'' theory. Although the Tao web diagram for a general 5d $T_N$ Tao theory may not have a direct connection to a Type IIA brane system, we will propose a 6d description of the 5d $T_N$ Tao theory by matching the global symmetry as well as the number of certain multiplets between the two theories. 

The organization of this paper is as follows. In section \ref{sec2:6dspN}, we first start the discussion of the 5d dualities that arise from the simplest 6d setup corresponding to \eqref{6dquiver1} with $n=1$. We then move onto the next simplest example that is \eqref{6dquiver2} with $n=1$ in section \ref{sec:6dSUNA}. In this case, we will see the appearance of the three types of the 5d dualities as well as an exotic example. In section \ref{sec:general}, we extend our analysis to 5d dualities for general 6d quiver theories of \eqref{6dquiver1} and \eqref{6dquiver2} and also the ones after various Higgsings. In section \ref{sec:speicalcases}, we focus on a special case where the Type IIA brane configuration has eight 
D8-branes and an $O8^-$-plane on top of each other. In this case, we find that the dual 5d theories can become another class that has not been obtained in the other sections.  
In section \ref{sec:TNTao}, we propose a 6d description of the 5d $T_N$ Tao theory and we find that it is not generically written by a Type IIA brane setup. 
In Appendix \ref{IntroTao} 
we give a brief introduction of Tao web diagrams. \\

While we are writing this paper, a related work \cite{Zafrir:2015rga} has appeared recently.

%% file: section2.tex

\section{6d \texorpdfstring{\boldmath $Sp(N)$}{Sp(N)} gauge theory with
 \texorpdfstring{\boldmath $N_f = 2N+8$}{Nf} and one tensor multiplet}\label{sec2:6dspN}
Various brane constructions in Type IIA string theory for 6d anomaly free theories were done in \cite{Hanany:1997gh,Brunner:1997gf} in terms of an O8 orientifold, D8-branes, D6-branes, and NS5-branes.  In this section, we analyze a circle compactification of a 6d $Sp(N)$ theory with $2N+8$ hypermultiplets in the fundamental representation and a tensor multiplet whose brane configuration has an $O8^-$ orientifold. The T-duality along the $S^1$ can give a 5d $SU(N+2)$ with $2N+8$ flavors as shown in 
\cite{Hayashi:2015fsa}, However, we claim that the same brane configuration 
also yields a 5d $Sp$ gauge theory with the same rank and 
the same number of flavors. There was an interesting observation \cite{Gaiotto:2015una} that the superconformal indices for seemingly different 6d theories are equivalent: one is 5d $SU(N+2)$ gauge theory with $N_f$ flavors and the Chern-Simons (CS) level $\kappa = N+4 - \frac{N_f}{2}$, and the other is 5d $Sp(N+1)$ gauge theory with $N_f$ flavors. 
Thus, it implies that the two theories are dual to each other. We here attempt to give an account for this equivalence 
from the perspective of different pattern of the resolution of  $O7^-$ orientifold planes in Type IIB string theory. As the  
decoupling limit of flavors yields another 5d theories of 5d UV fixed point, such a dual picture still holds for resulting 5d theories 
with less flavors if one decouples exactly the same flavors in the both pictures. 
Depicted in the $(p,q)$ 5-brane web diagram, the resulting 5d theories also have an S-dual picture.

Consider the 6d $\mathcal{N}=(1,0)$ $Sp(N)$ gauge theory with $N_f = 2N+8$ flavors and one tensor multiplet. The UV fixed point for this theory is known as 6d $(D_{N+4}, D_{N+4})$ minimal conformal matter theory. We compactify this theory on a circle to obtain 5d theories. Depending on the choice of 
Wilson lines, 
we either get 5d $SU(N+2)_0$ theory with 
$N_f=2N+8$ flavors and the zero CS level\footnote{From here on, we use the following short hand notation for denoting the CS level of the $SU(n)$ theory: $SU(n)_\kappa$ is an $SU(n)$ theory of the CS level $\kappa$.}, or 5d $Sp(N+1)$ theory with the same number of flavors. Although the introduction of the Wilson line breaks the global symmetry, in the symmetric phase, both have the same global symmetry as that of its 6d ``mother" theory, $SO(4N+16)$. We note that, as discussed in \cite{Hayashi:2015fsa}, the 5d global symmetry has an additional $U(1)_I$ factor 
associated to the KK modes.  
We discuss the two 5d theories using 
$(p,q)$ 5-brane web diagrams in details.

\subsection{5d \texorpdfstring{$SU(N+2)_0$}{SU(N+2)} theory with \texorpdfstring{$N_f=2N+8$}{Nf=2N+8} flavors}
\label{subsec:su(N+2)Nf}

\begin{figure}
\begin{center}
\includegraphics[width=11cm]{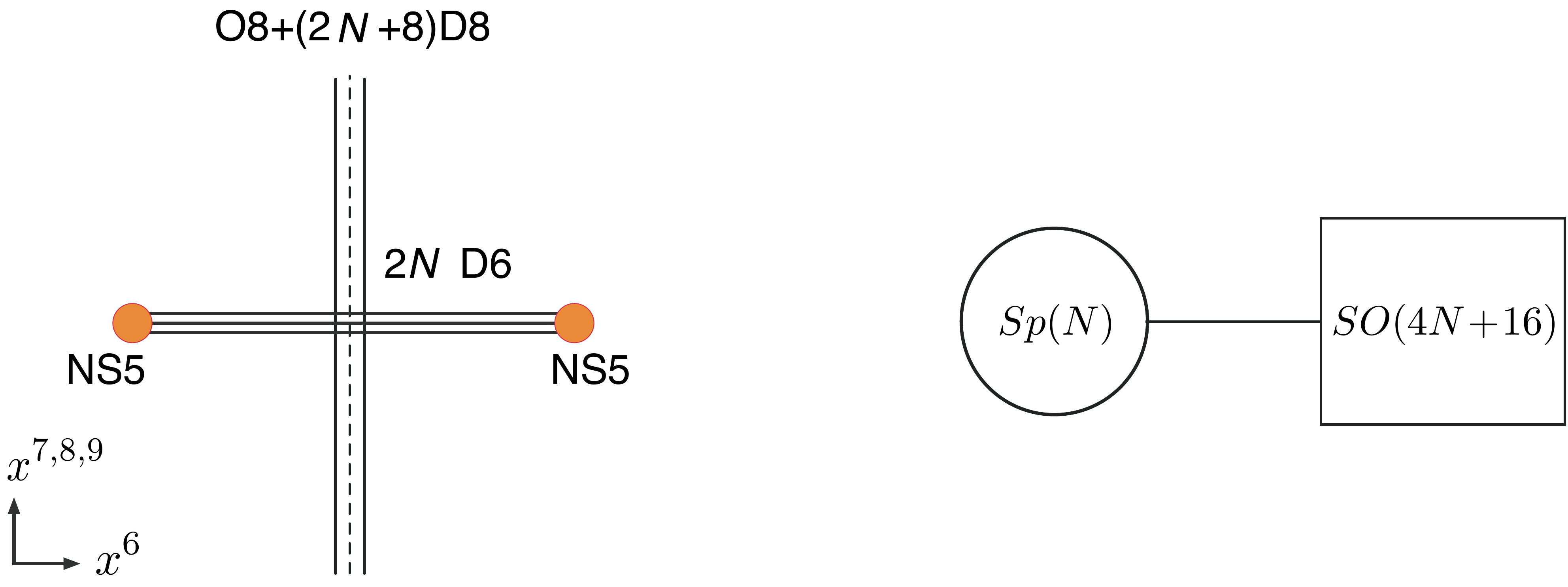}
 \end{center}
\caption{Left: Type IIA brane realization of the $(D_{N+4}, D_{N+4})$ minimal conformal matter in the tensor branch. Right: The quiver diagram of the 6d theory.} 
\label{Fig:6dspN}
\end{figure}
In Type IIA description, 6d brane configuration \cite{Hanany:1997gh, Brunner:1997gf} for $\mathcal{N}=(1,0)$ $Sp(N)$ gauge theory with $N_f = 2N+8$ flavors and a tensor multiplet consists of $2N$ D6 branes (its mirror images included) suspended between a NS5 brane and its mirror NS5 brane through an orientifold plane $O8^-$ together with $(2N+8)$ D8 branes on top of $O8^-$, as depicted in Figure \ref{Fig:6dspN}. The brane configuration in the ten-dimensional spacetime in Type IIA string theory is summarized in Table \ref{Tb:TypeIIAbrane}. 
\begin{table}[t]
\begin{center}
\begin{eqnarray}
 \begin{array}{c|c c c c  c  c |c|  c c c}
 & 0 & 1 & 2 & 3 & 4 & 5 & 6 & 7 & 8 & 9\\
 \hline
 \text{D6-brane} & \times & \times & \times & \times & \times & \times & \times &&& \\
 \text{NS5-brane} & \times & \times & \times & \times & \times & \times &  &&& \\
 \text{D8-brane} & \times & \times & \times & \times & \times &\times& &\times&\times&\times \\
  \text{O8-plane} & \times & \times & \times & \times & \times &\times & &\times&\times&\times 
 \end{array} \nonumber
 \end{eqnarray}
\end{center}
\caption{The brane configuration that yields a 6d theory on a tensor branch of a 6d SCFT in Type IIA string theory.}
\label{Tb:TypeIIAbrane}
\end{table}

\begin{figure}
\begin{center}
\includegraphics[width=15cm]{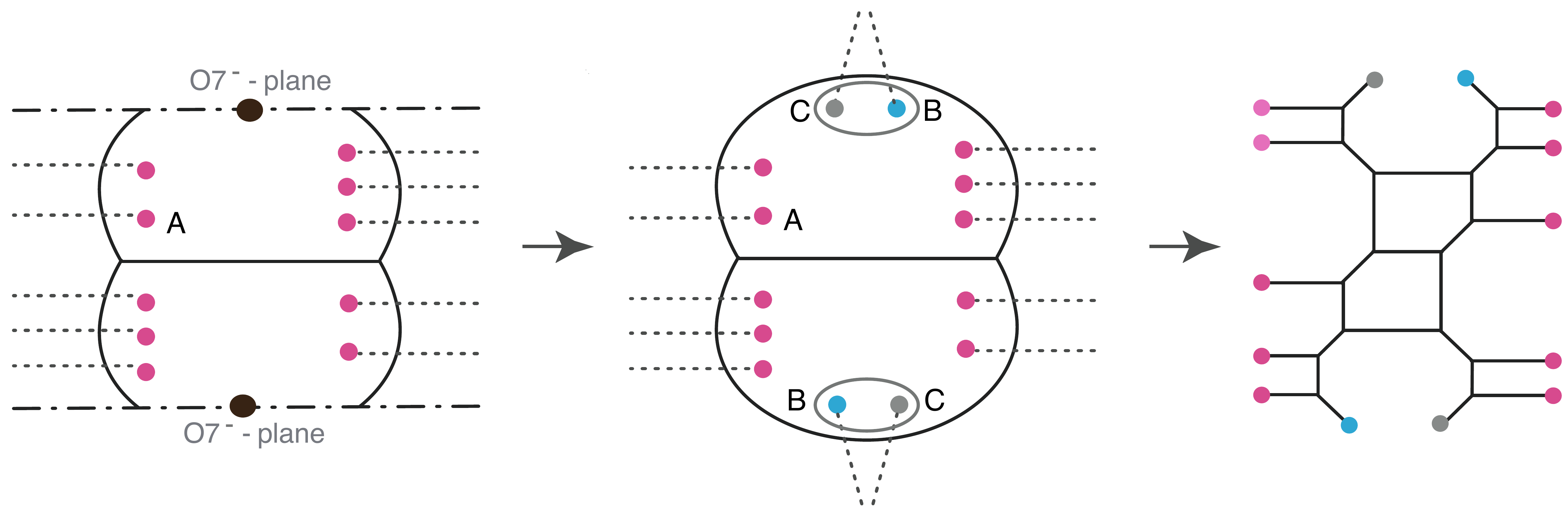}
 \end{center}
\caption{Type IIB brane descriptions for 6d ${\mathcal N}=(1,0)$ $Sp(N)$ gauge theory with one tensor multiplet and $2N+8$ flavors in the fundamental representation, which yields ${\mathcal N}=1$ $SU(N+2)$ gauge theory with the number of flavors. The horizontal direction is $x^6$ and the vertical direction is the T-dualized direction. For simplicity, $N=1$. Left: The brane configuration with two $O7^-$ planes.  A dashed line denotes the branch cut of 7-brane, while a dash-dot line denotes the mirror reflection cut of an $O7^-$ plane. 
Middle: The brane configuration with two $O7^-$ planes which are non-perturbatively resolved. Here we denote $\mathbf{A}$ for a D7 brane, $\mathbf{B}$ for a $[1,-1]$ 7-brane, and $\mathbf{C}$ for a $[1,1]$ 7-brane.  Right: Pulling out 7-branes gives rise to a web configuration for $SU(3)$ gauge theory with $N_f=10$ flavors and zero CS level.} 
\label{Fig:IIB}
\end{figure}

To have the 5d picture, we compactify one of the worldvolume direction of the NS5-brane that are parallel to the worldvolume of D6-branes, for example $x^5$, on $S^1$ with a Wilson line, and then take T-dual along the compactified direction. The resulting 5-brane configuration involves $N$ D5-branes, two NS5-branes, the $(2N+8)$ D7-branes and two $O7^-$-planes which are maximally separated apart along the T-dualized circle. The Wilson line putting $N$ D5-branes away from the $O7^-$-planes may break the 6d $Sp(N)$ gauge group to 5d $U(N)$. 
The non-perturbative effect in string theory can resolve the $O7^-$-planes into a pair of two $[p,q]$ 7-branes subject to the monodromy condition that the monodromy of the $O7^-$-plane with four D7-branes is the minus of the identity 
\cite{Sen:1996vd}. For instance, $O7^-$ can be resolved into 
a pair of $[1,1]$ and $[1,-1]$ 7-branes. 
As one resolves the $O7^-$ planes, the fundamental domain of the 5-brane web diagram that is confined by ``the mirror reflection 
lines'' is expanded to the two-dimensional plane 
and the boundary made by NS5 branes with the mirror reflection 
lines forms ``5-brane loops'' \cite{DeWolfe:1999hj}. (See Figure \ref{Fig:IIB}.)

Given such a $(p, q)$ 5-brane configuration with 7-branes and 5-brane loops, one has a 5-dimensional gauge theory description by pulling all the 7-branes to infinity, while keeping the gauge coupling finite as pulling out all the 7-branes. 5-branes are naturally generated as a result of the Hanany-Witten effect such that a 7-brane is attached to the end of an external 5-brane. Hence, $2N+8$ D7-branes give rise to semi-infinite $2N+8$ D5-branes yielding the flavor branes of the theory. Due to the charge conservation at a junction of the web diagram, 5-brane loops induce two more color D5-branes. This means that the resulting 5d gauge theory in this process has $N+2$ color branes, implying that the rank of the resulting 5d gauge theory is increased by one. For this, one can also argue that the 6d tensor multiplet contributes 
the 5d vector multiplet giving one more rank to the 5d gauge group. It is worth noting that the obtained 5d theory has zero CS level. As the CS level can be measured as the angle difference of two vertically inclined 5-branes playing a role of NS5-brane in the web configuration, the angle between the two upper external 5-branes and the angle between  the two lower external 5-branes are 
the same 
as the one between as shown in Figure \ref{Fig:IIB}, confirming the zero CS level of the 5d theory. Therefore, the resultant 5d theory is $\mathcal{N}=1$ $SU(N+2)_0$ gauge theory with $N_f=2N+8$ flavors.

\begin{figure}
\begin{center}
\includegraphics[width=15cm]{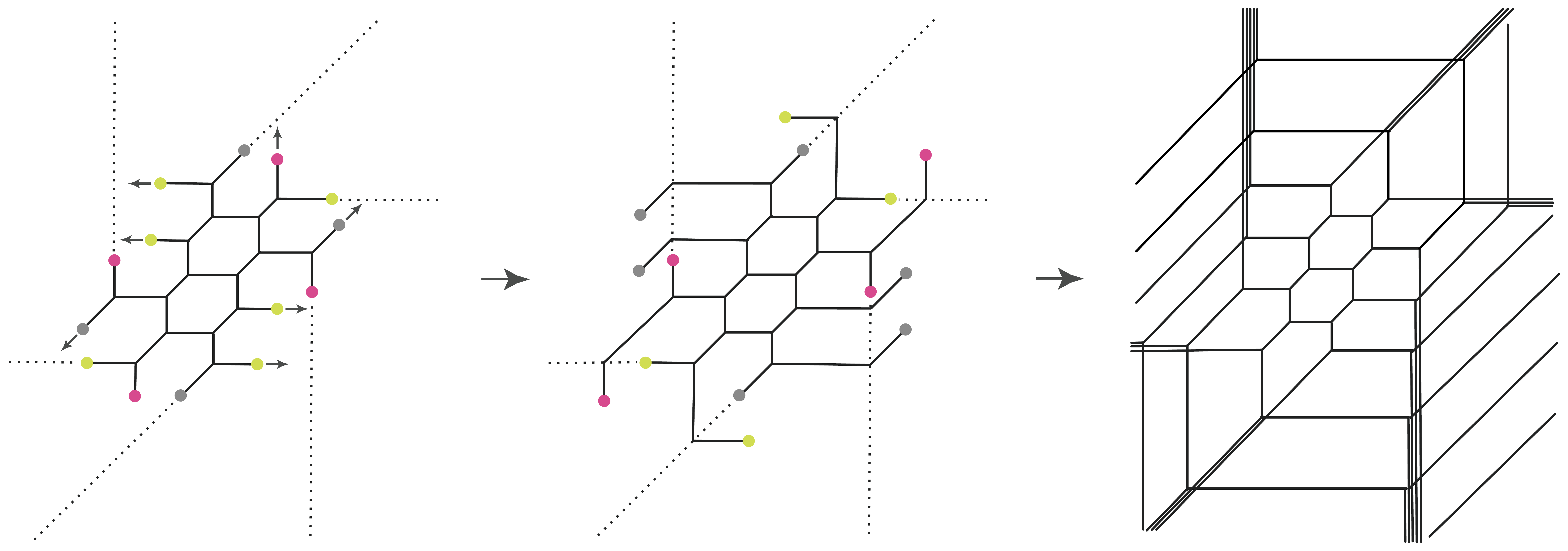}
\end{center}
\caption{Tao diagram for 5d $\mathcal{N}=1$ $SU(3)$ gauge theory with $N_f=10$ flavors which is a circle compactification of 6d $\mathcal{N}=(1,0)$ $Sp(1)$ gauge theory with one tensor multiplet and $10$ flavors. The rightmost brane configuration on Figure \ref{Fig:IIB} can be reorganized to be the leftmost web configuration given here, using 7-brane monodromies. Pulling out 7-branes passing through the branch cut of other 7-branes gives a Tao wed diagram on the rightmost.}
\label{Fig:su3tao}
\end{figure}

The web diagram for this case was already studied by the authors in \cite{Hayashi:2015fsa} and is named as a Tao web diagram. Such Tao diagram obtained by pulling all the 7-branes outside the 5-brane loops has the necessary structures to be interpreted as a circle compactification of the 6d theory. It is of the shape of an infinitely repeated spiral web
with a constant period, as given in Figure \ref{Fig:su3tao}. This period can be expressed in terms of the parameter of the 5d theory, the instanton factor, which is an expected property for a circle compactification of 6d theory. The spectrum coming from infinitely 
expanding 5-branes are naturally identified as the KK spectrum of the compactification. For a short introduction, see Appendix \ref{IntroTao} or \cite{Kim:2015jba, Hayashi:2015fsa}.

\subsection{5d \texorpdfstring{ $Sp(N+1)$}{Sp(N+1)} theory with \texorpdfstring{ $N_f=2N+8$}{Nf=2N+8} flavors}
\label{subsec:sp(N+1)Nf}
While the quantum resolution of both $O7^-$ planes leads to the 5d $SU(N+2)$ gauge theory with $2N+8$ flavors, one may resolve only one of the two $O7^-$ planes which arise from T-duality of the $O8^-$ plane in Type IIA brane picture of 6d $Sp(N)$ theory. In this process of reducing the 6d theory to 5d, the Wilson line is introduced such that all the D5-branes are located 
above the unresolved $O7^-$ plane, to retain an $Sp$ gauge group.  As discussed earlier, a resolution of the $O7^-$ plane into a pair of two 7-branes induces an extra color D5 brane when 7-branes are pulled out to infinity.
Hence, the resulting 5d theory is $\mathcal N=1$ $Sp(N+1)$ gauge theory with the same number of flavors, $N_f=2N+8$. The increase of the rank of the gauge group can be again due to the contribution of one 6d tensor multiplet.
\begin{figure}
\begin{center}
\includegraphics[width=15cm]{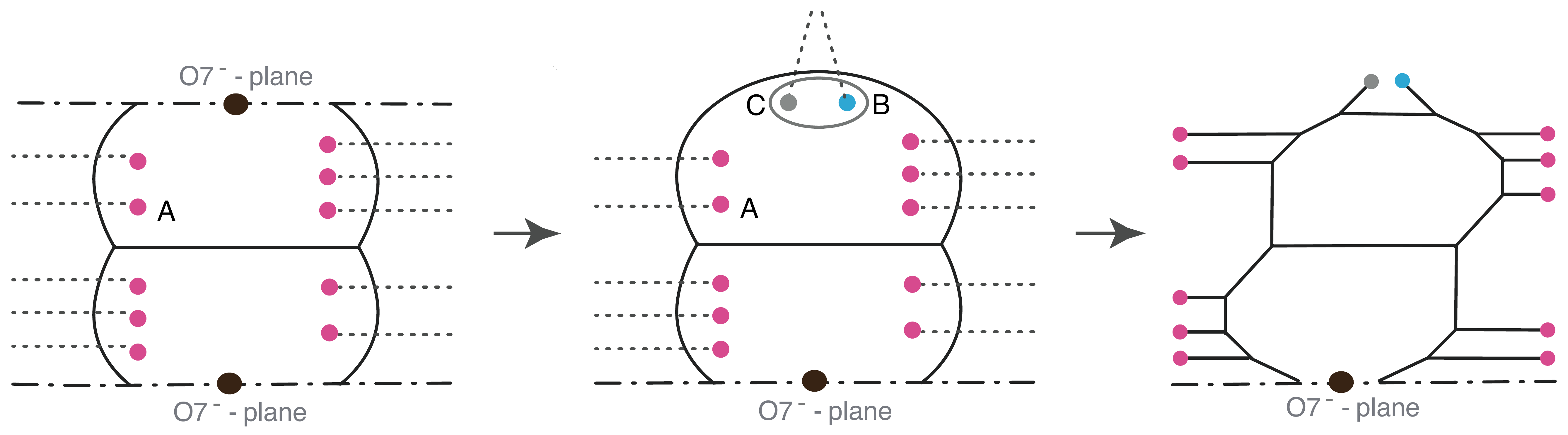}
\end{center}
\caption{Type IIB brane descriptions for 6d ${\mathcal N}=(1,0)$ $Sp(N)$ gauge theory with one tensor multiplet and $2N+8$ flavors in the fundamental representation, which yields ${\mathcal N}=1$ $Sp(N+1)$ gauge theory with the number of flavors. For simplicity, $N=1$. Left: The brane configuration with two $O7^-$ planes.  Middle: The brane configuration with only one of two $O7^-$ planes is resolved. Right: The resulting $Sp(2)$ gauge theory with $N_f=10$ flavors.
}
\label{Fig:O7ToSp}
\end{figure}
A brane setup for the 5d $Sp(N+1)$ gauge theory in the Coulomb branch is given in Figure \ref{Fig:O7ToSp}.

As $\pi_4(Sp(n))=\mathbb{Z}_2$, there is a discrete theta parameter for $Sp(n)$ gauge theory. This parameter can be interpreted as two inequivalent resolutions of the $O7^-$ plane if the $Sp(n)$ gauge theory has no flavors. When the $Sp(n)$ gauge theory has flavors, the effect of the discrete theta parameter can be absorbed by a redefinition of mass parameters and the difference is not physical \cite{Intriligator:1997pq}. Therefore, one can choose whatever splitting of the $O7^-$-plane. For example, one can resolve the $O7^-$ plane with a pair of $[1,1]$ and $[1,-1]$ 7-branes as chosen earlier, and one can also resolve it with a pair of $[2,1]$ and $[0,1]$ 7-branes.




\subsection{Equivalence between Sp an SU theories and their flavor decoupling limits}
\label{subsec:equivalence.ch2}

In the preceding subsections, we have discussed a circle compactification of 6d $\mathcal N=(1, 0)$ $Sp(N)$ gauge theory with $N_f=2N+8$ flavor and one tensor multiplet in two ways with respect to the resolution of the $O7^-$-planes as well as the Wilson lines: one of which leads to the 5d $SU(N+2)$ theory involving the resolution of two $O7^-$-planes, and the other leads to the 5d $Sp(N+1)$ theory involving the resolution of only one $O7^-$-plane. 
Both have the same number of flavors as the 6d theory, as the flavors are associated with D7-branes which come from a T-dual of D8-branes in the brane picture of the 6d theory. Note 
that global symmetry for both theories is $SO(4N+16)\times U(1)_I$ where $SO(4N+16)$ is the global symmetry of 6d theory and $U(1)_I$ comes from the circle compactification.

As both have the same UV completion as the 6d SCFT, $(D_{N+4}, D_{N+4})$ minimal conformal matter theory, and thus the same global symmetry, we claim that these two theories are dual to each other at the UV. Hence, some physical observables like index functions would be equivalent. For instance, it would be interesting to see the partition functions, or elliptic genera, of both theories do agree or not.

Interestingly, the superconformal index for the 5d $SU(N+2)$ theory and the $Sp(N+1)$ theory with less flavors $N_f\le 2N+7$ which have the 5d UV fixed point,
was computed \cite{Gaiotto:2015una}, showing the equivalence of the two indices. In the web diagram, 
this equivalence with the flavors $N_f\le 2N+7$ can be easily explained if one considers the flavor decoupling of the 5d theories from the critical flavors. Namely, one can take the mass decoupling limit by deforming the mass of a flavor to an extreme value of either $\infty$ or $-\infty$. 

We note that in order to hold the 5d dualities for the less flavor cases, the CS level of $SU(N+2)$ theory should be either maximum or minimum. This comes from the fact that we need to decouple the flavors exactly in the same way between the 5d $SU(N+2)$ gauge theory and the 5d $Sp(N+1)$ gauge theory. For example, from the Figure \ref{Fig:O7ToSp}, we can take the mass of the flavors only to $+\infty$ for the $Sp(N+2)$ gauge theory. Therefore, we should also take the same flavor decoupling limit for the 5d $SU(N+2)$ gauge theory. 

\subsection{S-dualities}
\label{subsec:Sdual.ch2}
In the $(p, q)$ 5-brane web diagram, the S-duality structure of the theory is manifest as one rotates the $(p, q)$ 5-brane web 
by 90 degrees, namely, D5-branes become NS5-branes and NS5-branes become D5-branes. A simple example would be the S-duality of the $SU(3)$ theory which yields the quiver theory of $SU(2)\times SU(2)$. 
\begin{figure}
\begin{center}
\includegraphics[width=13cm]{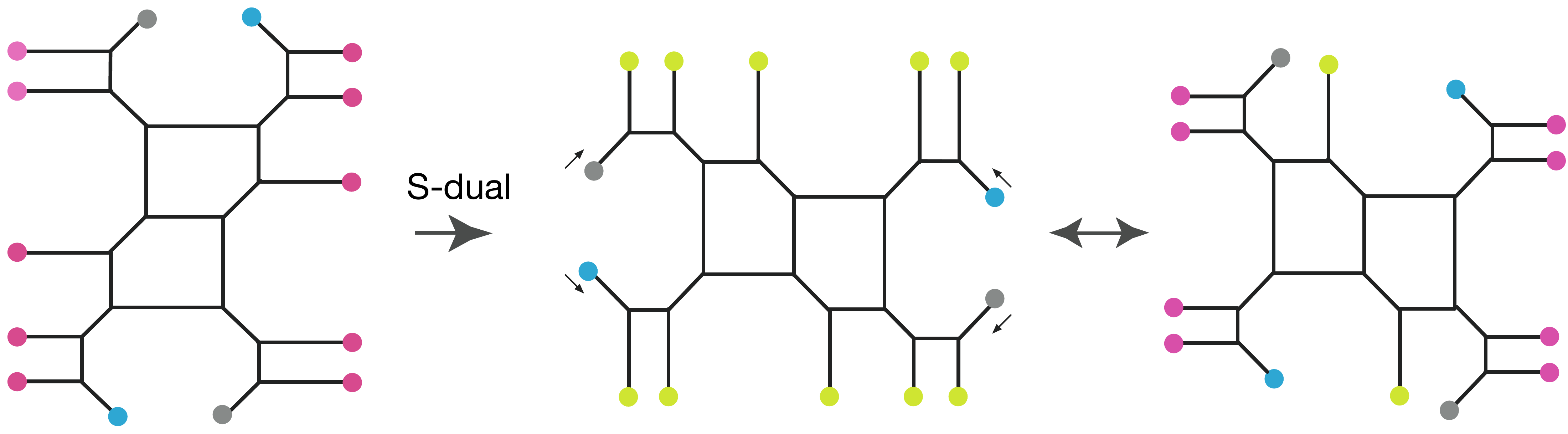}
\end{center}
\caption{The S-dual of $SU(3)$ theory with $N_f=10$ flavor is $SU(2)\times SU(2)$ quiver theory with $N_f=4$ flavors to each $SU(2)$ theory of the quiver.
}
\label{Fig:su2su2}
\end{figure}
For instance, the brane configuration on the right of Figure \ref{Fig:IIB} describes a 5d $SU(3)$ theory with $N_f=10$ flavors, and then its S-dual theory is the $[4]-SU(2)\times SU(2)-[4]$ quiver theory which has four flavors\footnote{The middle configuration of Figure \ref{Fig:su2su2} has two ``$SU(1)$''s linked to each $SU(2)$ instead of four flavors. Such $SU(1)$ has the instanton factor which can be used to express the period of the Tao web diagram. On the other hand, using 7-brane monodromy analysis, it can be shown that such $SU(1)$ factor together with a bi-fundamental hypermultiplet connecting between the $SU(1)$ and $SU(2)$ also represents two flavors.}
(denoted in the square bracket ``$[~]$'') 
on each gauge group as in Figure \ref{Fig:su2su2}. In such Tao diagrams of Figure \ref{Fig:IIB}, as discussed, the constant period of spiral shape is expressed in terms of the instanton factor of the gauge theory. It holds for the S-dual picture, the constant period is expressed in terms of instantons factors of the quiver theory \cite{Hayashi:2015fsa}. This S-duality generalizes to higher rank cases so that the S-dual of the 5d $SU(N+2)$ gauge theory with $N_f = 2N+8$ flavors is given by the $SU(2)$ quiver theory
\begin{align}
[4] -\underbrace{SU(2) - \cdots - SU(2)}_\text{$N+1$ nodes} - [4].
\end{align}
The period for these Tao diagram is expressed in terms of a product of the instanton factors of each gauge node. 

We summarize the 5d theories obtained by compactifying the 6d $Sp(N)$ gauge theory with $N_f = 2N+8$ and one tensor multiplet, on a circle:
\begin{enumerate}[(i)]
\item $SU(N+2)-[\,2N+8\,]$,  ~$SU(N+2)$ gauge theory with $N_f = 2N+8$ flavors,
\item $Sp(N+1)-[\,2N+8\,]$, ~$Sp(N+1)$ gauge theory with $N_f = 2N+8$ flavors,
\item $[4] - SU(2) - \cdots - SU(2) - [4]$, a quiver theory with $N+1$ $SU(2)$ gauge nodes and 4 flavors are attached to each boundary node. 
\end{enumerate}
The (i) and (ii) are obtained by resolving either both two $O7^-$ planes or only one $O7^-$ plane when reducing to 5d, and (iii) is S-dual of (i). Therefore, (iii) is also (S-)dual to (ii).

\paragraph{Decoupling flavors}
As discussed in \cite{Hayashi:2015fsa}, one can decouple flavors of a given Tao diagram, which yields 5d supersymmetric gauge theory which has the 5d UV fixed point. In $Sp(N+1)$ or $SU(N+2)$ gauge theories, we can decouple the same number of flavors. The same idea applies to the $SU(2)\times SU(2)$ quiver theory. 
The resulting flavor 
decoupled theories from $SU(N+1)$, $Sp(N+1)$, and the $SU(2)\times SU(2)$ quiver theories are all dual to each other, and are expected to have the same 5d UV fixed point.



\begin{figure}
\begin{center}
\includegraphics[width=13cm]{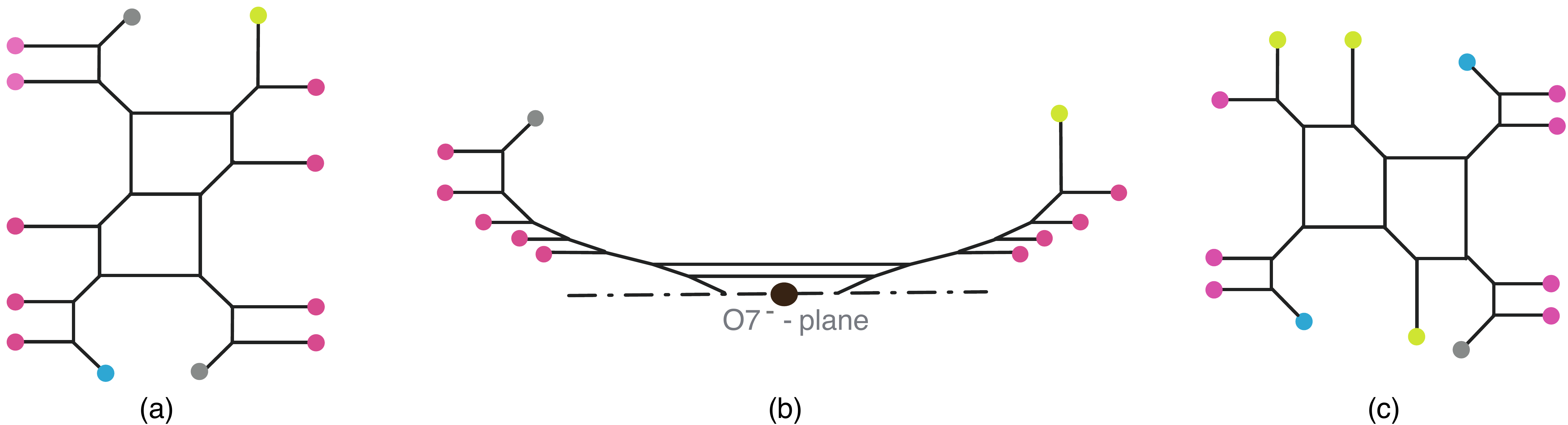}
\end{center}
\caption{Mass decoupling limit. (a) $SU(3)_\frac{1}{2}$ theory with $N_f=9$ flavors and the CS level $\frac12$;  (b) $Sp(2)$ theory with $N_f=9$ flavors and $k=0$; (c) $[3]-SU(2)-SU(2)-[4]$ quiver theory.
}
\label{Fig:massdef9F}
\end{figure}

%% file: section3.tex
\section{6d \texorpdfstring{$SU(N)$}{SU} gauge theory with \texorpdfstring{$N_f = N+8$}{Nf} and \texorpdfstring{$N_a = 1$}{Na}}
\label{sec:6dSUNA}

In this section, we consider six-dimensional $SU(N)$ gauge theory with $N_f = N+8$ flavors and $N_a = 1$ hypermultiplet in antisymmetric 
representation together with one tensor multiplet. 
This theory has $SU(N+8) \times U(1)$ anomaly free global symmetry.

When $N$ is small, there can be the enhancement. 
When $N=3$, the anti-symmetric representation is equivalent to the (anti-)fundamental representation. 
Therefore, this theory is interpreted as SU(3) gauge theory with 12 flavors.
In this case, the global symmetry $SU(11) \times U(1)$ actually enhances to $SU(12)$.
When $N=4$, anti-symmetric representation is the real representation.
Therefore, we can consider the $SU(2)$ global symmetry acting on two half hypermultiplets.
In this case, the global symmetry $SU(12) \times U(1)$ actually enhances to $SU(12) \times SU(2)$. The discussion below is similar to that in section \ref{sec2:6dspN}. Depending on resolving both or only one $O7^-$-plane(s), we obtain two different 5d descriptions. We also discuss various 5d dual descriptions corresponding to this 6d theory.

\subsection{5d \texorpdfstring{$SU\times SU$}{SUSU} gauge theory with flavors}\label{subsec:SUSU}

\begin{figure}
\centering
\begin{minipage}{0.4\hsize}
\includegraphics[width=5cm]{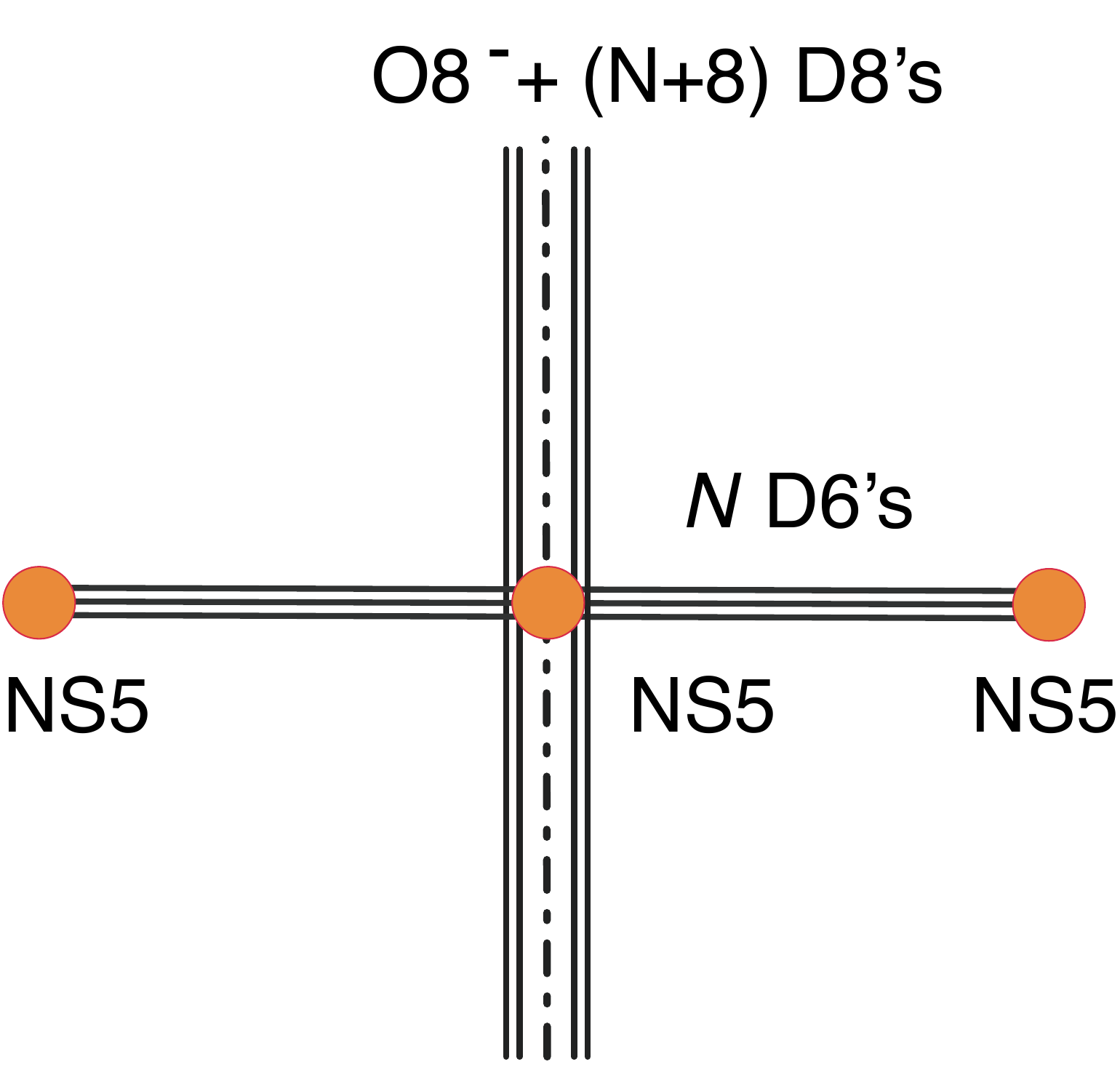}
\end{minipage}
\begin{minipage}{0.4\hsize}
\includegraphics[width=5cm]{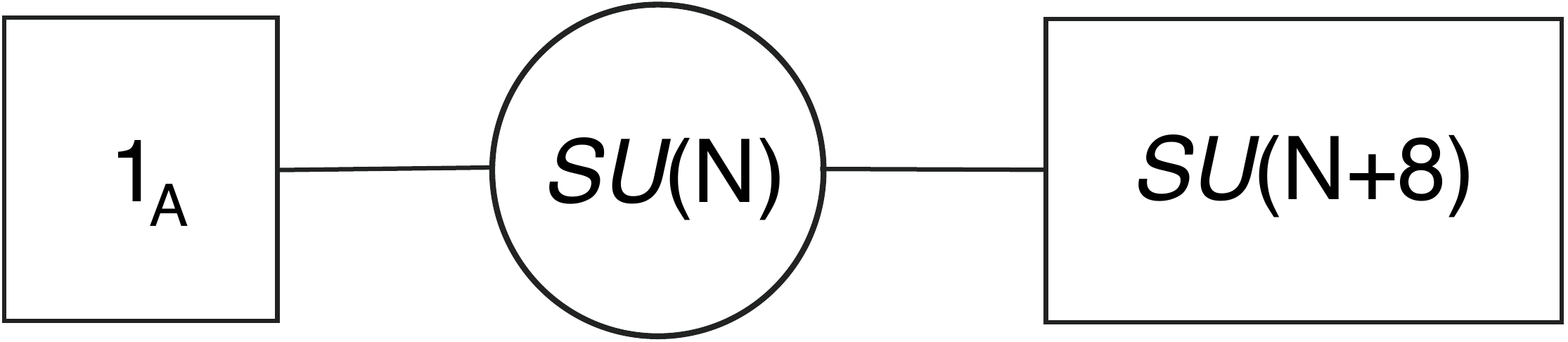}
\end{minipage}
\caption{Left: Type IIA brane set up for 6d \texorpdfstring{$SU(N)$}{SU} gauge theory with \texorpdfstring{$N_f = N+8$}{Nf} and \texorpdfstring{$N_a = 1$}{Na}. Right: Corresponding quiver diagram.} 
\label{Fig:6dSUa}
\end{figure}

6d $SU(N)$ gauge theory with $N_f = N+8$, $N_a = 1$ and with one tensor multiplet
is realized by the brane set up depicted in Figure \ref{Fig:6dSUa}.
Compared to Figure \ref{Fig:6dspN}, there is an additional NS5-brane on top of the $O8^-$ plane.
Moreover, odd number of D6-branes are also allowed in this case.

\begin{figure}
\centering
\includegraphics[width=7cm]{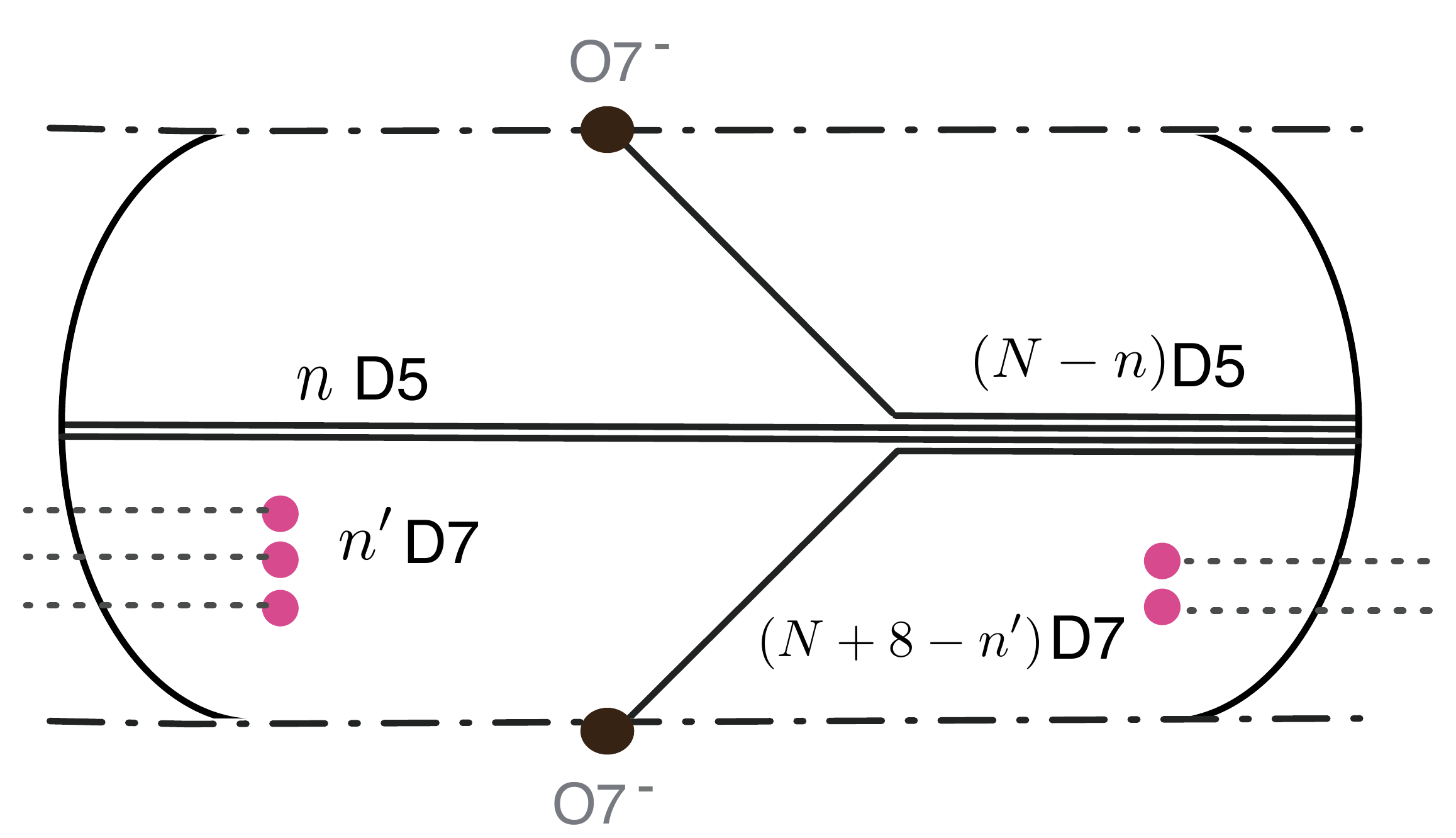}
\caption{A T-dual description of Figure \ref{Fig:6dSUa}.} 
\label{Fig:5SUSU}
\end{figure}

We compactify one of the directions which NS5-brane extends and take T-duality along this direction.
Then, we obtain the IIB brane setup depicted in Figure \ref{Fig:5SUSU},
where we chose the fundamental region to be the half of the compactified $S^1$ direction.
Compared to \ref{Fig:IIB}, we have additional $(p,q)$ 5-brane connecting two $O7^-$ planes,  
where $n$ D5-branes are placed on the left hand side and $(N-n)$ D5-branes are place on the right hand side of this 5-brane.
We note that this ``distribution'' labeled by $n$ depends on the Wilson line of the $SU(N)$ gauge group which we introduce when we compactify on $S^1$.
Accordingly, the charge of the 5-branes attached to the $O7^-$ planes depends on this distribution. Likewise, the D7-branes which are T-dual of the D8-branes are also distributed in analogous way into $n'$ and $N+8-n'$.

\begin{figure}
\centering
\includegraphics[width=8cm]{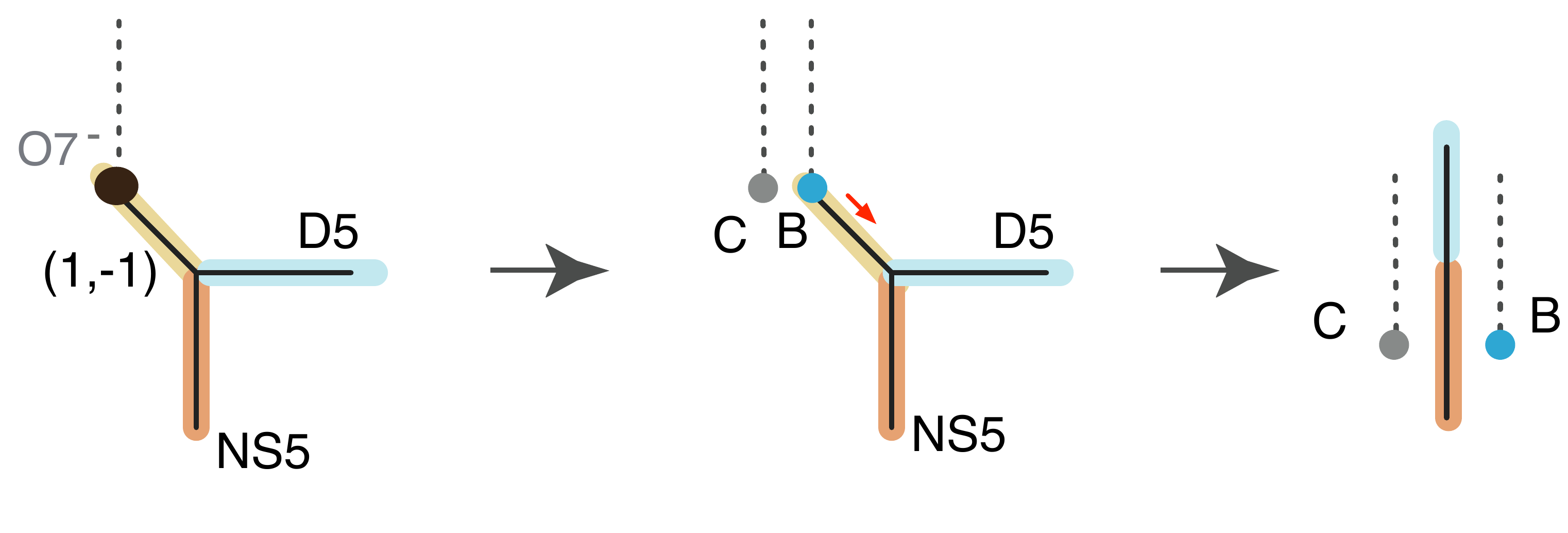}
\caption{Resolution of the $O7^-$ plane attached to $(1,-1)$ 5-brane. A $[1,-1]$ 7-brane is denoted by ${\bf B}$ and a $[1,1]$ 7-brane is denoted by ${\bf C}$.} 
\label{O7dec}
\end{figure}

Here, we would like to consider the quantum resolution for one of the $O7^-$ planes based on \cite{Sen:1996vd}.
The decomposition in this case is, however, more non-trivial than that on discussed in section \ref{sec2:6dspN}, because the $O7^-$ planes here are attached to the 5-branes. As discussed in \cite{Bergman:2015dpa}, one needs to properly choose
the charge of the resolved two 7-branes in such a way that one of the 7-branes should be attached to 
the 5-brane that is attached to the $O7^-$-plane before the resolution. 
For example, when an $O7^-$ plane is attached to $(1,-1)$ 5-brane as depicted on the left of Figure \ref{O7dec}, the $O7^-$ plane 
can be decomposed into ${\bf B}=[1,-1]$ 7-brane and ${\bf C}=[1,1]$ 7-brane so that 
 ${\bf B}$ 7-brane is attached to the $(1,-1)$ 5-brane.\footnote{It is also possible for one to resolve the $O7^-$-plane into $[3,-1]$ and $[1,-1]$ 7-branes as the total monodromy is the same as that of an $O7^-$-plane. In this case, however, one can rearrange the 7-branes so as to yield the same configuration with $\bf B$ and $\bf C$ 7-branes.}
Once the $O7^-$-plane is resolved, 7-branes induce new configurations. Imagine, for example, a 5-brane junction composed of an NS5-brane, a D5-brane and a $(1,-1)$ 5-brane which is connected to ${\bf B}$ 7-brane as in the middle of Figure \ref{O7dec}.
By moving ${\bf B}$ 7-brane along the direction of the $(1,-1)$ 5-brane using Hanany-Witten transition,
one can transform the configuration into a brane configuration where the ${\bf B}$ 7-brane becomes fee so that it is not attached to the 5-brane. As a result, ${\bf B}$ and ${\bf C}$ 7-branes are separated by the NS5-brane as depicted on the right of Figure \ref{O7dec}. This kind of transforming procedure is useful in obtaining various five-dimensional (5d) field theory descriptions after resolving $O7^-$-planes.

%

\begin{figure}
\centering
\includegraphics[width=14cm]{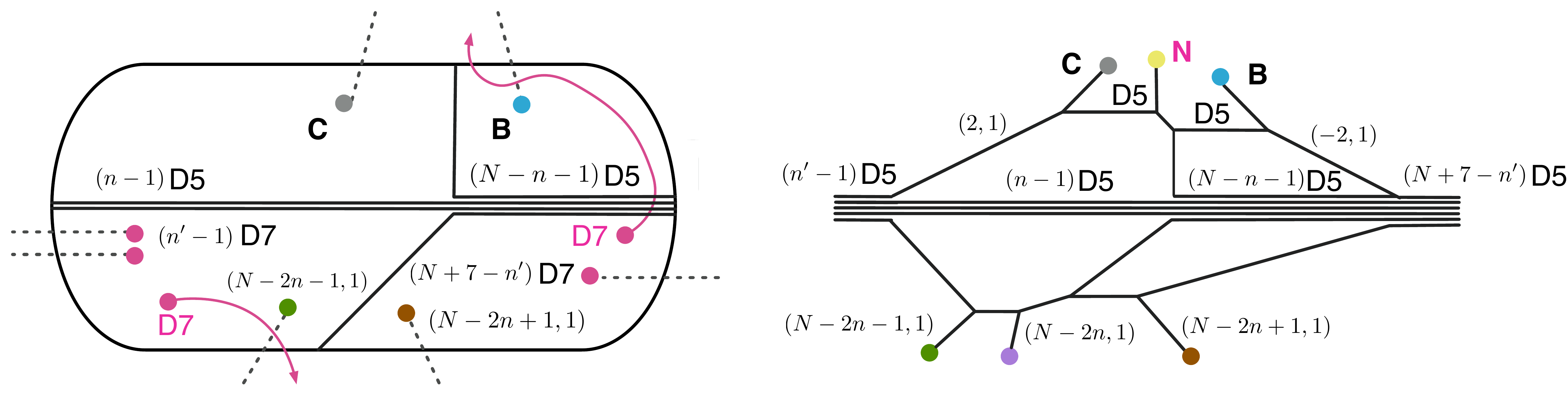}
\caption{Left: The two $O7^-$ planes in Figure \ref{Fig:5SUSU} is resolved. Right: Web-diagram obtained by moving all the 7-branes outside.} 
\label{Fig:5dquiv}
\end{figure}

As discussed in section \ref{sec2:6dspN}, the quantum resolution of $O7^-$-planes generates new 5-branes to make a closed 5-brane loop. 
The resulting brane configuration is depicted on the left side of Figure \ref{Fig:5dquiv}, where two $O7^-$-planes are resolved into two 7-branes and one of which  is attached to a 5-brane is moved along the direction of the charge of the 5-brane to be a free 7-brane as described. To obtain a 5d gauge theory description for this brane configuration, one can move around one of the D7-branes passing through D5-branes as well as 7-branes toward outside the 5-brane loop. (see the arrows attached to D7-branes in Figure \ref{Fig:5dquiv}.) 
Then it leads to a 5-brane web diagram depicted on the right side of Figure \ref{Fig:5dquiv}.
We note that a 5d field theory description is possible only when the newly generated 5-branes are D5-brane, provided that the two distribution parameters, the number of the color D5-branes $n$ and the number of the flavor D7-branes $n'$, are constrained by $n'=3n+4-N$. 

With a suitable Wilson line for the $SU(N_f)$ flavor group, the brane configuration on the right of Figure \ref{Fig:5dquiv} gives rise to the 5d quiver gauge theory
\begin{eqnarray}
[3n+3-N] - SU(n+1)_0 - SU(N+1-n)_0 - [2N+3-3n],
\label{eq:5dSUSU}
\end{eqnarray}
where as the distribution parameter, the integer $n$ should be chosen in such a way that all the flavors and ranks of the gauge groups of this theory should be positive. With such $n$, we claim that any resulting 5d quiver gauge theory is a possible 5d description for 6d $SU(N)$ gauge theory with $N_f = N+8$, $N_a = 1$ and with one tensor multiplet compactified on $S^1$. 
In other words, the theories with different $n$ should have the identical 6d UV fixed point. We denote this equivalence as ``{\it distribution duality}.''
We will discuss this distribution duality more in detail in section \ref{subsec:distribution}.

%

\subsection{5d \texorpdfstring{$SU(N+1)$}{SU} gauge theory with \texorpdfstring{$N_f = N+7$}{Nf} and \texorpdfstring{$N_a = 1$}{Na}}\label{5dNa=1}
\begin{figure}
\centering
\includegraphics[width=14cm]{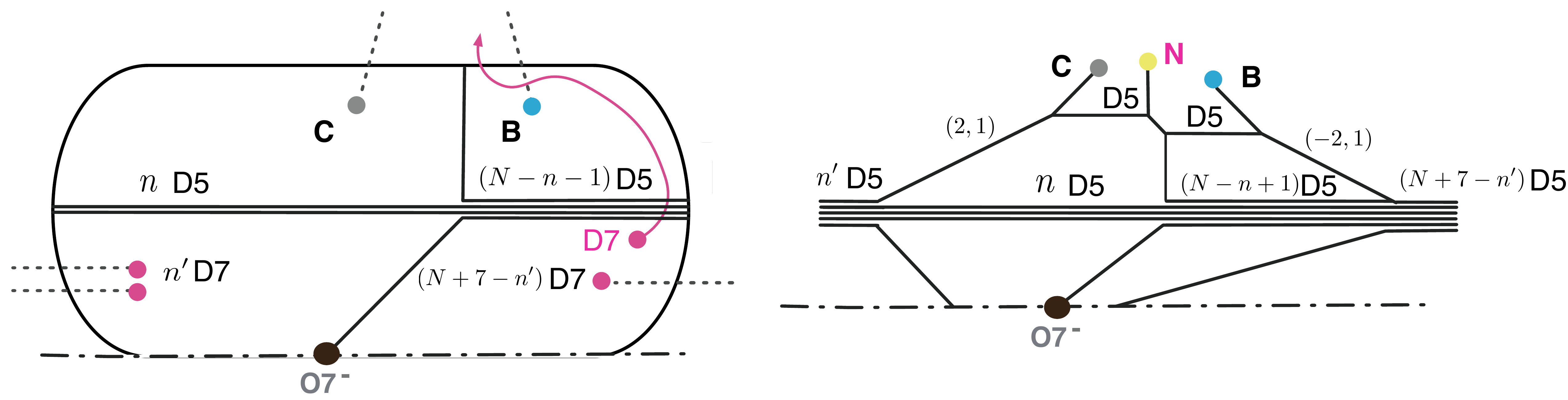}
 \caption{5d $SU(N+1)_0$ gauge theory with $N_f = N+7$ and $N_a = 1$} 
\label{Fig:5dSUa}
\end{figure}

Analogous to section \ref{subsec:sp(N+1)Nf}, we can decompose only one $O7^-$-plane out of the two $O7^-$-planes in Figure \ref{Fig:5SUSU}. Suppose we resolve the upper $O7^-$-plane from \ref{Fig:5SUSU}, then we get the configuration on the left of Figure \ref{Fig:5dSUa}. As discussed in the previous subsection, we can take a D7-brane through to locate between ${\bf B}$ and ${\bf C}$ 7-branes, which convert the $[p,q]$ charge of the D7-brane to be ${\bf N}=[0,1]$. This leads to the configuration on the right of  Figure \ref{Fig:5dSUa}. 
This configuration corresponds to adding the flavor branes to the one discussed in \cite{Bergman:2015dpa}.

The resulting 5d description is a 5d $SU(N+1)_0$ gauge theory with $N_f = N+7$ flavor and $N_a = 1$ antisymmetric hypermultiplet coming from the 5-brane connected to unresolved $O7^-$-plane. Notice that the number of fundamental hypermultiplet is reduced by one, since we used up a D7-brane to convert to a $[0,1]$ 7-brane which is essential to have a brane configuration with an antisymmetric hypermultiplet.
Therefore, we conclude that 
6d $SU(N)$ gauge theory with $N_f = N+8$, $N_a = 1$ and with one tensor multiplet compactified on $S^1$ is also described by 5d $SU(N+1)_0$ gauge theory with $N_f = N+7$ flavors and $N_a = 1$.

\subsection{Equivalence of the two 5d descriptions and their flavor decoupling limits}\label{subsec:mass dec}

In the previous subsections, we discussed two different 5d descriptions for the 6d $SU(N)$ gauge theory with $N_f = N+8$, $N_a = 1$ and one tensor multiplet, which are summarized as 
\begin{enumerate}[(i)]
\item 5d $[3n+3-N] - SU(n+1)_0 - SU(N+1-n)_0 - [2N+3-3n]$;
\item 5d $SU(N+1)_0$ gauge theory with $N_f = N+7$ and $N_a = 1$.
\end{enumerate}
These two descriptions are dual in the sense that they have the same 6d UV fixed point.
Furthermore, by considering the flavor decoupling limit by taking some of the mass parameters to be infinity,  we will again obtain the dual theories, which then have again the identical 5d fixed point.

For instance, in Figure \ref{Fig:5dquiv}, let us take $n_L$ flavors out of the left flavors of the quiver theory (i), $[3n+3-N]$, and $n_R$ flavors out of the right flavors of (i), $[2N+3-3n]$. By giving heavy masses to $n_L$ and $n_R$ flavors, corresponding to moving these $n_L$ and $n_R$ flavor branes upward in Figure \ref{Fig:5dquiv}, we can decouple $n_L$ and $n_R$ flavors from the brane configuration, which reduces the left and right flavors to be $3n+3-N-n_L$ and $2N+3-3n-n_R$, respectively. This also alter the CS levels of $SU(n+1)\times SU(N+1-n)$ to be shifted by $n_L$ and $n_R$ respectively. After the flavor decoupling, we get  
\begin{align}\label{i-1}
{\rm 5d}~~ [3n+3-N-n_L] - SU(n+1)_{n_L} - SU(N+1-n)_{n_R} - [2N+3-3n-n_R].
\end{align}

In a similar fashion, we can decouple some flavors of (ii). 
Say, we give the positive infinite masses to $n_L$ flavors from the left $n'=3n+4-N$ flavors and $n_R$ flavors from the right $N+7-n'$ flavors of the brane configuration on the right of Figure \ref{Fig:5dSUa}. This also shifts the CS level of $SU(N+1)$ to be $n_L-n_R$ and hence reduces the theory to be 
\begin{align}\label{ii-1}
 {\rm 5d}~~ SU(N+1)_{n_L-n_R}\quad {\rm with}~ N_f = N+7-(n_L+n_R)~{\rm and}~ N_a = 1. 
\end{align}
As the way we decouple the flavors from (i) and (ii) are same, we expect that 
two 5d theories \eqref{i-1} and \eqref{ii-1} obtained from the flavor decoupling should have the identical 5d UV fixed point, giving rise to another non-trivial 5d duality.

\subsection{S-dualities and mass parameters}\label{subsec:SUA-Sdual}
\begin{figure}
\centering
\begin{minipage}{0.3\hsize}
\includegraphics[width=5cm]{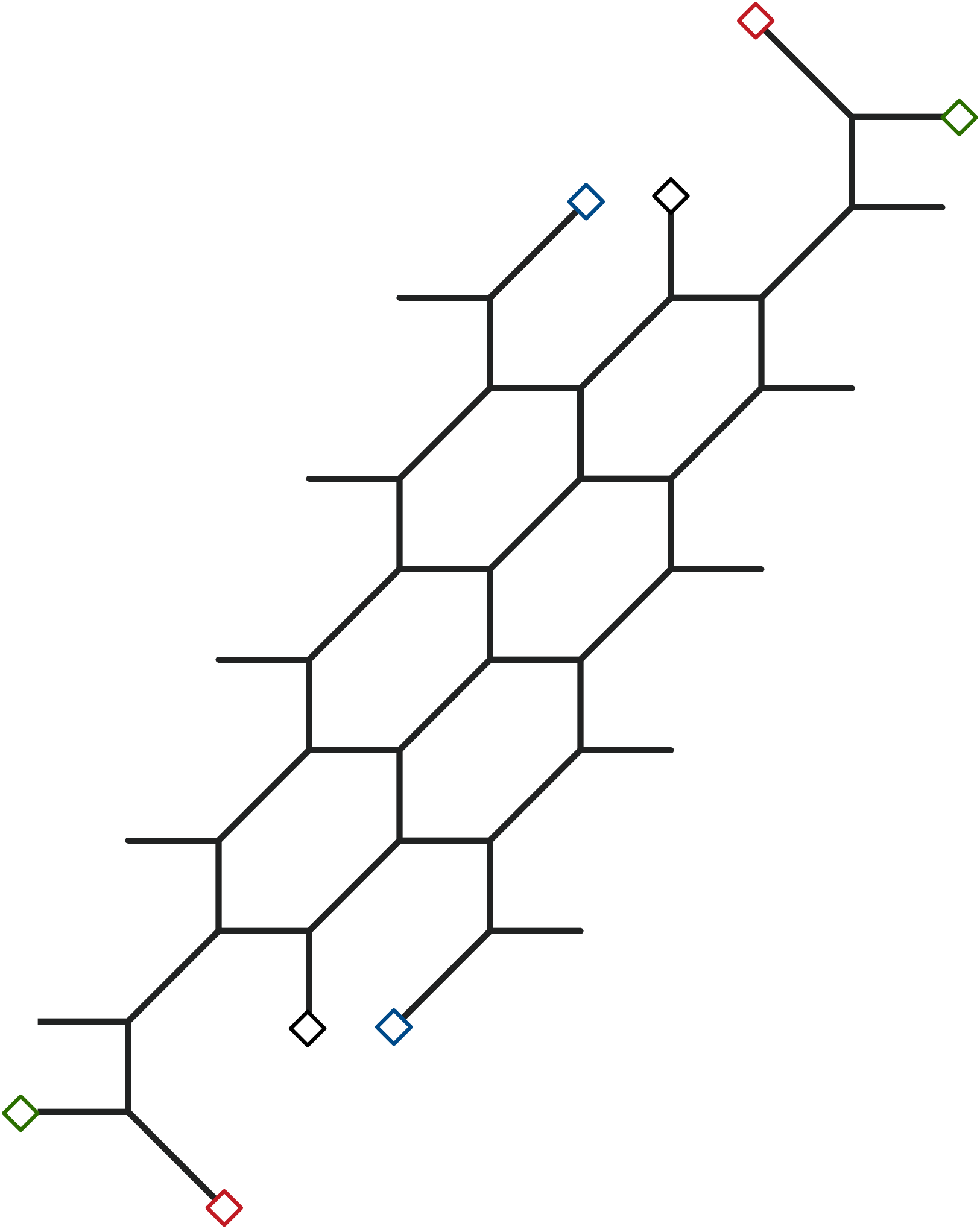}
\end{minipage}
\begin{minipage}{0.36\hsize}
\includegraphics[width=6cm]{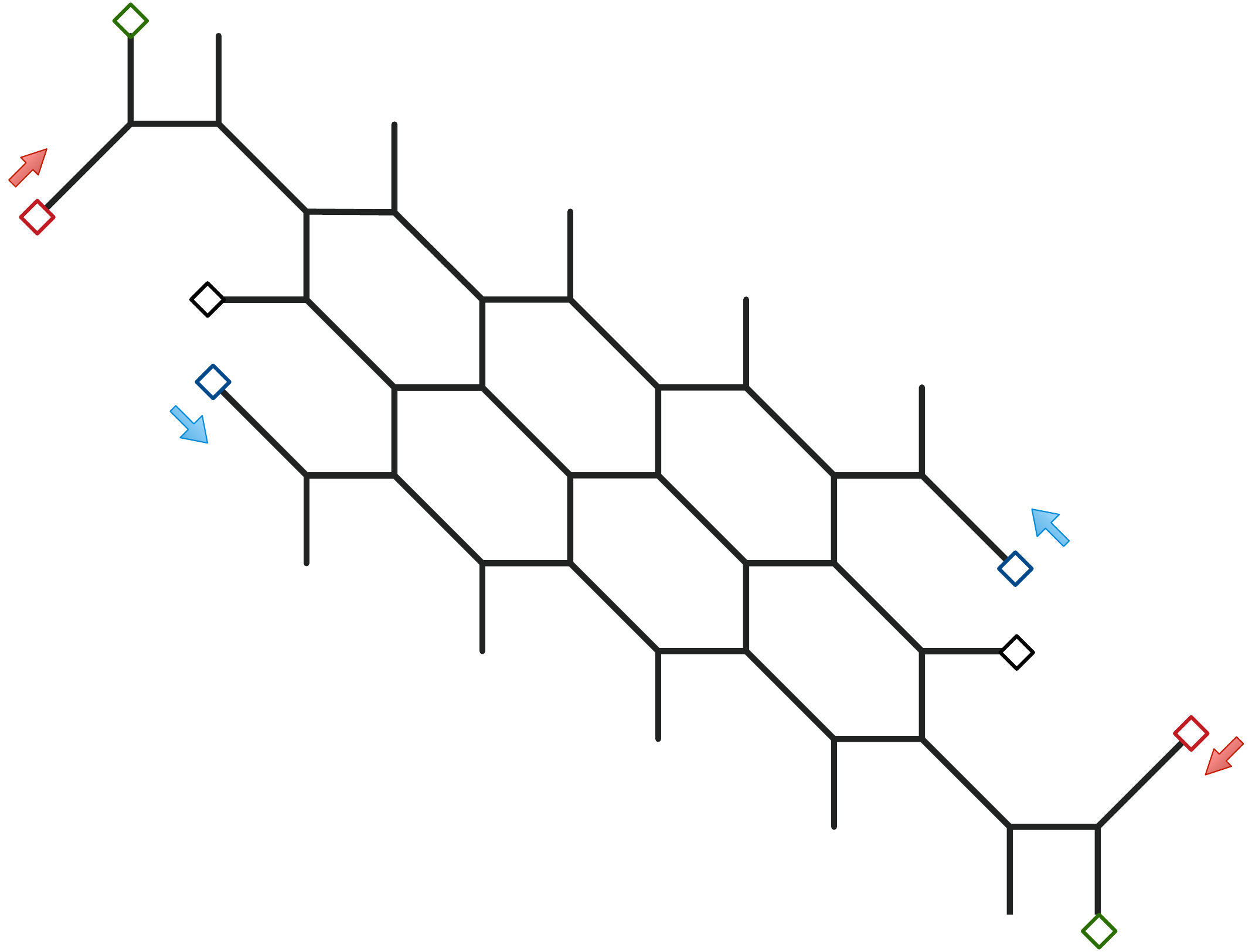}
\end{minipage}
\begin{minipage}{0.30\hsize}
\includegraphics[width=5cm]{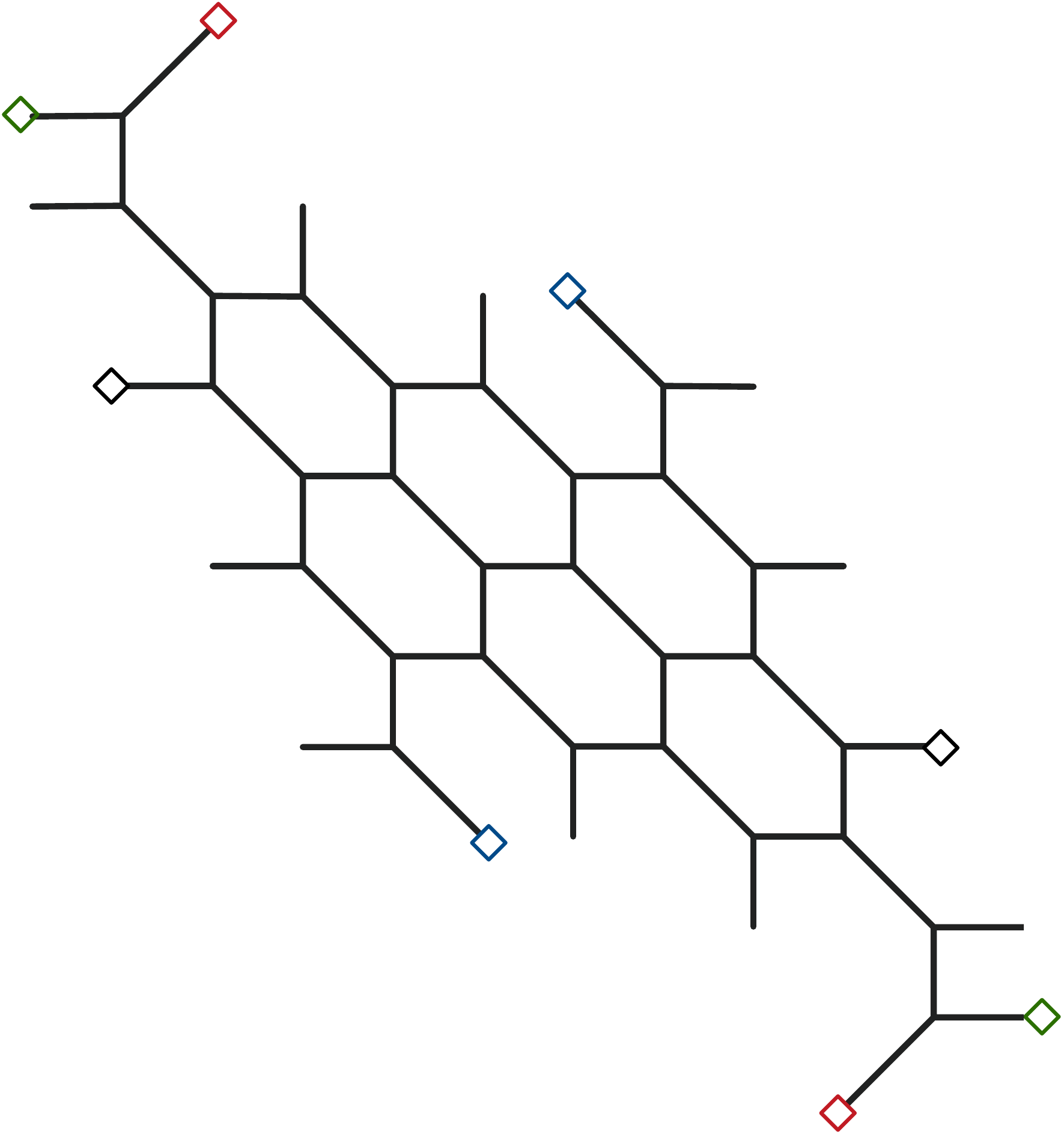}
\end{minipage}
\caption{S-transformation of the 5d quiver gauge theory $[6] - SU(4)_{0} - SU(4)_{0} - [6]$.}
\label{Fig:Sdual90}
\end{figure}

By using S-duality, we can obtain more 5d descriptions which have the identical UV fixed point.
In this subsection, we consider S-dual description for the quiver theory discussed in section \ref{subsec:SUSU}. We discuss the dual descriptions with a specific example, and generalization is then straightforward. 
Take $N=6$ and $m=3$, then the 5d quiver gauge theory \eqref{eq:5dSUSU} becomes
\begin{eqnarray}\label{6446quiver}
[6] - SU(4)_{0} - SU(4)_{0} - [6].
\end{eqnarray}
The corresponding web diagram is depicted on the left diagram of Figure \ref{Fig:Sdual90}.

Parallel to section \ref{subsec:Sdual.ch2}, we first consider the S-duality transformation 
exchanging D5-branes and NS5-branes to each other,
which corresponds to a 90-degree rotation of the diagram.
We denote this type of S-duality as ``S-transformation''.
Acting the S-transformation to the left diagram on Figure \ref{Fig:Sdual90},
we obtain the middle diagram of Figure \ref{Fig:Sdual90}. In this middle diagram, we move $[1,1]$ 7-branes and $[1,-1]$ 7-branes along with the Hanany-Witten transition, then
we obtain the following 5d quiver gauge theory
\begin{eqnarray}
[5] - SU(3)_{0} - SU(3)_{0} - SU(3)_{0} - [5],
\label{333}
\end{eqnarray}
where the corresponding web diagram depicted on the right of Figure \ref{Fig:Sdual90}.

%

\begin{figure}
\centering
\begin{minipage}{0.32\hsize}
\includegraphics[width=5cm]{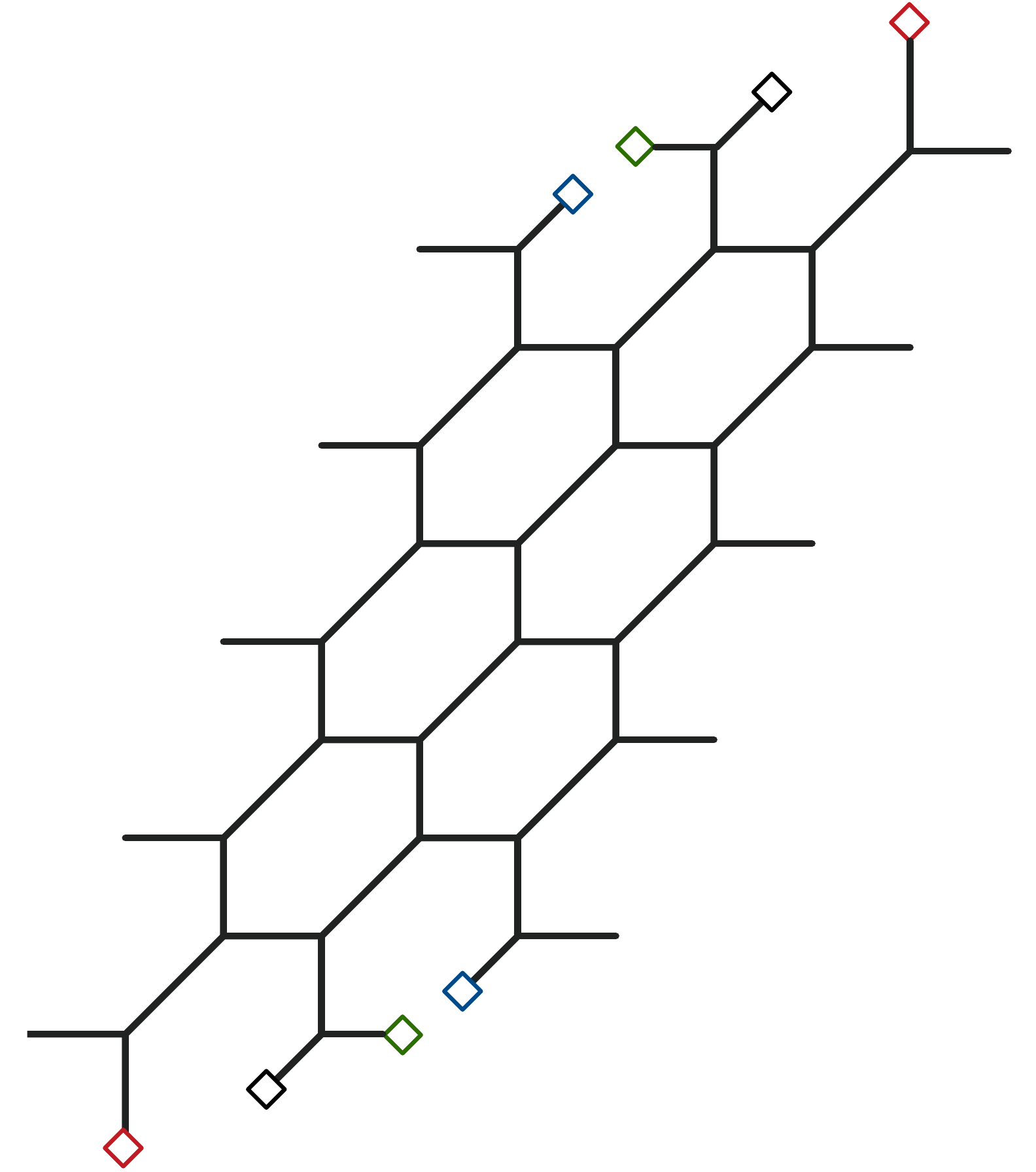}
\end{minipage}
\begin{minipage}{0.32\hsize}
\includegraphics[width=5cm]{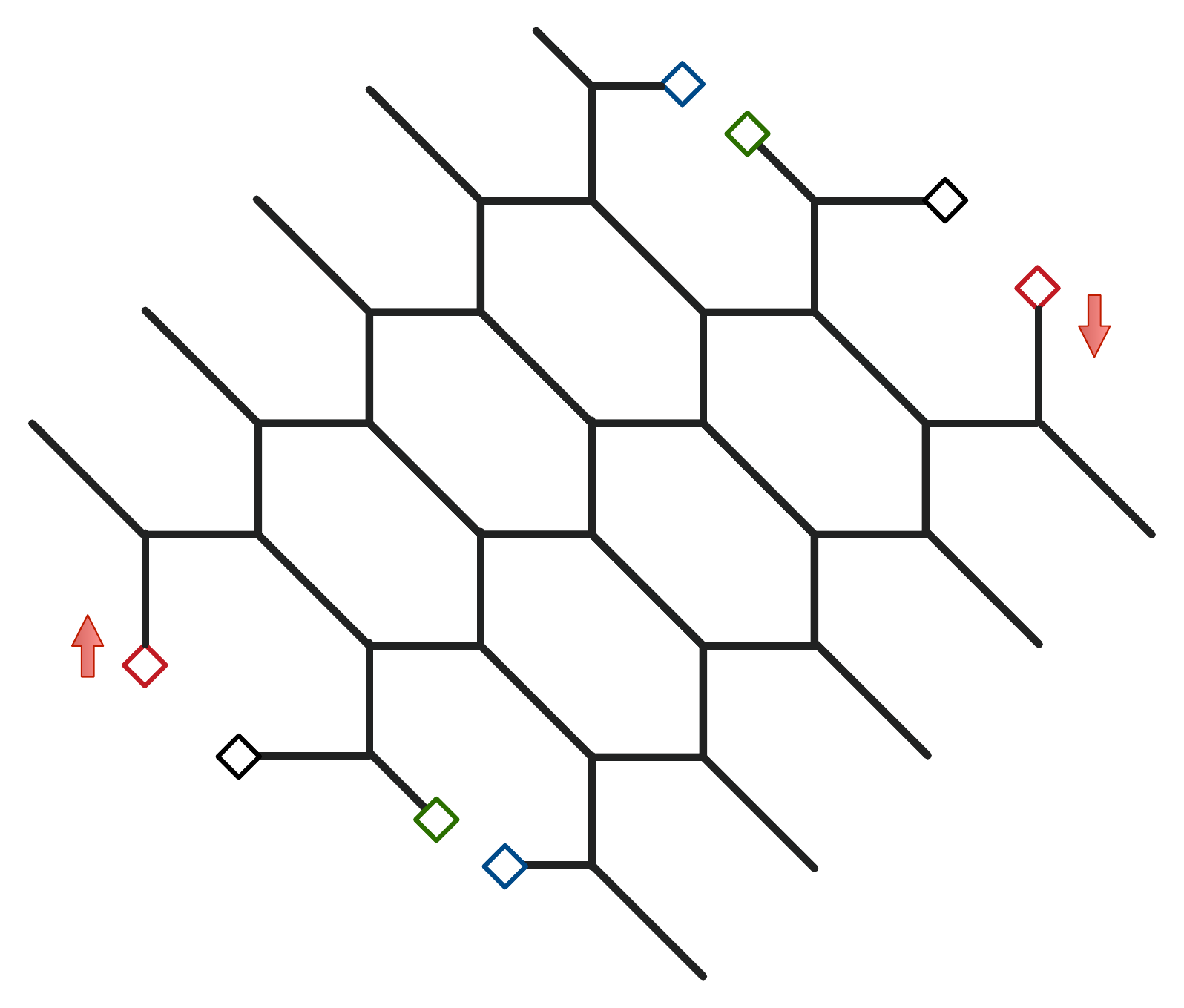}
\end{minipage}
\begin{minipage}{0.32\hsize}
\includegraphics[width=5cm]{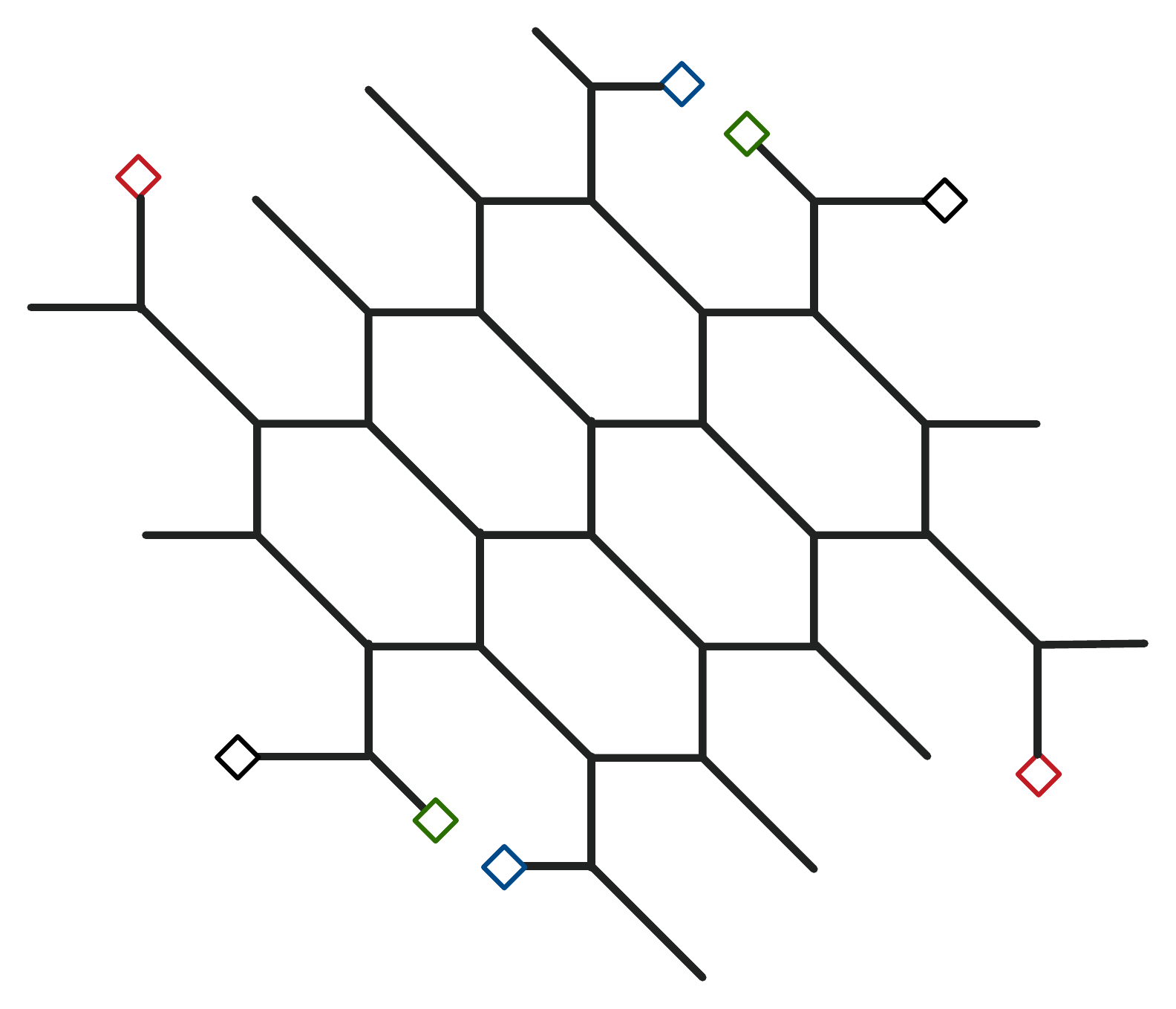}
\end{minipage}
\caption{STS-transformation of the 5d quiver gauge theory $[6] - SU(4)_{0} - SU(4)_{0} - [6]$.}
\label{Fig:Sdual45}
\end{figure}

Here, we can consider another type of S-duality,
which transforms D5 branes, NS5 branes, and $(1,1)$ 5-branes 
into $(1,-1)$ 5-branes, NS5 branes, and D5 branes, respectively.
We denote this as ``STS transformation'', which roughly corresponds to a 45-degree rotation of the diagram.
We start from the same theory \eqref{6446quiver}, but for convenience we consider the web diagram on the left of Figure \ref{Fig:Sdual45}, which is 
related to the left diagram of Figure \ref{Fig:Sdual90} by 
a simple Hanany-Witten transition to move the right top and left bottom D7-branes to the NS5 brane inward.
After taking the STS transformation and then moving $[0,1]$ 7-branes upward or downward, we obtain the 5d quiver gauge theory
\begin{eqnarray}
[3] - SU(2)
- {\overset{\overset{\text{\large$[1]$}}{\textstyle\vert}}{SU(3)_0}} 
- {\overset{\overset{\text{\large$[1]$}}{\textstyle\vert}}{SU(3)_0}}
 - SU(2) - [3], 
 \label{2332}
 \end{eqnarray}
 which is different from the usual S-daulized theory \eqref{333}.
In this way, we can obtain several 5d descriptions by different types of S-duality transformation.

\begin{figure}
\centering
\includegraphics[width=4cm]{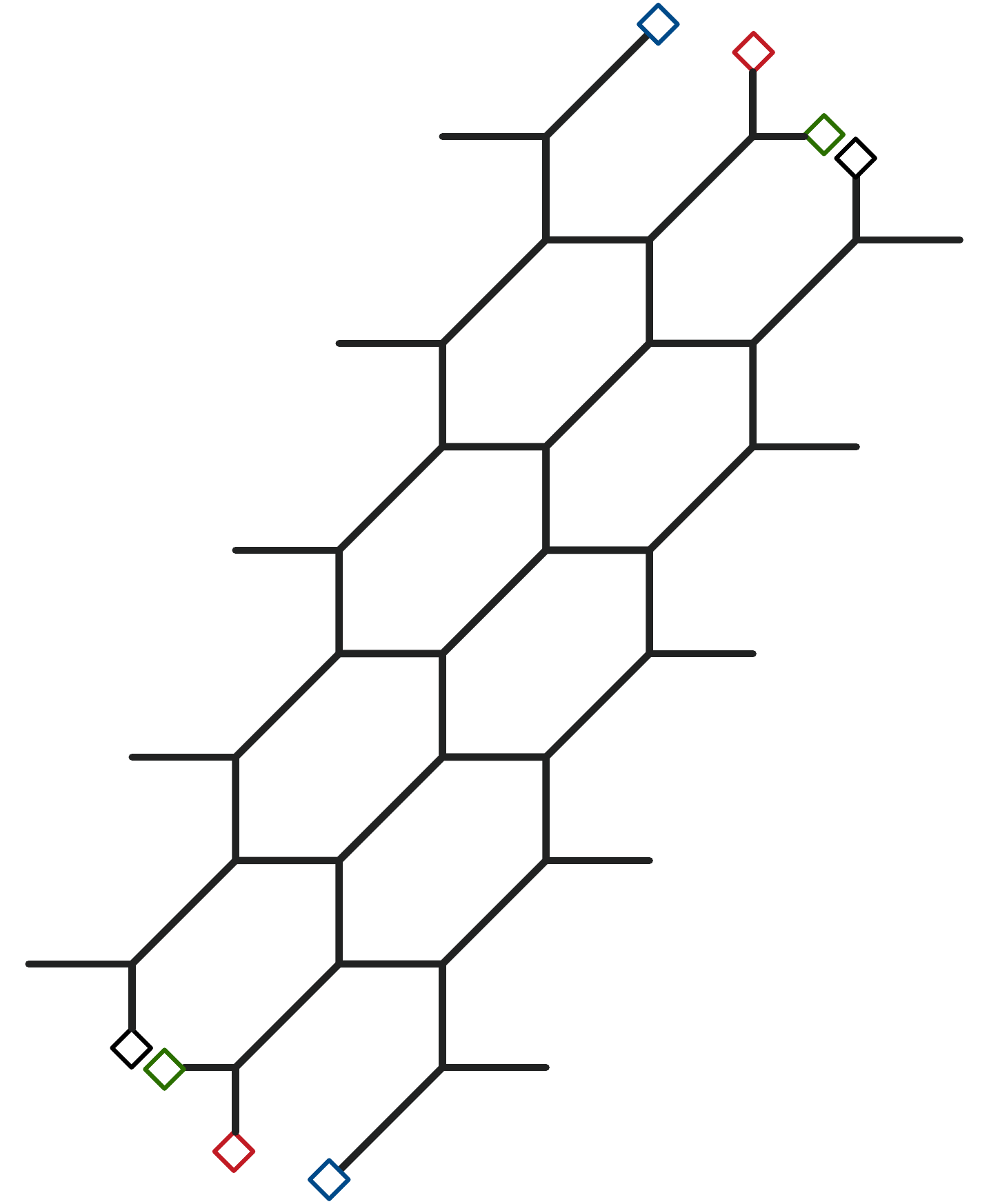}
\caption{Same theory with different mass parameter}
\label{Fig:shifted}
\end{figure}

We remark here that a given gauge theory may have web diagrams withe different ranges of mass parameters. For example, 
instead of starting from the left of Figure \ref{Fig:Sdual90} or Figure \ref{Fig:Sdual45},
we could have started from the diagram on Figure \ref{Fig:shifted}, which describe the same theory but with different mass parameters.
The existence of different web configuration leads to an interesting yet another dual description. Namely, although the original 5d description is identical up to the value of the mass parameters, one may obtain different results under the S-duality transformation.
For instance, it is straightforward to see that the S-transformation yields 
5d quiver gauge theory \eqref{2332} rather than \eqref{333}.
Analogously,  
the STS-transformation gives the theory \eqref{333} rather than \ref{2332}. 
It is worth emphasizing that 
depending on the value of the mass parameters of the flavors,
the same 5d gauge theory can be mapped to different gauge theories 
under the identical S-duality transformation.

\subsection{Distribution duality}\label{subsec:distribution}

In section \ref{subsec:SUSU}, we discussed that 5d quiver gauge theories (\ref{eq:5dSUSU}) with different values of $n$ have an identical 6d UV fixed point, referred to the distribution duality. We devote ourselves in this subsection to discuss the distribution duality in more details from the viewpoint of type IIB 5-brane web diagram.


\begin{figure}
\centering
\includegraphics[width=15cm]{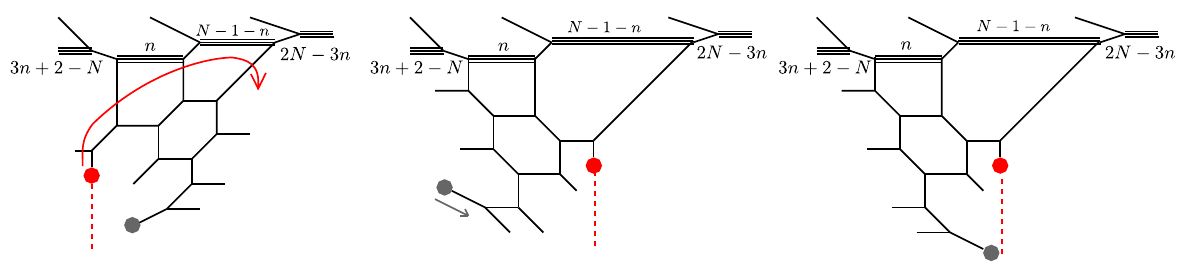}
\caption{Derivation of the distribution duality.}
\label{Fig:Distr}
\end{figure}

We start from the left diagram in Figure \ref{Fig:Distr},
which describes the 5d quiver gauge theory (\ref{eq:5dSUSU}).
Then, we move 
the $[0,1]$ 7-brane painted in red along the arrow in the leftmost diagram in Figure \ref{Fig:Distr}.
In this process, the charges of 5-branes in the lower part of the diagram are modified because they went across the monodromy cut generated by the $[0,1]$ 7-brane. 
This changes the structure of the web diagram so that the resulting diagram gives rise to the following 5d quiver gauge theory 
\begin{eqnarray}
[3n+6-N]-SU(n+2)_0-SU(N-n)_0-[2N-3n].
\label{eq:5dSUSU2}
\end{eqnarray}
We find that this theory is the one obtained by shifting $n$ by 1 in the quiver gauge theory (\ref{eq:5dSUSU}).
The repetition of this procedure yields that the quiver gauge theories (\ref{eq:5dSUSU}) and (\ref{eq:5dSUSU2}) with an arbitrary $n$ are dual to each other as long as all the numbers of flavor and the rank of the gauge groups are positive.
%


By considering the flavor decoupling limit discussed in section \ref{subsec:mass dec}, 
it is also straightforward to show that this distribution duality holds also for the 
theories with 5 dimensional UV fixed points.
Since the process depicted in Figure \ref{Fig:Distr} does not depend on 
the upper part of the diagram, the discussion is parallel even after we 
decouple $n_1$ and $n_2$ flavors coupling to the two gauge groups.
We see that the following two theories are again related by the analogous procedure
\begin{itemize}
\item 5d $[3n+3-N-n_1]-SU(n+1)_{n_1}-SU(N+1-n)_{n_2}-[2N-3n+3-n_2]$ 
\item 5d $[3n+6-N-n_1]-SU(n+2)_{n_1}-SU(N-n)_{n_2}-[2N-3n-n_2]$ 
\end{itemize}
We can further consider moving the flavor branes upward.
Thus, this class of theories with all the possible value $n$ are related by the distribution duality.

\begin{figure}
\centering
\includegraphics[width=12cm]{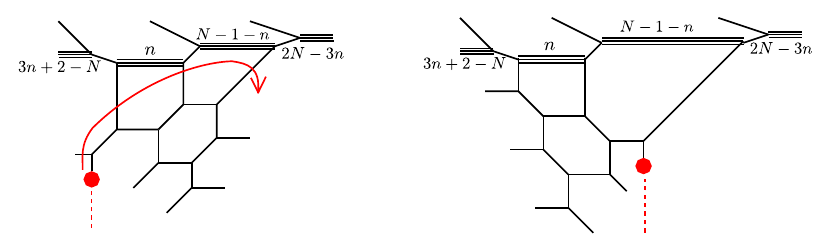}
\caption{Derivation of the distribution duality 2.}
\label{Fig:Distr2}
\end{figure}

Up to here, we have discussed the flavor decoupling limit, which corresponds to moving the flavor branes upward in Figure \ref{Fig:Distr}.
We can also consider another type of flavor decoupling limit corresponding to moving flavor branes downward.

As an example, suppose that we move one of  the flavor branes coupled to $SU(N+1-n)$ downward in the left of Figure \ref{Fig:Distr} to downward.
Then, we obtain the left diagram in Figure \ref{Fig:Distr2}.
By considering the analogous procedure, we find the following two are related 
\begin{itemize}
\item 5d $[3n+3-N] - SU(n+1)_{0} - SU(N+1-n)_{-\frac{1}{2}} - [2N-3n+2]$ 
\item 5d $[3n+5-N] - SU(n+2)_{-\frac{1}{2}} - SU(N-n)_{0} - [2N-3n]$ 
\end{itemize}
Contrary to the previous case, they are not related by shifting $n$ any more.
However, this duality keeps the total number of flavors, total number of gauge ranks,
and total number of CS levels.
In this sense, this is also the analogue of the distribution duality.

\begin{figure}
\begin{center}
\includegraphics[width=12cm]{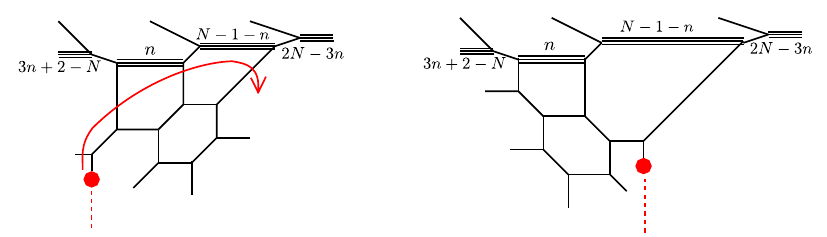}
 \end{center}
 \caption{Derivation of the distribution duality 3.} 
\label{Fig:Distr3}
\end{figure}

It would be also straightforward to see the following duality from Figure \ref{Fig:Distr3} :
\begin{itemize}
\item 5d $[3n+3-N] - SU(n+1)_{0} - SU(N+1-n)_{-1} - [2N-3n+1]$ 
\item 5d $[3n+4-N] - SU(n+2)_{-1} - SU(N-n)_{0} - [2N-3n]$ 
\end{itemize}

It would be also possible to combine them with the flavor decoupling limit
 to move $n_1$ and $n_2$ flavor branes upward.
Therefore, the discussion here can be summarized as the following duality:
\begin{itemize}
\item 5d $[3n+3-N-n_1] - SU(n+1)_{n_1} - SU(N+1-n)_{n_2-\frac{k}{2}} - [2N-3n+3-k-n_2]$ 
\item 5d $[3n+6-N-k-n_1] - SU(n+2)_{n_1-\frac{k}{2}} - SU(N-n)_{n_2} - [2N-3n-n_2]$ 
\end{itemize}
for $k=0,1,2$.

\begin{figure}
\centering
\includegraphics[width=10cm]{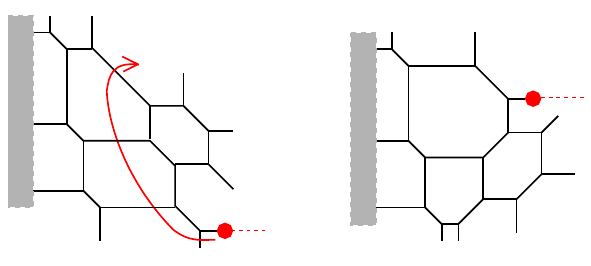}
\caption{Distribution duality in the S-dual frame}
\end{figure}

Although the processes to move the $[0,1]$ 7-brane as in Figure \ref{Fig:Distr}, \ref{Fig:Distr2}, or \ref{Fig:Distr3}
may look exotic, it looks often quite natural when we see this in the S-dual frame.
Take Figure \ref{Fig:Distr3} as an example and consider the S-transformation for both.
Then, we observe that both diagrams correspond to the identical 5d theory in the form
\begin{eqnarray}
\cdots 
- {\overset{\overset{\text{\large$[1]$}}{\textstyle\vert}}{SU(3)}} 
-SU(2)-[1].
\end{eqnarray}

The deformation of moving the $[0,1]$ 7-brane in Figure \ref{Fig:Distr3}
is translated to moving the D7-brane.
This D7-brane corresponds to the flavor charged under the $SU(3)$ gauge group
and moving it simply corresponds to changing the mass of this flavor.
Therefore, the ``distribution duality'', which change the distribution of the rank of the gauge group in the original frame,
is actually just the mass deformation in the S-dual frame.
This is essentially the same observation seen in section \ref{subsec:SUA-Sdual} that 
depending on the value of the mass parameters of the flavors,
the same 5d gauge theory can be mapped to different gauge theories 
under the identical S-duality transformation.

\subsection{Chain of duality and exotic example}\label{subsec:chain}

\begin{figure}[htbp]
\centering
\includegraphics[width=12.5cm]{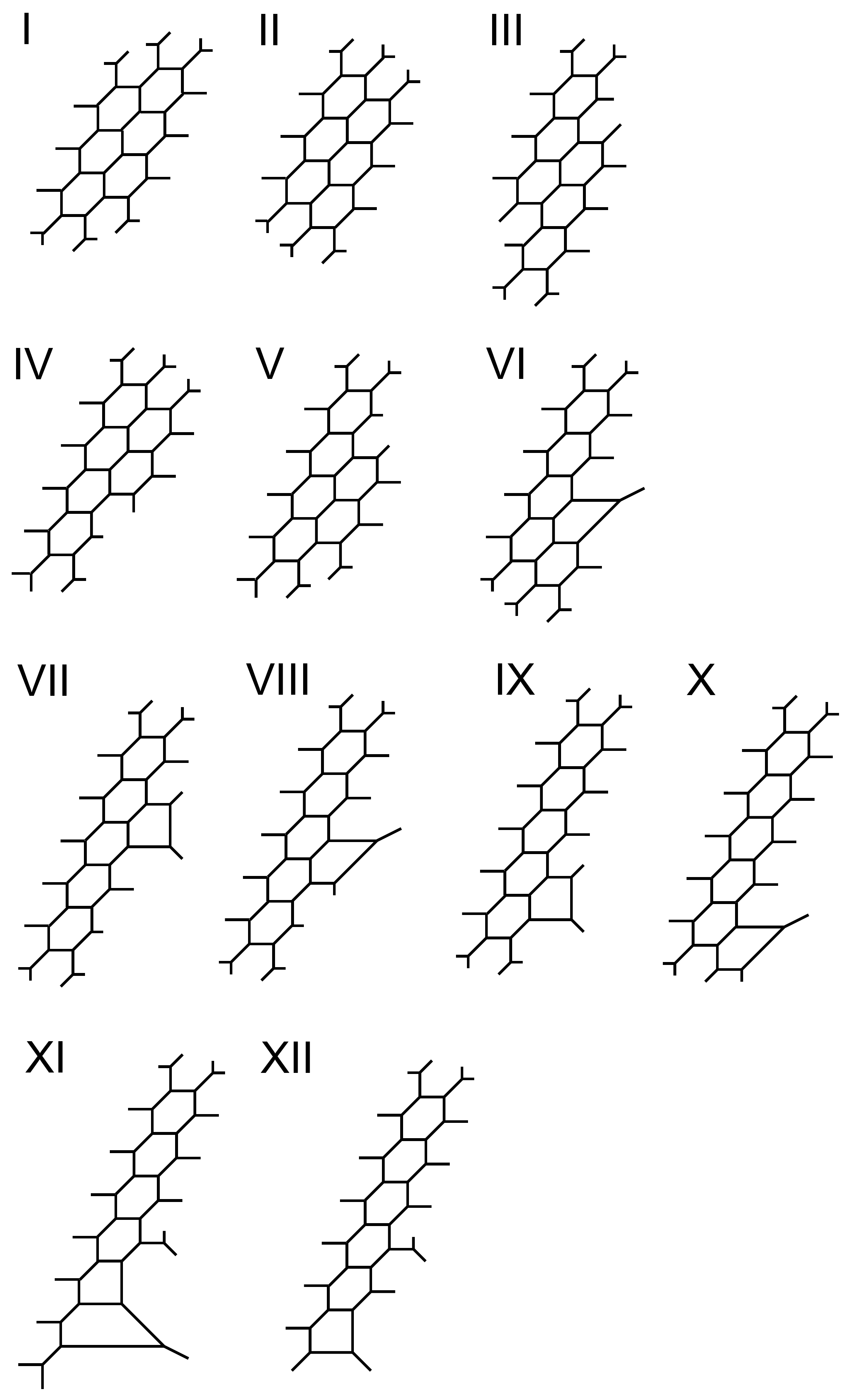}
\caption{Chain of dualities.}
\label{chain}
\end{figure}

Combining what we studied in section \ref{subsec:SUA-Sdual}
and section \ref{subsec:distribution},
we demonstrate that varous 5d quiver gauge theories have identical 6d UV fixed point.

We again consider the 5d theories corresponding to 
6d $SU(6)$ with $N_f=14$ and $N_a=1$ and with a tensor multiplet.
In Figure \ref{chain}, we write various web diagrams,
which are related to each other either by mass deformation 
or the distribution duality, which is sometimes interpreted also as the mass deformation in the S-dual frame.

In the original frame, we see that 
\begin{eqnarray}
{\sf I, II, III}
&:& [6] - SU(4)_{0}-SU(4)_{0} -[6]
\nonumber \\
{\sf IV, V, VI}
&:& [9] - SU(5)_0-SU(3)_{0} -[3]
\nonumber  \\
{\sf VII, VIII, IX, X}
&:& [12] - SU(6)_0-SU(2)
\nonumber  \\
{\sf XI, XII}
&:&  [12] - SU(7)_{\frac32?} - ``SU(1)"
\end{eqnarray}
The last one includes $``SU(1)"$ factor and thus, does not have the standard field theory description
and moreover, does not look fit with the distribution duality discussed previously.
However, by using the analogous process to move one of the $[0,1]$ 7-brane
as in Figure \ref{Fig:Distr}, \ref{Fig:Distr2}, or \ref{Fig:Distr3},
we can show that diagram {\sf X} and diagram {\sf XI} are actually connected by such process.
We note that ${\sf II \leftrightarrow IV}$ and ${\sf VI \leftrightarrow VIII}$
are also related by the distribution dualities.

In the S-dual frame, the gauge theory description is possible only for the following:  
\begin{eqnarray}
{\sf I} &:& [5] - SU(3)_0 - SU(3)_0 - SU(3)_0 - [5]
\nonumber  \\
{\sf II, IV} 
&:& [3] - SU(2)
- {\overset{\overset{\text{\large$[1]$}}{\textstyle\vert}}{SU(3)_0}} 
- {\overset{\overset{\text{\large$[1]$}}{\textstyle\vert}}{SU(3)_0}} 
-SU(2) -[3]
\nonumber  \\ 
{\sf III}
&:& [3] - SU(2)-SU(2)-SU(3)_0-SU(2)- SU(2) - [3]
\nonumber  \\ 
{\sf V} &:& [3] - SU(2)-SU(2)-SU(3)_{\frac{1}{2}}-SU(3)_{0} - [5]
\nonumber  \\
{\sf VII}
&:&  [3] - SU(2)-SU(2)-SU(3)_{1}-SU(2)- SU(2) - [3]
\nonumber  \\
{\sf VIII}
&:&  [3] - SU(2)-SU(2)-SU(2)-SU(2)- SU(3)_{1} - [3]
\nonumber  \\
{\sf XII}
&:& [3] - SU(2)-SU(2)-SU(2)-SU(2)
- {\overset{\overset{\text{\large$[1]$}}{\textstyle\vert}}{SU(2)}} 
-SU(2)
\end{eqnarray}
We omitted the case where the diagram does not have clear field theory interpretation.
In this way, we obtain various 5d quiver gauge theories which have identical 6d UV fixed point.

Here, we comment on the global symmetry.
These theories are all expected to have $SU(14)$ global symmetry at UV fixed point.
Indeed, we can explicitly check that 7-brane monodromy analysis for these diagrams all show this expected global symmetry.
When we apply the method \cite{Yonekura:2015ksa} using instanton operators,
some of these nicely shows affine $SU(14)$ symmetry and thus,
consistent with the claim that these 5d theories have 6d UV fixed point with $SU(14)$ global symmetry.
However, strange to say, application of \cite{Yonekura:2015ksa} to some of these 5d quiver gauge theories 
 fails to reproduce the expected affine structure.
For example, the S-dual description of diagram {\sf III}, 
the central gauge node $SU(3)$ has flavor $N_f < 2N$ and thus, does not show affine symmetry.
We claim that one-instanton analysis is not enough in this case.
It would be interesting to generalize the method with instanton operators
to include the higher instanton contribution for these case to check the 
expected affine global symmetry is obtained.

%% file: section4-1.tex
\bigskip
\section{5d dualities from 6d: Generalization}
\label{sec:general}

In this section, we consider a general 6d quiver theory constructed by NS5-branes, D6-branes, D8-branes an $O8^-$-plane introduced in \cite{Brunner:1997gf, Hanany:1997gh}. It is straightforward to repeat essentially the same analysis done in section \ref{sec2:6dspN} and \ref{sec:6dSUNA} for the general 6d quiver theories. We will see that the 5d reduction yields various 5d dualities. 

\subsection{6d \texorpdfstring{$Sp- \prod SU$}{sp-su} quiver theories}
\label{sec:6dSpSUquiver}

\begin{figure}
\begin{center}
\includegraphics[width=8cm]{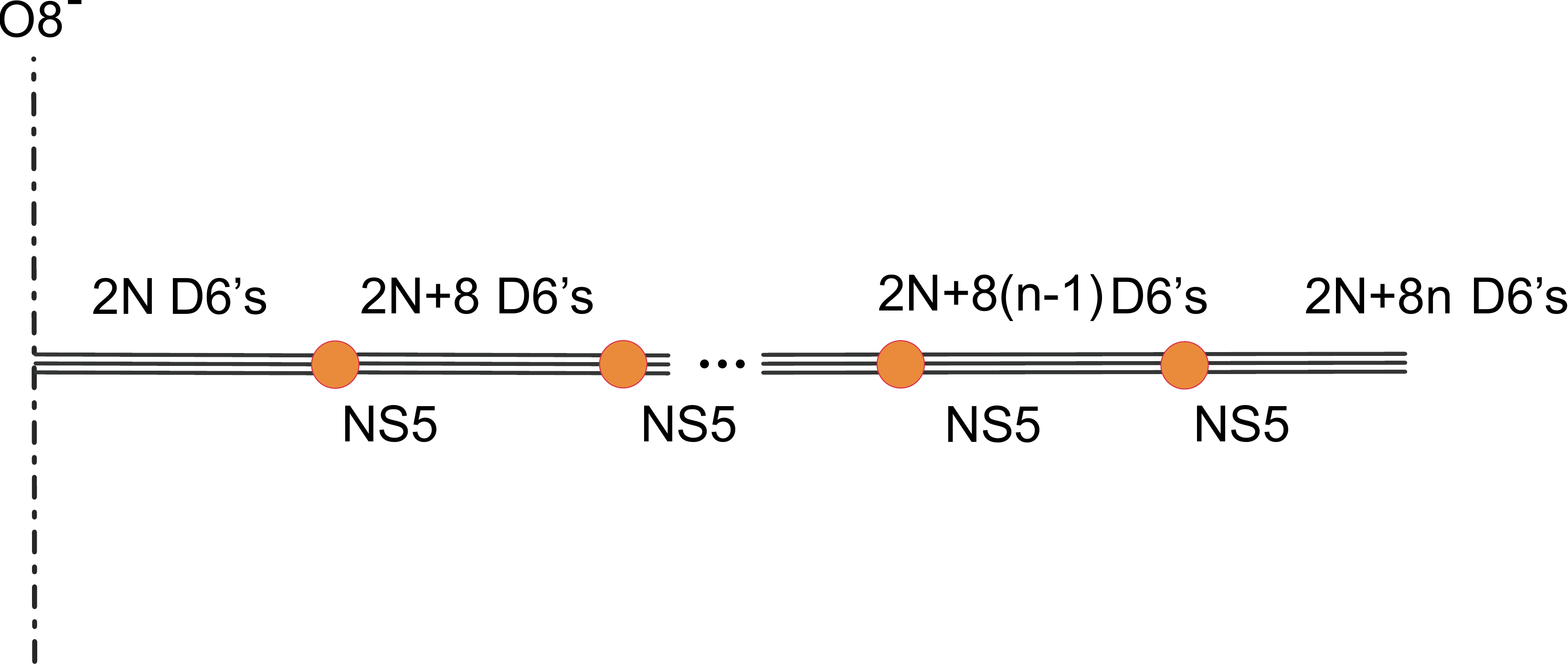}
\end{center}
\caption{Type IIA brane configuration for the 6d linear quiver theory \eqref{6dcanonical1}.} 
\label{Fig:6dquiver-1}
\end{figure}

We first focus on 6d quiver theories with one $Sp$ gauge node, which is a generalization of the 6d quiver considered in section \ref{sec2:6dspN}. The simple generalization of the setup in section \ref{sec2:6dspN} realizes a 6d linear quiver 
\begin{equation}
6d \; Sp(N) - SU(2N+8) - SU(2N+16) - \cdots - SU(2N+8(n-1)) - [2N+8n], \label{6dcanonical1}
\end{equation}
and its Type IIA brane configuration is given in Figure \ref{Fig:6dquiver-1}. The case with $n=1$ falls in to the case in section \ref{sec2:6dspN}. Hence, we concentrate on the cases with $n > 1$.  The global symmetry is generically $SU(2N+8n) \times U(1)$. The number of the tensor multiplets is $n$ and the total rank of the gauge groups is $N(2n-1) + 4n^2-5n+1$.

\begin{figure}
\begin{center}
\includegraphics[width=15cm]{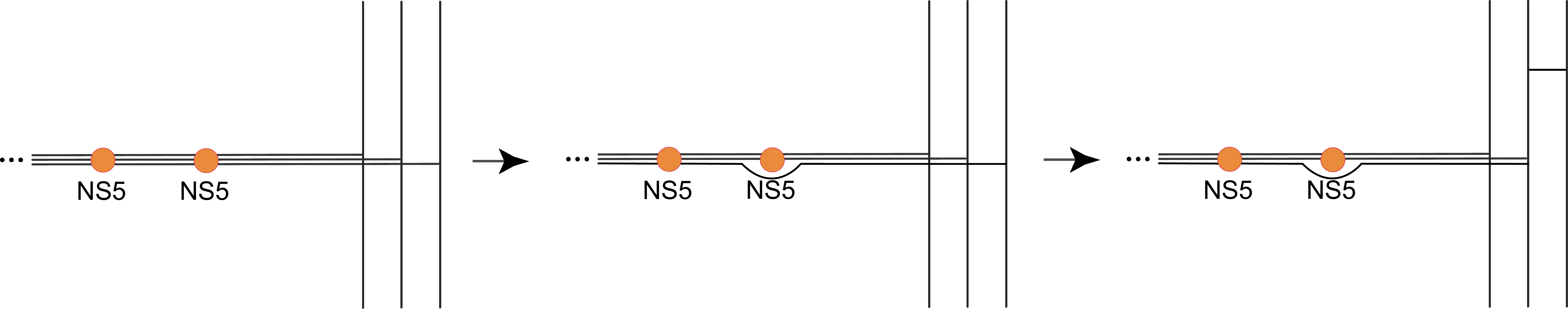}
\end{center}
\caption{A process of the 6d Higgsing.} 
\label{Fig:Higgs}
\end{figure}

We can further generalize the 6d quiver theory by Higgsing. The flavor symmetry of the brane construction of the canonical 6d quiver \eqref{6dcanonical1} is associated to a symmetry on semi-infinite $2N+8n$ D6-branes at the end. It is also possible to let one semi-infinite D6-brane become finite and end on one D8-brane. Since each D6-brane end on one D8-brane, we have $2N+8$ D8-branes in total. In this picture, each D6-brane at the end connect an NS5-brane with a D8-brane. A Higgs branch of the 6d theory arises by attaching one D6-brane on two or more D8-branes. Since the condition of preserving supersymmetry implies that only one D6-brane can connect an NS5-brane with a D8-brane \cite{Hanany:1996ie}, the D6-brane should connect to other D6-brane between NS5-branes by jumping some NS5-branes as in Figure \ref{Fig:Higgs}.
Then, there appears fractionated D6-branes between D8-branes. The motion of the D6-branes between D8-branes describes the vev of the corresponding hypermultiplets. After sending the fractionated D6-branes into infinity, we have decoupled D8-branes and hence the flavor symmetry is reduced. Due to the appearance of the jumping of D6-branes over NS5-branes, the ranks of some gauge groups are also reduced.

\begin{figure}
\begin{center}
\includegraphics[width=8cm]{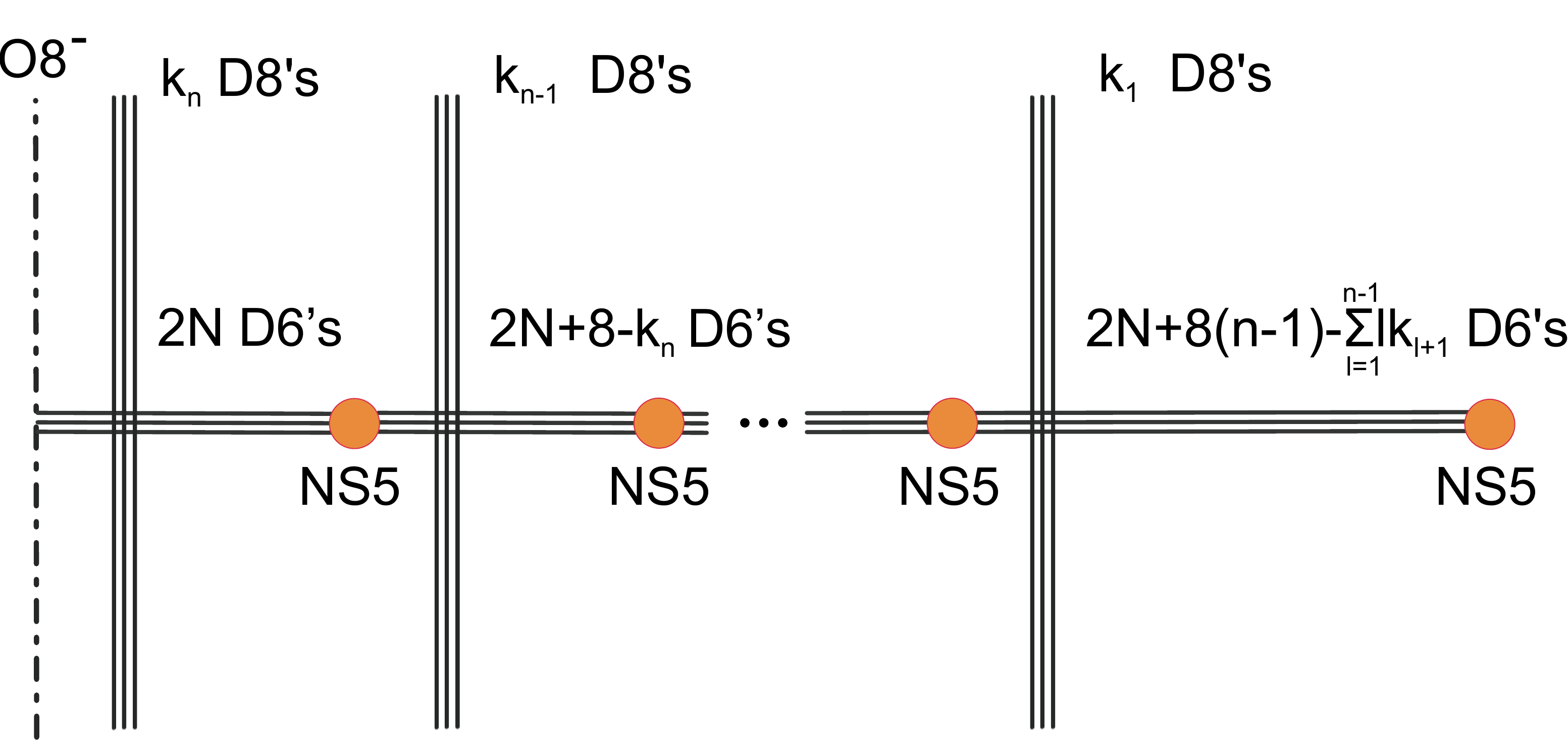}
\end{center}
\caption{Type IIA brane configuration for the 6d linear quiver theory \eqref{6dHiggs1}.} 
\label{Fig:6dquiver-2}
\end{figure}

In general, such a Higgsing is classified by a Young diagram as in \cite{Gaiotto:2014lca}. When one represents a Young diagram by a vector with non-increasing numbers, each component represent the number of D6-branes ending on D8-brane. The total number of the boxes in the Young diagram is $2N+8n$. Therefore, the brane configuration in Figure \ref{Fig:6dquiver-1} corresponds to a Young diagram $[1, \cdots, 1]$ with $2N+8n$ entries of $1$'s. A general Higgsing corresponds to a Young diagram $[n, \cdots, n, n-1, \cdots, n-1, \cdots, 2, \cdots, 2, 1, \cdots, 1]$ where the number of $l$ is $k_l$ with a condition $\displaystyle\sum_{l=1}^{n}l\cdot k_l = 2N+8n$. The 6d gauge theory content can be read off by moving D8-branes to the left in the brane configuration until no D6-branes end on the D8-branes. Depending on the Young diagram, various fundamental hypermultiplets are introduced to some middle gauge nodes in the 6d quiver. Hence, after a Higgsing corresponds to the Young diagram $[n, \cdots, n, n-1, \cdots, n-1, \cdots, 2, \cdots, 2, 1, \cdots, 1]$, we obtain the following 6d quiver theory at low energies 
\begin{equation}
6d \; {\overset{\overset{\text{\large$[k_{n}]$}}{\textstyle\vert}}{Sp(N) }} - {\overset{\overset{\text{\large$[k_{n-1}]$}}{\textstyle\vert}}{SU(2N+8-k_n)}} - 
\cdots - 
{\overset{\overset{\text{\large$[k_1]$}}{\textstyle\vert}}{SU(2N+8(n-1) - \sum_{l=1}^{n-1}lk_{l+1})}}. \label{6dHiggs1}
\end{equation}
Type IIA brane configuration is depicted in Figure \ref{Fig:6dquiver-2}.

We will consider a circle compactification of the 6d quivers of \eqref{6dcanonical1} and \eqref{6dHiggs1}, and see the 5d descriptions of the theories as well as the 5d dualities. 

\subsubsection{5d \texorpdfstring{$SU$}{SU}  quviers}
\label{subsubsec:5dsuquiver1}

\begin{figure}
\begin{center}
\includegraphics[width=15cm]{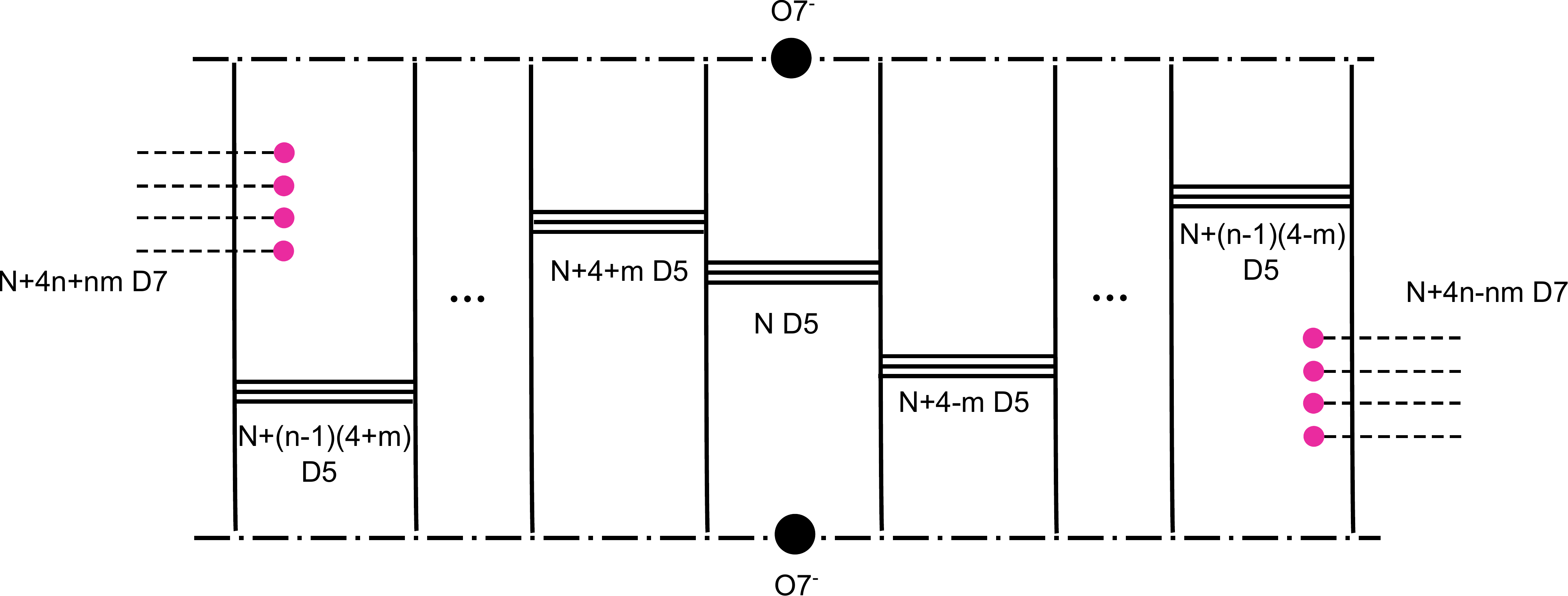}
\end{center}
\caption{Type IIB brane configuration after performing a T-duality to Figure \ref{Fig:6dquiver-1}.} 
\label{Fig:Brane6dcanonical1}
\end{figure}

We first focus on an $S^1$ compactification of the canonical 6d quiver theory \eqref{6dcanonical1}. The T-duality along the $S^1$ gives us a 5-brane configuration with two $O7^-$-planes as in Figure \ref{Fig:Brane6dcanonical1}.
After the T-duality, we have the distribution ambiguity discussed in section \ref{subsec:SUSU}. The number of D5-branes in the middle column is  always $N$, which corresponds to the first gauge node of the 6d quiver \ref{6dcanonical1}.
The $N+8$ color D5 branes originated from the 2nd gauge node in 6d quivers \ref{6dcanonical1} 
are distributed into the next columns, which are left and right to the center, respectively.
The number of the D5-branes in these columns can change by the distribution ambiguity. 
We label the ambiguity by a non-negative number $m$, and set the number of the D5-branes in the left column to the middle to be $N+4+m$ and the number of the D5-branes in the right column to the middle to be $N+4-m$. 

The $N+8i$ color D5 branes corresponding to the $(i+1)$-th gauge node in 6d quivers \ref{6dcanonicalA} are distributed into the $i$-th left columns and $i$-th right columns from the center.
Here, we need to care about the condition that the resulting 5-brane web diagram after pulling all the 7-branes outside has a 5d gauge theory interpretation. In fact, this condition may determine the number of D5-branes in the column next to the column which is next to the middle one. By repeating this analysis, it turns out that there is only one degree of freedom for the ambiguity parameterized by $m$. The resulting distribution is depicted in Figure \ref{Fig:Brane6dcanonical1}. 

\begin{figure}
\begin{center}
\includegraphics[width=10cm]{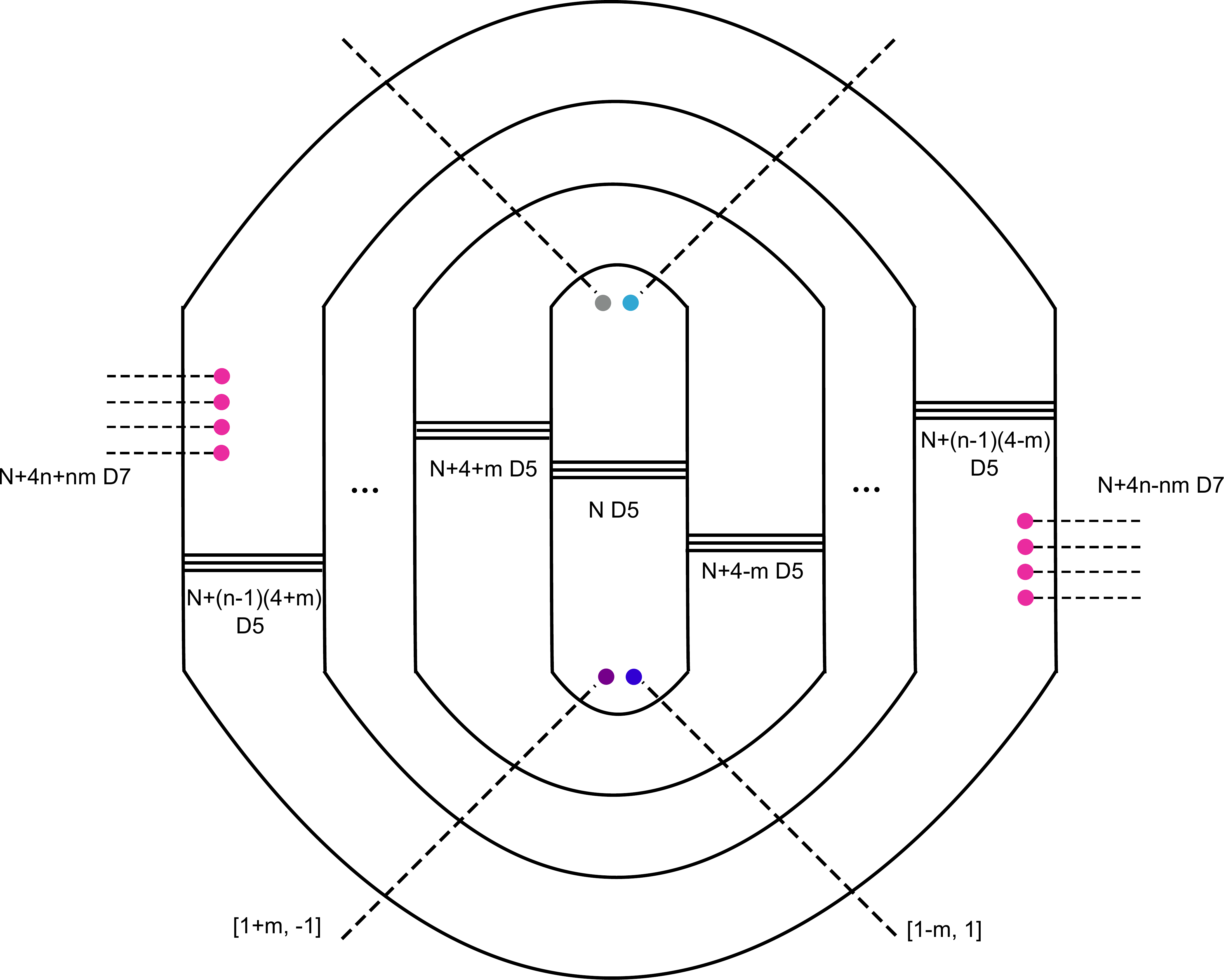}
\end{center}
\caption{The 5-brane web diagram after splitting two $O7^-$-planes compared to Figure \ref{Fig:Brane6dcanonical1}. Note that the slope of the lines is schematically depicted and does not represent the corresponding 5-brane charge precisely.} 
\label{Fig:5braneloop}
\end{figure}

\begin{figure}
\begin{center}
\includegraphics[width=10cm]{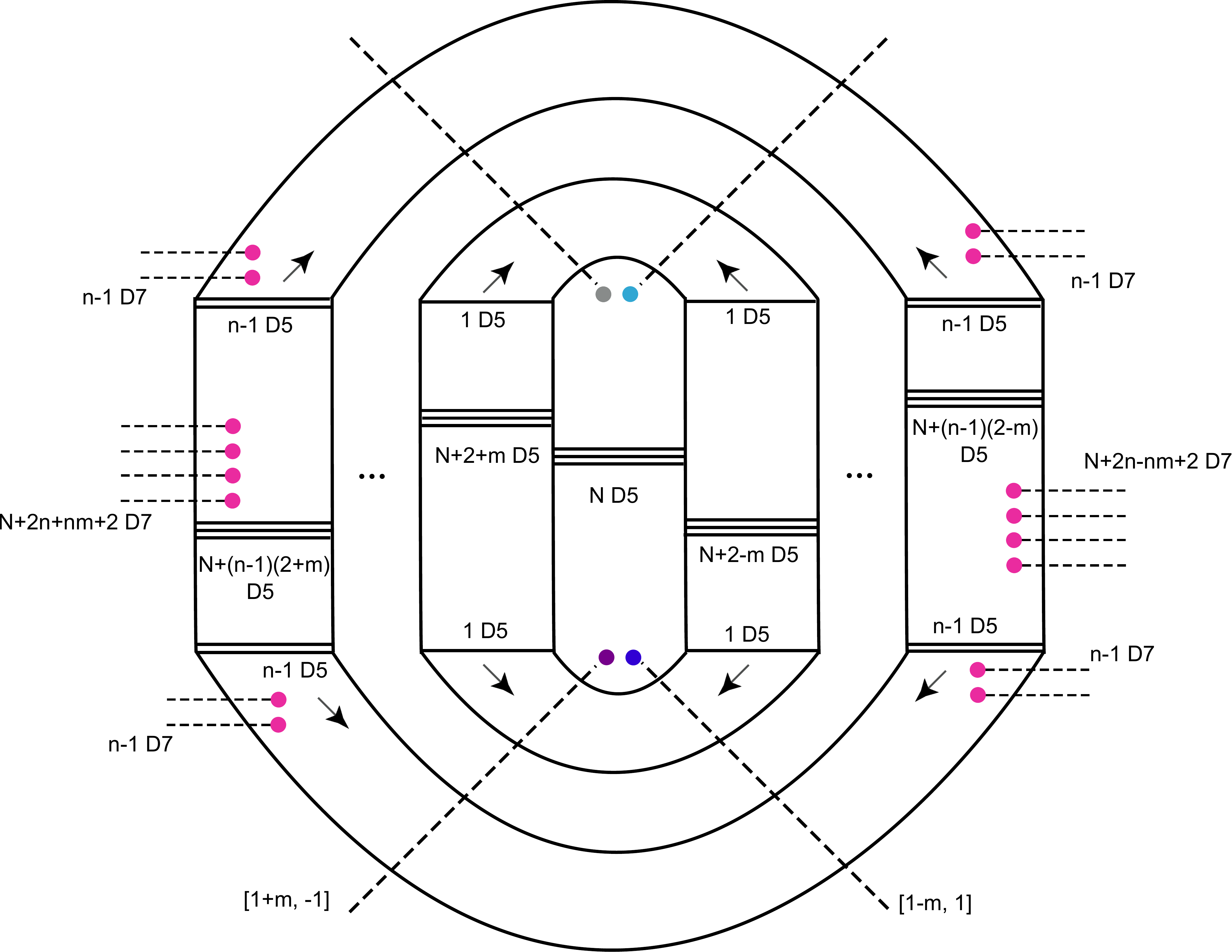}
\end{center}
\caption{The motion of the branch cuts of ${\bf B}, {\bf C}$ and ${\bf X}_{[1+m, -1]}\; {\bf X}_{[1-m,1]}$. Relatively, it can be realized by moving some amount of D5-branes and D7-branes. The D5-branes and the D7-branes that move are indicated by arrows in this figure.} 
\label{Fig:5braneloop2}
\end{figure}

\begin{figure}
\begin{center}
\includegraphics[width=10cm]{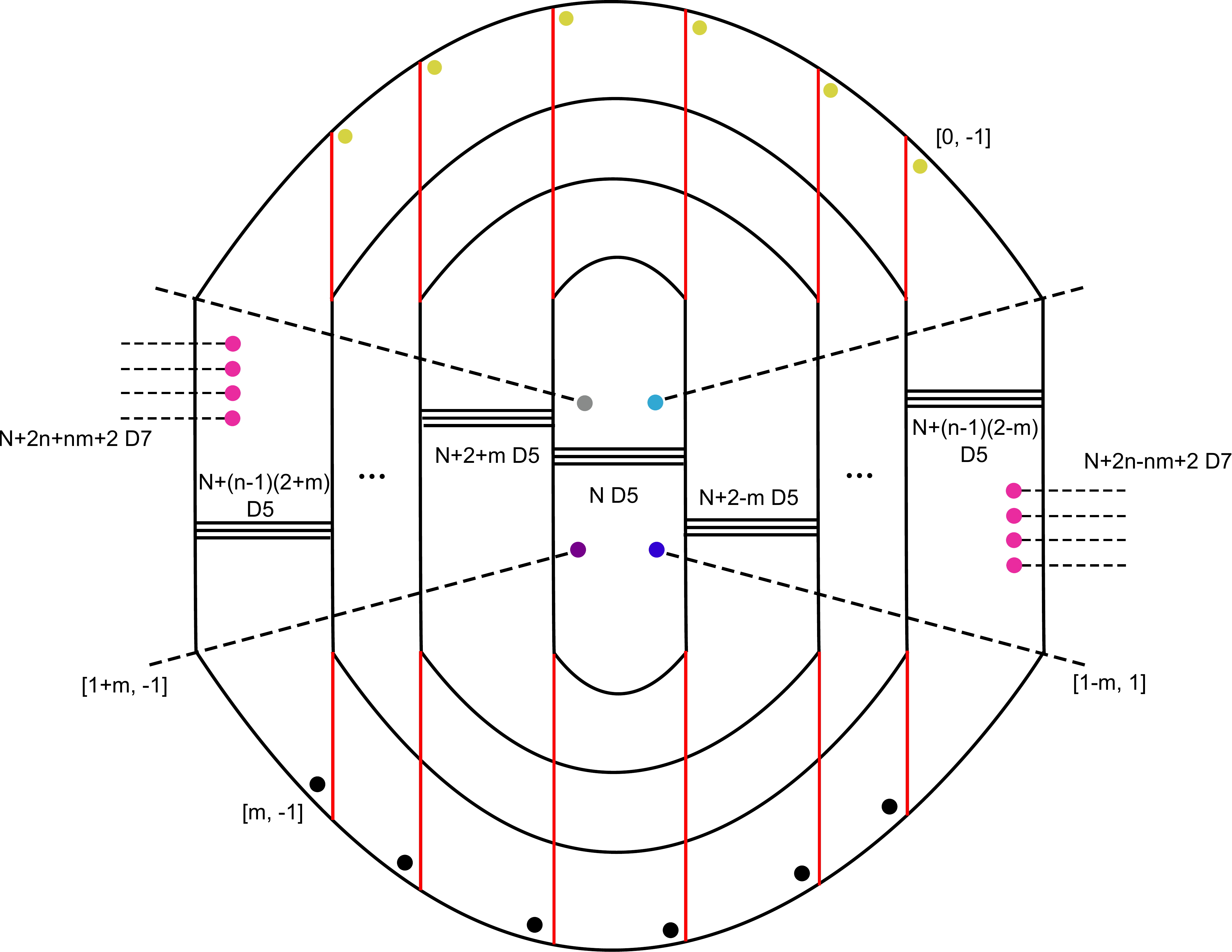}
\end{center}
\caption{The schematic diagram after moving the branch cuts of ${\bf B}, {\bf C}$ and ${\bf X}_{[1+m, -1]}\; {\bf X}_{[1-m,1]}$ 7-branes from Figure \ref{Fig:5braneloop}. The red lines arise from D5-branes which cross the branch cuts of ${\bf B}, {\bf C}$ and ${\bf X}_{[1+m, -1]}\; {\bf X}_{[1-m,1]}$ 7-branes. The $[0,1]$ 7-branes denoted by the yellow circles in the upper part of the diagram arise from D7-branes which cross the branch cuts of ${\bf B}, {\bf C}$ 7-branes. On the other hand, the $[m,-1]$ 7-branes denoted by the black circles in the lower part of the diagram arise from D7-branes which cross the branch cuts of  ${\bf X}_{[1+m, -1]}\; {\bf X}_{[1-m,1]}$  7-branes. } 
\label{Fig:5braneloop3}
\end{figure}

Next task is the resolution of the $O7^-$-planes. As in section \ref{sec2:6dspN}, we can consider either splitting both $O7^-$-planes or splitting only one of the two $O7^-$-planes. We consider the former case first. The condition that the final 5-brane web after pulling out all the 7-branes should have a 5d gauge theory interpretation also constrains the splitting type of the $O7^-$-planes. The relative difference between the two splitting type is not important, and hence we fix the splitting type of the upper $O7^-$-plane to be ${\bf B}\; {\bf C}$. Then, the splitting type of the lower $O7^-$-plane is determined to be ${\bf X}_{[1+m, -1]}\; {\bf X}_{[1-m,1]}$ in order that the resulting 5-brane web diagram after pulling all the 7-branes outside has a 5d gauge theory interpretation. The resolution creates the $n$ 5-brane loops as in Figure \ref{Fig:5braneloop}.
Then, we move the branch cuts of ${\bf B}, {\bf C}$ and ${\bf X}_{[1+m, -1]}\; {\bf X}_{[1-m,1]}$ so that some of the D5-branes and D7-branes cross the branch cuts as in Figure \ref{Fig:5braneloop2}. 
In the upper part of the diagram, D7-branes and D5-branes cross the branch cuts of the {\bf B} and {\bf C} 7-branes in a counterclockwise and clockwise direction respectively. Then the D7-branes become $[0,1]$ 7-branes, and the D5-branes become NS5-branes. On the other hand, in the lower part of the diagram,  D7-branes and D5-branes cross the branch cuts of the  ${\bf X}_{[1+m, -1]}\; {\bf X}_{[1-m,1]}$ 7-branes in a counterclockwise and clockwise direction respectively. Then, the D7-branes and the D5-branes become $[m, -1]$ 7-brane and $(m, -1)$ 7-brane respectively. After crossing the branch cuts, the 5-brane loops are divided by $2n$ vertical lines as in Figure \ref{Fig:5braneloop3}. This is the origin of 5d quiver theories with $2n-1$ gauge nodes. 

\begin{figure}
\begin{center}
\includegraphics[width=8cm]{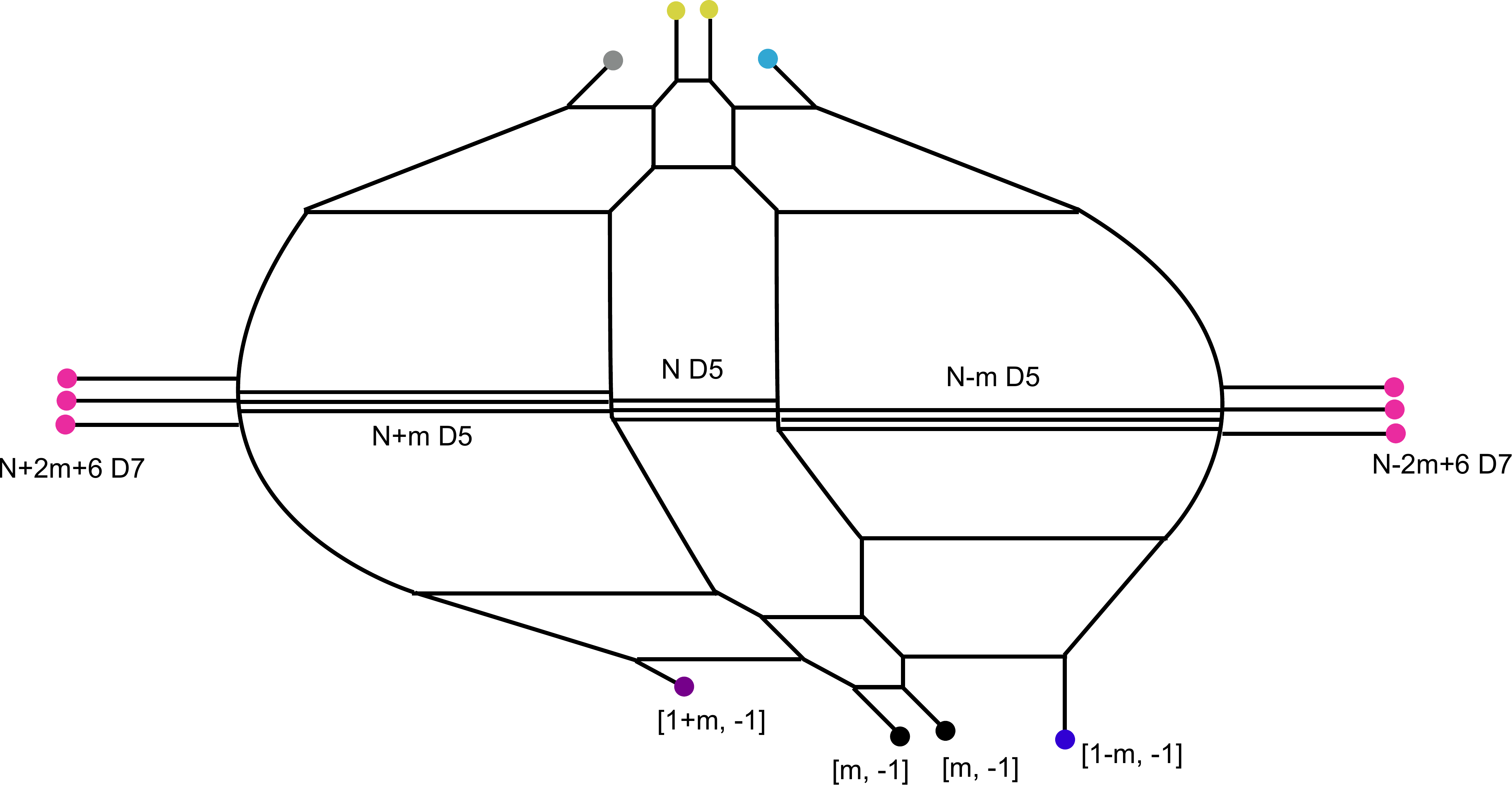}
\end{center}
\caption{The 5-brane web diagram after pulling out all the 7-branes outside of the 5-brane loops from the one in Figure \ref{Fig:5braneloop2}. For simplicity, we write a web diagram of $n=2$.} 
\label{Fig:Brane6dcanonical2}
\end{figure}

After pulling out all the 7-branes, the 5-brane web diagram becomes the one in Figure \ref{Fig:Brane6dcanonical2}.
 The 5d gauge theory realized by the 5-brane web is then
 \begin{eqnarray}
{\overset{\overset{\text{\large$[L_1]$}}{\textstyle\vert}}{SU(N_1)  }} - SU(N_{2}) - \cdots - SU(N_{n-1}) - SU(2N+2n) - SU(M_{n-1}) - \cdots  - SU(M_{2}) - {\overset{\overset{\text{\large$[R_1]$}}{\textstyle\vert}}{SU(M_1)  }}\nonumber\\
\label{5dsuquiverfromSpquiver}
\end{eqnarray}
where 
\begin{eqnarray}
N_l &=& N+2n+(n-l)m,\qquad \text{for}\quad 1 \leq l \leq n-1,\\
M_l &=&  N+2n -(n-l)m,\qquad \text{for}\quad 1 \leq l \leq n-1,\\
L_1 &=& N+2n+2+nm,\\
R_1&=&N+2n+2-nm.
\end{eqnarray}
The parameters are constrained such that the rank of each gauge group or the number of the flavors should be greater than zero at least. 

The number of the Coulomb branch moduli can be easily counted and it is $(2n-1)(N+2n-1)$ which is the sum of the number of the tensor multiplets and the number of the vector multiplets in the Cartan subalgebra of the 6d quiver theory \eqref{6dcanonical1}. The global symmetry analysis by 7-branes should recover $SU(2N+8n)$ since the very first 5-brane web gives $2N+8n$ D7-branes on top of each other\footnote{Note that a D7-brane is mutually local to an $O7^-$-plane. D7-branes do not change its change when it crosses the branch cut of an $O7^-$-plane.}.

The different choice of $m$ in \eqref{5dsuquiverfromSpquiver} gives a different-looking 5d gauge theory. However, we claim that they are dual to each other, and it is the distribution duality.

\subsubsection{5d \texorpdfstring{$Sp- \prod SU$}{sp-su} quivers}
\label{subsubsec:5dspsuquiver}

\begin{figure}
\begin{center}
\includegraphics[width=10cm]{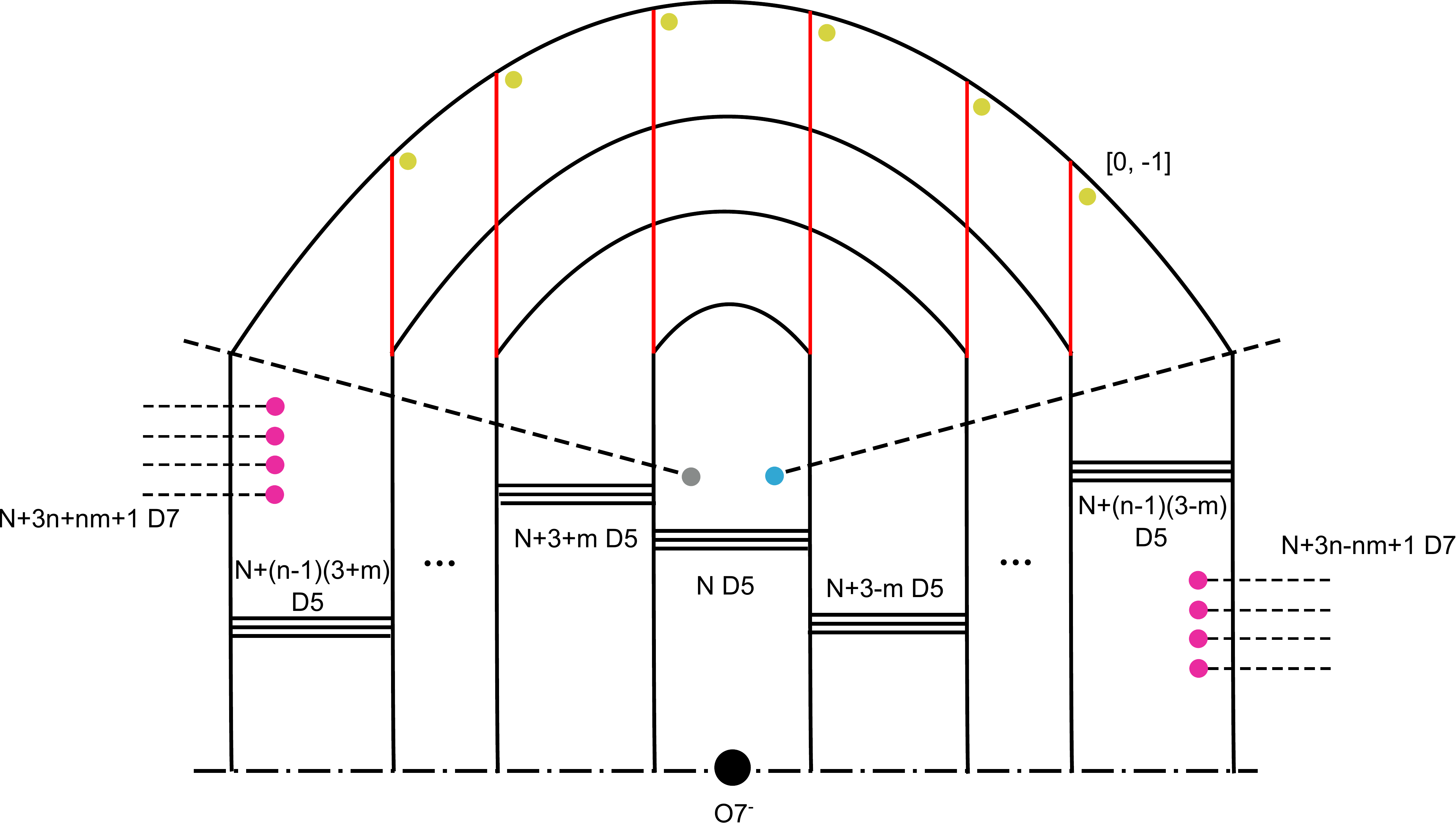}
\end{center}
\caption{The schematic diagram after moving the branch cuts of ${\bf B}, {\bf C}$ and ${\bf X}_{[1+m, -1]}\; {\bf X}_{[1-m,1]}$ 7-branes in the case when one resolves one of the $O7^-$-planes in Figure \ref{Fig:Brane6dcanonical1}. The motion corresponds to the upper ones depicted in Figure \ref{Fig:5braneloop2}.} 
\label{Fig:5braneloop4}
\end{figure}

We then consider the case of splitting only one of the two $O7^-$-planes. The quantum resolution splits the $O7^-$-plane into {\bf B} and {\bf C} 7-branes. The analysis is essentially the same as in section \ref{subsubsec:5dsuquiver1}. About the motion of the branch cuts of 7-branes, we can only consider the one corresponding to the upper transformations in Figure \ref{Fig:5braneloop2}, the resulting 5-brane configuration is depicted in \ref{Fig:5braneloop4}. 

\begin{figure}
\begin{center}
\includegraphics[width=8cm]{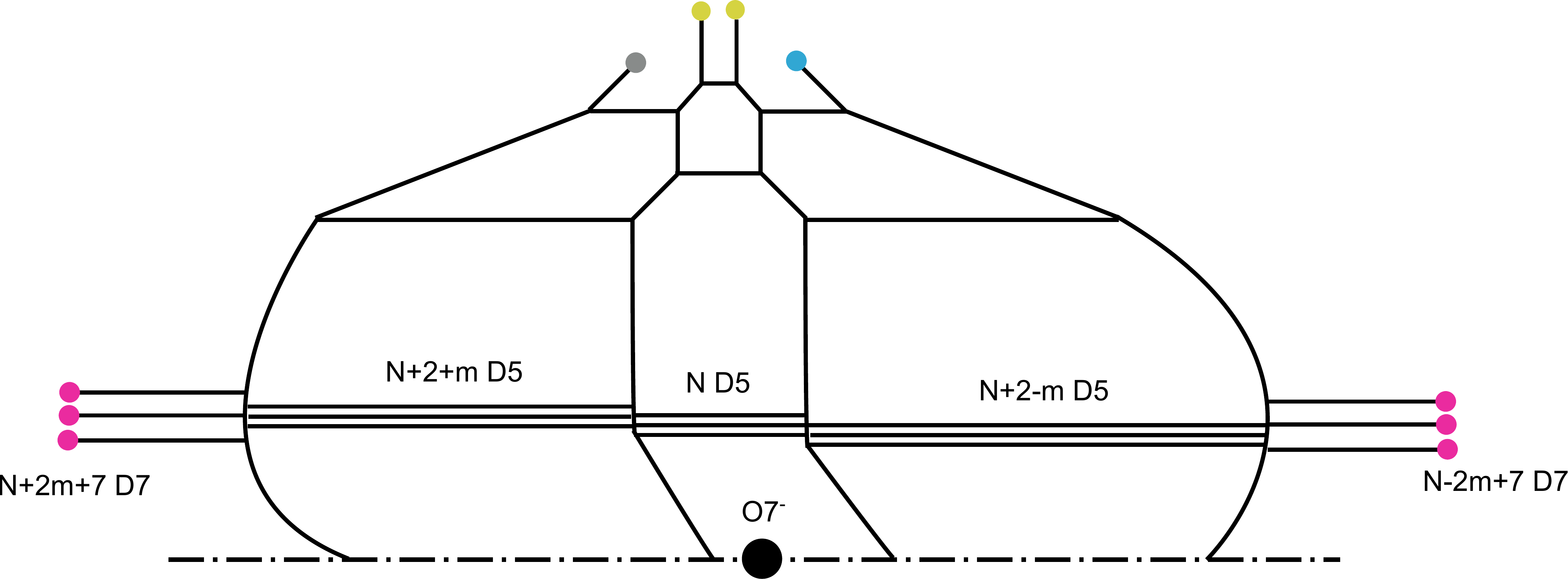}
\end{center}
\caption{The 5-brane web diagram realizing the 5d $Sp(N+2) - SU(2N+8) - [2N+14]$. We only write the web diagram for the case with $n=2$ for simplicity. This web diagram can be obtained by pulling all the 7-branes outside the 5-brane loops from the configuration depicted in Figure \ref{Fig:5braneloop4} with $n=2$.} 
\label{Fig:Brane6dcanonical3}
\end{figure}

After pulling out all the 7-branes, the final 5-brane web diagram is given by Figure \ref{Fig:Brane6dcanonical3}.
Note that in this case the distribution ambiguity does not matter. Due to the presence of the $O7^-$-plane, the column which originally has $N+k(4+m)$ D5-branes is connected to the column which originally has $N+k(4-m)$ D5-branes for $1 \leq k \leq n-1$. Therefore, the resulting 5d theory is 
\begin{equation}
Sp(N+n) - SU(2N+2n+4) - SU(2N+2n+8) - \cdots - SU(2N+6n-4) - [2N+6n+2].\label{5dspsuquiverfromSpquiver}
\end{equation}
Due to the process of obtaining (\ref{5dspsuquiverfromSpquiver}), the 7-brane analysis should give the $SU(2N+8n)$ flavor symmetry. The number of the Coulomb branch moduli is again $(2n-1)(N+2n-1)$, which is consistent with the sum of the number of the tensor multiplets and the number of the vector multiplets in the Cartan subalgebra of the 6d quiver theory \eqref{6dcanonical1}.

\subsubsection{5d dualities}
\label{subsubsec:5ddualsuquiver}

From the same 6d quiver theory of \eqref{6dcanonical1}, we have obtained two types of 5d quiver theories. One type has only $SU$ gauge nodes and the other type has one $Sp$ gauge node with other $SU$ gauge nodes. Those two theories are described by essentially the same 5-brane web diagrams and we claim that they are dual to each other. This is the generalization of the claim in section \ref{subsec:equivalence.ch2}. 

As in section \ref{subsec:Sdual.ch2}, it is possible to generate further dual 5d theories by acting the $SL(2, \mathbb{Z})$ duality on \eqref{5dsuquiverfromSpquiver}. For example, let us consider a case where $m=0$, which yields, 
\begin{equation}
[N+2n+2] - SU(N+2n)  - \cdots -  SU(N+2n) - \cdots  -  SU(N+2n) - [N+2n+2],  \label{5dsuspecialquiverfromSpquiver}\\
\end{equation}
where it has $2n-1$ gauge nodes. The S-duality or the $90$ degrees rotation gives 
\begin{equation}
[2n+2] - SU(2n)  - \cdots -  SU(2n) - \cdots  -  SU(2n) - [2n+2],  \label{5dsuspecialquiverfromSpquiver.d1}\\
\end{equation}
 where it has $N+2n-1$ gauge nodes. It is in fact more interesting to see the TST-duality or the $45$ degrees rotation. The resulting theory is 
 \begin{equation}
{\overset{\overset{\text{\large$[3]$}}{\textstyle\vert}}{SU(2) }} - SU(3) - \cdots - {\overset{\overset{\text{\large$[1]$}}{\textstyle\vert}}{SU(N+2n) }}-  SU(N+2n) - \cdots  -  SU(N+2n)- {\overset{\overset{\text{\large$[1]$}}{\textstyle\vert}}{SU(N+2n) }}- \cdots -  SU(3) - {\overset{\overset{\text{\large$[3]$}}{\textstyle\vert}}{SU(2) }},  \label{5dsuspecialquiverfromSpquiver.d2}\\
\end{equation}
where it has $2n-3$ $SU(N+2n)$ gauge nodes. In particular, when $n=2$, the 5d quiver \eqref{5dsuspecialquiverfromSpquiver.d2} has a peculiar form  
 \begin{equation}
{\overset{\overset{\text{\large$[3]$}}{\textstyle\vert}}{SU(2) }} - SU(3) - \cdots - {\overset{\overset{\text{\large$[2]$}}{\textstyle\vert}}{SU(N+2n) }}- \cdots -  SU(3) - {\overset{\overset{\text{\large$[3]$}}{\textstyle\vert}}{SU(2) }},  \label{5dsuspecialquiverfromSpquiver.d3}\\
\end{equation}
 which can be regarded as gauging $SU(N+2n)$ in the $SU(N+2n+2)$ flavor symmetry of two the $T_{N+2n}$ Tao theories which we will discuss in detail in section \ref{sec:TNTao}.
 
 As in section \ref{sec2:6dspN}, it is possible to decouple some flavors from the 5d theories whose UV completion is the 6d SCFT. After decoupling some flavors, the 5d theory has  a 5d UV fixed point. Suppose we have dual 5d theories whose UV completion is the same 6d SCFT, decoupling the same flavors in the both theories leads to another dual 5d theories whose UV completion is a same 5d SCFT. Working on explicit examples is quite straightforward and can be done in parallel to the analysis in section \ref{subsec:equivalence.ch2} and \ref{subsec:Sdual.ch2}.

\subsubsection{Higgsed cases}\label{subsubsec:genHiggsSp}

Let us then move on to a circle compactification of the Higgsed 6d quiver theories \eqref{6dHiggs1}. After performing the T-duality along the $S^1$, we again obtain a brane configuration with 5-branes and two $O7^-$-planes. The difference from the cases in section \ref{subsubsec:5dspsuquiver} is that we now have D7-branes in various columns, depending on the number of the fundamental hypermultiplets attached to some gauge nodes in the 6d quiver theory \eqref{6dHiggs1}. In this case, there is another ambiguity of distributing D7-branes in addition to the distribution of D5-branes which we saw in section \ref{subsubsec:5dsuquiver1}. Due to this distribution ambiguity of D7-branes, one can allocate $k_{n-l}$ D7-branes for $1 \leq i \leq n-1$, which originate from the $k_{n-l}$ D8-branes in the Type IIA brane configuration in Figure \ref{Fig:6dquiver-2}, to the $i$-th left column and the $i$-th right column from the center column. Again, the number of D7-branes in the middle column is fixed to be $k_n$. We choose the branch cuts of D7-branes in the columns left from the center extend in the left direction, and the branch cuts of D7-branes in the columns right from the center extend in the right direction. Furthermore, we also assume that the branch cuts of $k_n$ D7-branes extend in the left direction. 

\begin{figure}
\begin{center}
\includegraphics[width=15cm]{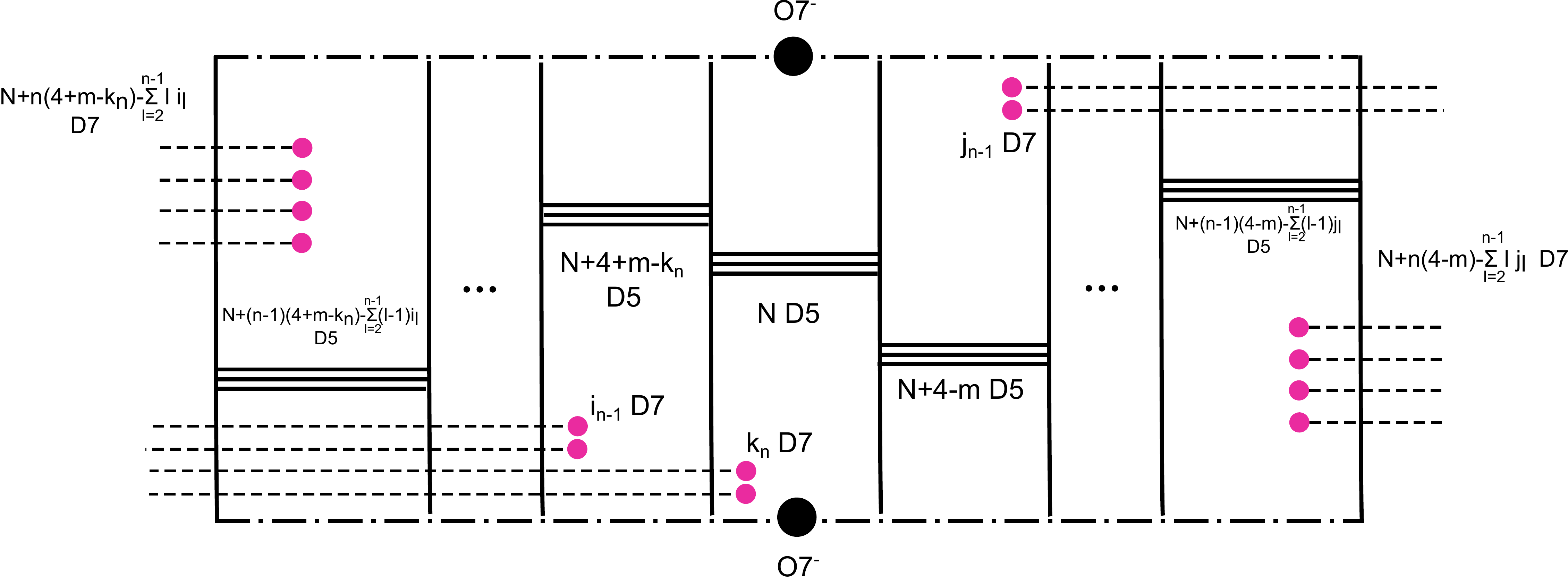}
\end{center}
\caption{Type IIB brane configuration after performing a T-duality to Figure \ref{Fig:6dquiver-2}. We defined non-negative numbers $i_l, j_l$ such that they satisfy $i_l + j_l = k_l$ for $l=1, \cdots, n-1$.} 
\label{Fig:Brane6dHiggsed1}
\end{figure}

The requirement that the final 5-brane web diagram admits a 5d gauge theory interpretation constrains the number of D5-branes in each column except for the center. The number of D5-branes in the middle column is again always $N$. Compared to the case in section \ref{subsubsec:5dspsuquiver}, we have the branch cuts of D7-branes in the columns that extend in either left or right direction. The branch cuts also affect the number of D5-barnes in each column which gives a web diagram admitting a 5d gauge theory description at the final stage after pulling all the 7-branes outside. It turns out that the 5-brane configuration depicted in Figure \ref{Fig:Brane6dHiggsed1} yields a 5-brane web which admits a 5d gauge theory interpretation. 

In order to go from the brane configuration in Figure \ref{Fig:Brane6dHiggsed1} to a 5-brane web yielding a 5d gauge theory, we take two steps as in section \ref{subsubsec:5dsuquiver1} and \ref{subsubsec:5dspsuquiver}. The first step is we resolve either two or one of the $O7^-$-planes. When we split two $O7^-$-planes, we fix the splitting type of the upper $O7^-$-plane into {\bf B} and {\bf C} 7-branes, and the splitting type of the lower $O7^-$-plane is chosen to be ${\bf X}_{[1+m, -1]}\; {\bf X}_{[1-m,1]}$ so that the final 5-brane web admits a 5d gauge theory interpretation. When we split only one of the two $O7^-$-planes, the splitting type may be generically {\bf B} and {\bf C} 7-branes. The second step is moving the branch cuts of 7-branes which arise by the quantum resolution of two or one $O7^-$-plane. After the second step, the 5-brane loops are divided by vertical lines and the structure eventually leads to a 5d quiver theory. The procedure is completely parallel to the one in section \ref{subsubsec:5dsuquiver1} and \ref{subsubsec:5dspsuquiver}. 

\begin{figure}
\begin{center}
\includegraphics[width=12cm]{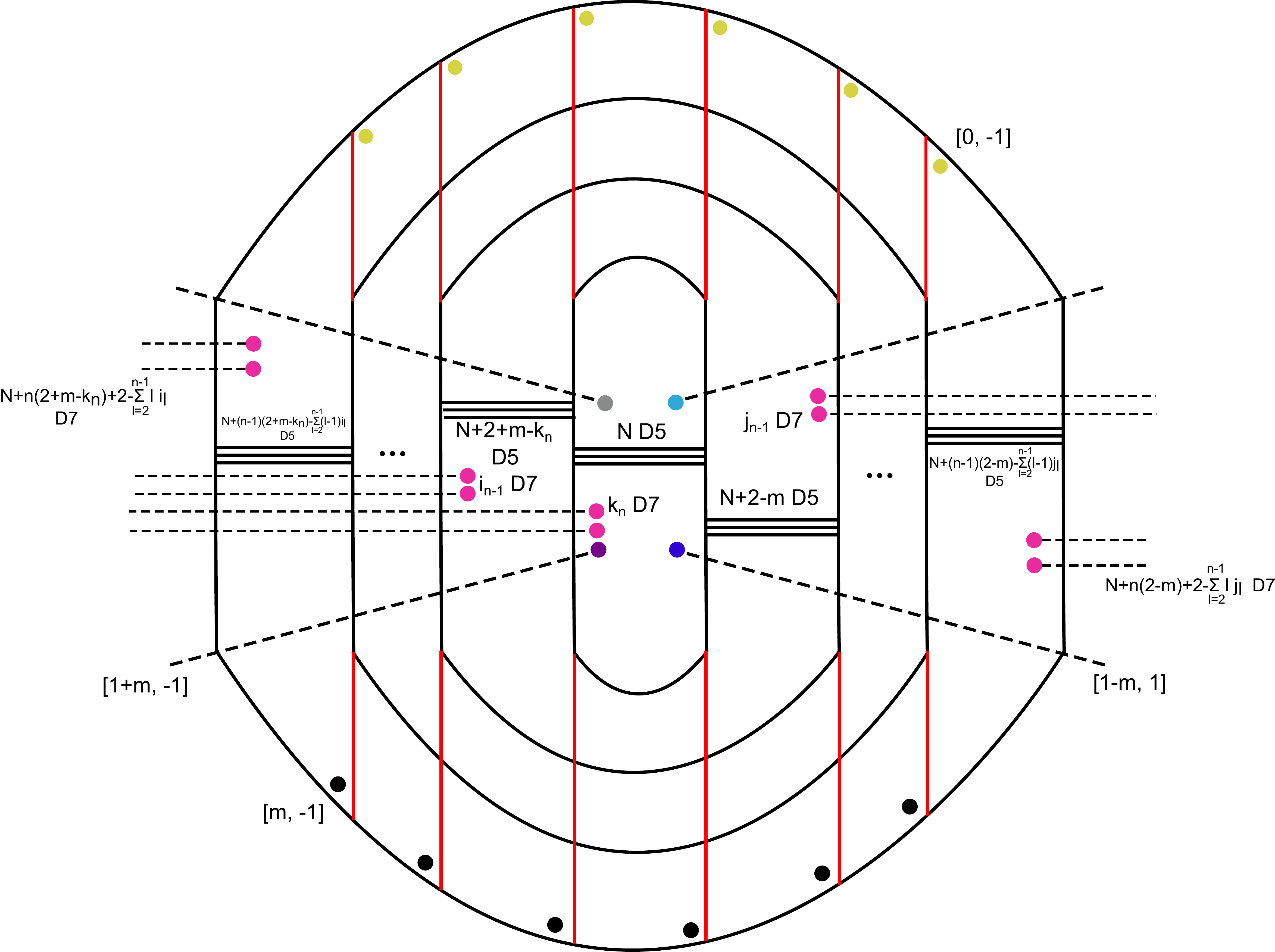}
\end{center}
\caption{Type IIB brane configuration after the two steps from the one in Figure \ref{Fig:Brane6dHiggsed1} when two $O7^-$-planes are resolved.} 
\label{Fig:5braneloopH1}
\end{figure}

When we split the two $O7^-$-planes, we obtain the 5-brane configuration in Figure \ref{Fig:5braneloopH1} after the two steps.
After pulling all the 7-branes from Figure \ref{Fig:5braneloopH1}, the final 5-brane web configuration gives rise to a 5d quiver theory with $SU$ gauge nodes with various flavors attached to each node. The explicit expression is 
 \begin{eqnarray}
{\overset{\overset{\text{\large$[L_1]$}}{\textstyle\vert}}{SU(N_1)  }} -{\overset{\overset{\text{\large$[L_2]$}}{\textstyle\vert}}{SU(N_2)  }} - \cdots - {\overset{\overset{\text{\large$[L_{n-1}]$}}{\textstyle\vert}}{SU(N_{n-1})  }} - {\overset{\overset{\text{\large$[k_n]$}}{\textstyle\vert}}{SU(xN+2n)  }}- {\overset{\overset{\text{\large$[R_{n-1}]$}}{\textstyle\vert}}{SU(M_{n-1})  }} - \cdots  -{\overset{\overset{\text{\large$[R_2]$}}{\textstyle\vert}}{SU(M_2)  }}  - {\overset{\overset{\text{\large$[R_1]$}}{\textstyle\vert}}{SU(M_1)  }}\nonumber\\
\label{5dsuquiverhiggsed}
\end{eqnarray}
where 
\begin{eqnarray}
N_p &=& N + 2n + (n-p)(m - k_n) - \sum_{l=p+1}^{n-1}(l-p)i_l\qquad \text{for} \quad 1 \leq p \leq n-1,\\
M_p &=& N + 2n - (n-p)m - \sum_{l=p+1}^{n-1}(l-p)j_l \qquad \text{for} \quad1 \leq p \leq n-1,\\
L_p &=& i_p, \qquad \text{for} \quad 2 \leq p \leq n-1,\\
L_1 &=& N+2n+n(m-k_n)-\sum_{l=2}^{n-1}l i_l +2,\\
R_p&=& j_p, \qquad \text{for} \quad 2 \leq p \leq n-1,\\
R_1 &=& N+2n-nm-\sum_{l=2}^{n-1}l j_l +2
\end{eqnarray}
 where $(i_l + j_l) = k_l$ for $l=1, \cdots, n-1$, and the rank of each gauge group as well as the number of the flavors should be larger than zero at least. 
 
\begin{figure}
\begin{center}
\includegraphics[width=12cm]{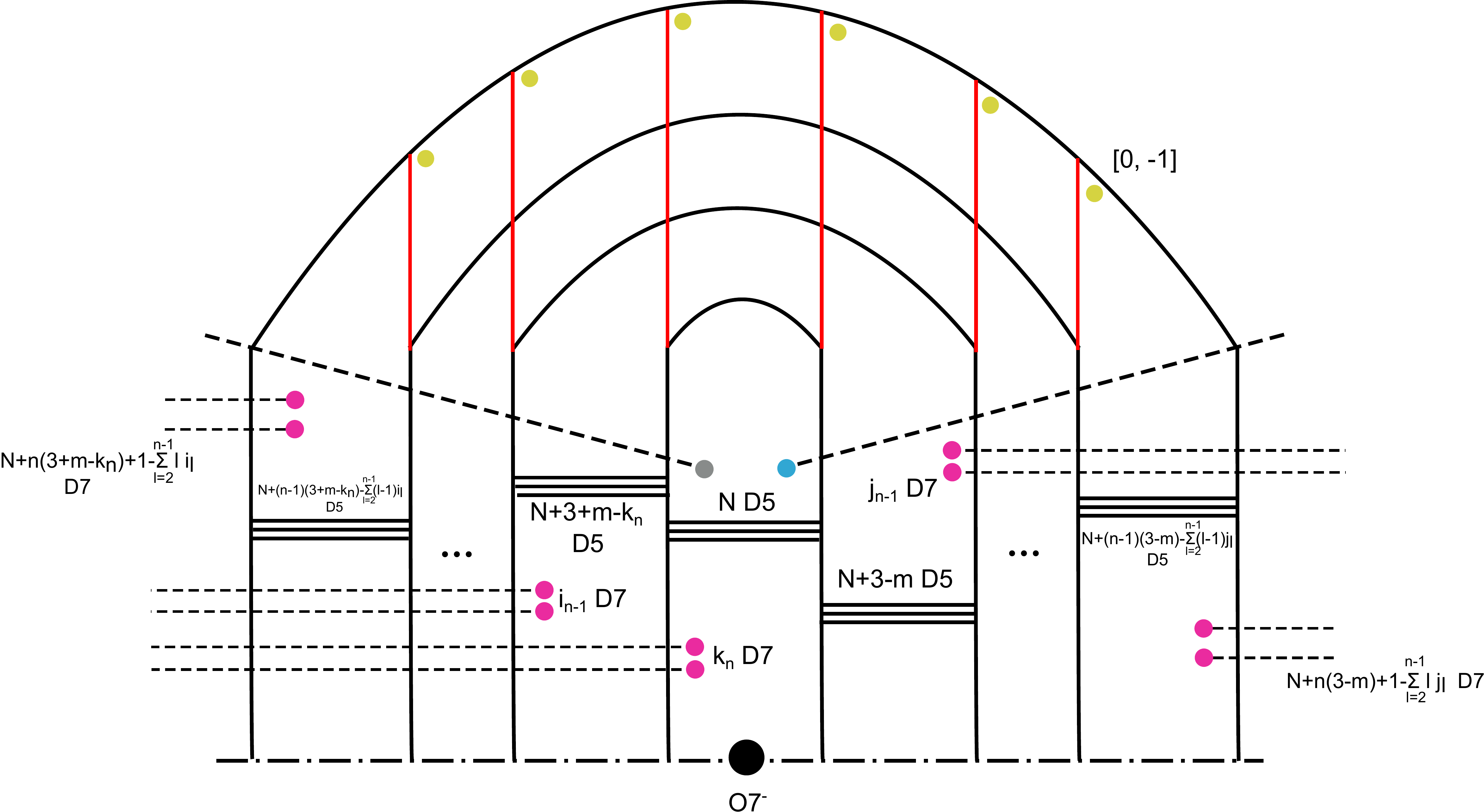}
\end{center}
\caption{Type IIB brane configuration after the two steps from the one in Figure \ref{Fig:Brane6dHiggsed1} when one $O7^-$-plane is resolved.} 
\label{Fig:5braneloopH2}
\end{figure}

When we split only one of the $O7^-$-planes, we obtain the 5-brane configuration in Figure \ref{Fig:5braneloopH2} after the two steps.
After pulling all the 7-branes from Figure \ref{Fig:5braneloopH2}, the final 5-brane web configuration gives rise to a 5d quiver theory with one $Sp$ gauge node and other $SU$ gage nodes with various flavors attached to each node. The explicit expression is 
\begin{eqnarray}
{\overset{\overset{\text{\large$ [k_{n}] $}}{\textstyle\vert}}{ Sp(N+n)  }}- 
{\overset{\overset{\text{\large$ [k_{n-1}]$}}{\textstyle\vert}}{SU(N_1)  }} &-& 
{\overset{\overset{\text{\large$[k_{n-2}]$}}{\textstyle\vert}}{SU(N_2)  }} - \cdots-
{\overset{\overset{\text{\large$[k_1-2n+2]$}}{\textstyle\vert}}{SU(N_{n-1})   }},
\end{eqnarray}
where 
\begin{equation}
N_p = 2N+2n+4p-\sum_{l=n-p}^{n-1}(l-n+p+1)k_{l+1},\qquad \text{for} \quad 1 \leq p \leq n-1.
\end{equation}
In this case, the distribution ambiguity of D5-branes and D7-branes does not matter since $i$-th left column from the center and the $i$-th right column from the center are the same column due to the orientifold action.  

Since the two types of the theories have the same UV completion as the 6d SCFT, they are dual to each other, which is a further generalization of the claim in section \ref{subsubsec:5ddualsuquiver}. Furthermore, the 5d $SU$ quiver theory \eqref{5dsuquiverhiggsed} has parameters associated to the distribution of D5-branes and D7-branes. Again, all the combinations descend from the same 6d theory, we argue that they are dual to each other. The $90$ degrees or $45$ degrees rotation of the 5d theories also give various 5d dual theories.

It is also possible to decouple some flavors from the 5d theories. Then, each 5d theory will have a5d UV fixed point. Decoupling exactly the same flavors from the dual 5d theories should give another dualities between 5d theories which has the same UV completion as a 5d SCFT.

%% file: section4-2.tex

\subsection{6d \texorpdfstring{$SU$}{SU} quivers with an antisymmetric hypermultiplet}
\label{subsec:6dSUquivgen}

\begin{figure}
\begin{center}
\includegraphics[width=8cm]{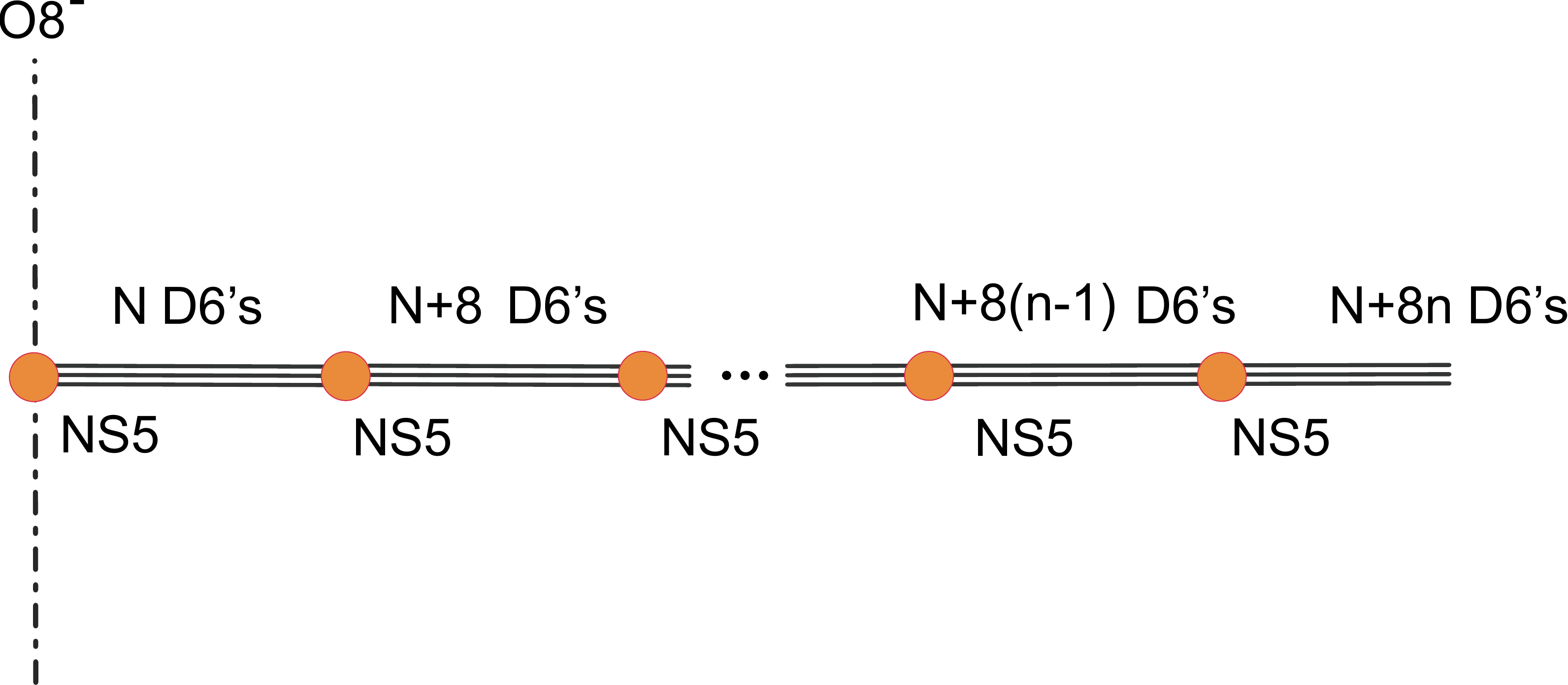}
\end{center}
\caption{Type IIA brane configuration for the 6d linear quiver theory \eqref{6dcanonicalA}.} 
\label{Fig:6dquiverA}
\end{figure}

In this subsection, we consider 6d $SU$ quiver gauge theories with the gauge node at the edge having
one hypermultiplet in the antisymmetric tensor representation, which is a generalization of what we studied in section \ref{sec:6dSUNA}.
We consider the generalization analogous to what we did in section \ref{sec:6dSpSUquiver}.
The simple generalization is the 6d linear quiver
\begin{eqnarray}
6d \; [1]_A- SU(N) - SU(N+8) - \cdots - SU(N+8(n-1)) - [N+8n], \label{6dcanonicalA}
\end{eqnarray}
The type IIA brane configuration is depicted in Figure \ref{Fig:6dquiverA}. 
Compared to Figure \ref{Fig:6dquiver-1}, we have an extra NS5-brane on top of $O8^-$-plane.
The global symmetry is generically $SU(N+8n) \times U(1)$.

\begin{figure}
\begin{center}
\includegraphics[width=8cm]{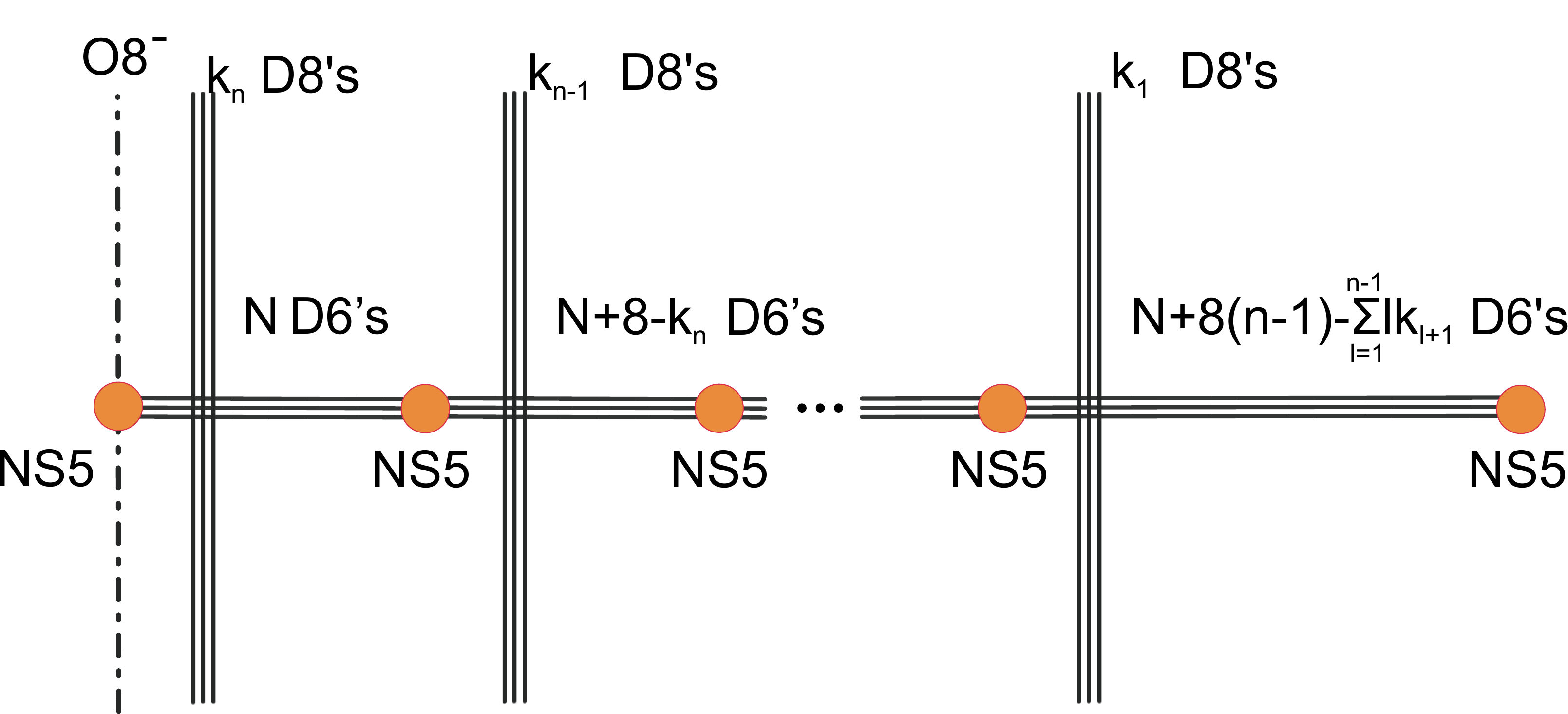}
\end{center}
\caption{Type IIA brane configuration for the 6d linear quiver theory \eqref{6dHiggsA}.} 
\label{Fig:6dquiverA-2}
\end{figure}

We can again further generalize this quiver by Higgsing which is induced exactly by the same mechanism discussed in section \ref{sec:6dSpSUquiver}.
By considering the Higgsing specified by the Young diagram $[n, \cdots, n, n-1, \cdots, n-1, \cdots, 2, \cdots, 2, 1, \cdots, 1]$ where the number of $l$ is $k_l$ with a condition $\sum_{l=1}^{n}lk_l = 2N+8n$,
we obtain the 6d quiver gauge theory
\begin{equation}
6d \quad \; 
{\overset{\overset{\text{\large$[k_{n}, 1_A]$}}{\textstyle\vert}}{SU(N) }} 
- {\overset{\overset{\text{\large$[k_{n-1}]$}}{\textstyle\vert}}{SU(N+8-k_n)}} - 
\cdots - 
{\overset{\overset{\text{\large$[k_1]$}}{\textstyle\vert}}{SU\Big(N+8(n-1) - \sum_{l=1}^{n-1}l\,k_{l+1}\Big)}}. \label{6dHiggsA}
\end{equation}
The brane setup for this theory is given in Figure \ref{Fig:6dquiverA-2}.
We will study these 6d quivers compactified on $S^1$ and consider their 5d descriptions 
as well as the 5d dualities.

\subsubsection{5d \texorpdfstring{$SU$}{SU} quivers}
\label{subsubsec:5dSUquivgen}

\begin{figure}
\begin{center}
\includegraphics[width=15cm]{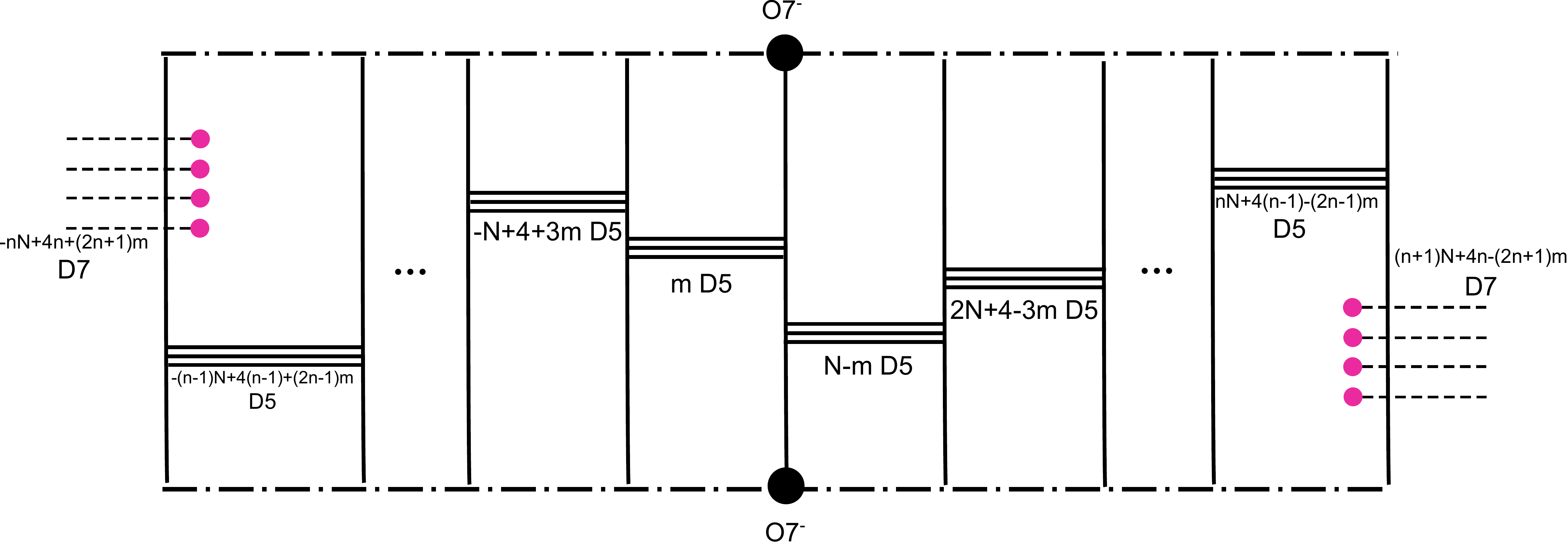}
\end{center}
\caption{Type IIB brane configuration after performing a T-duality to Figure \ref{Fig:6dquiverA}.} 
\label{Fig:Brane6dcanonicalA}
\end{figure}

First, we start with the 6d quiver gauge theory (\ref{6dcanonicalA}) on $S^1$,
T-duality along  the $S^1$ gives type IIB brane configuration.
Analogous to the brane setup discussed in section \ref{subsubsec:5dsuquiver1},
the distribution of the D5-branes should be considered.
There is no center column in this case and 
the $N+8(i-1)$ color D5 branes originated from the $i$-th gauge node in 6d quivers (\ref{6dcanonicalA})
are distributed into $i$-th left column and $i$-th right column from the center.
By imposing the condition that we should obtain a 5d gauge theory interpretation in the end, 
it turns out that the distribution ambiguity is again labeled by the single non-negative integer $m$.
The resulting distribution is depicted in Figure \ref{Fig:Brane6dcanonicalA}.

\begin{figure}
\begin{center}
\includegraphics[width=12cm]{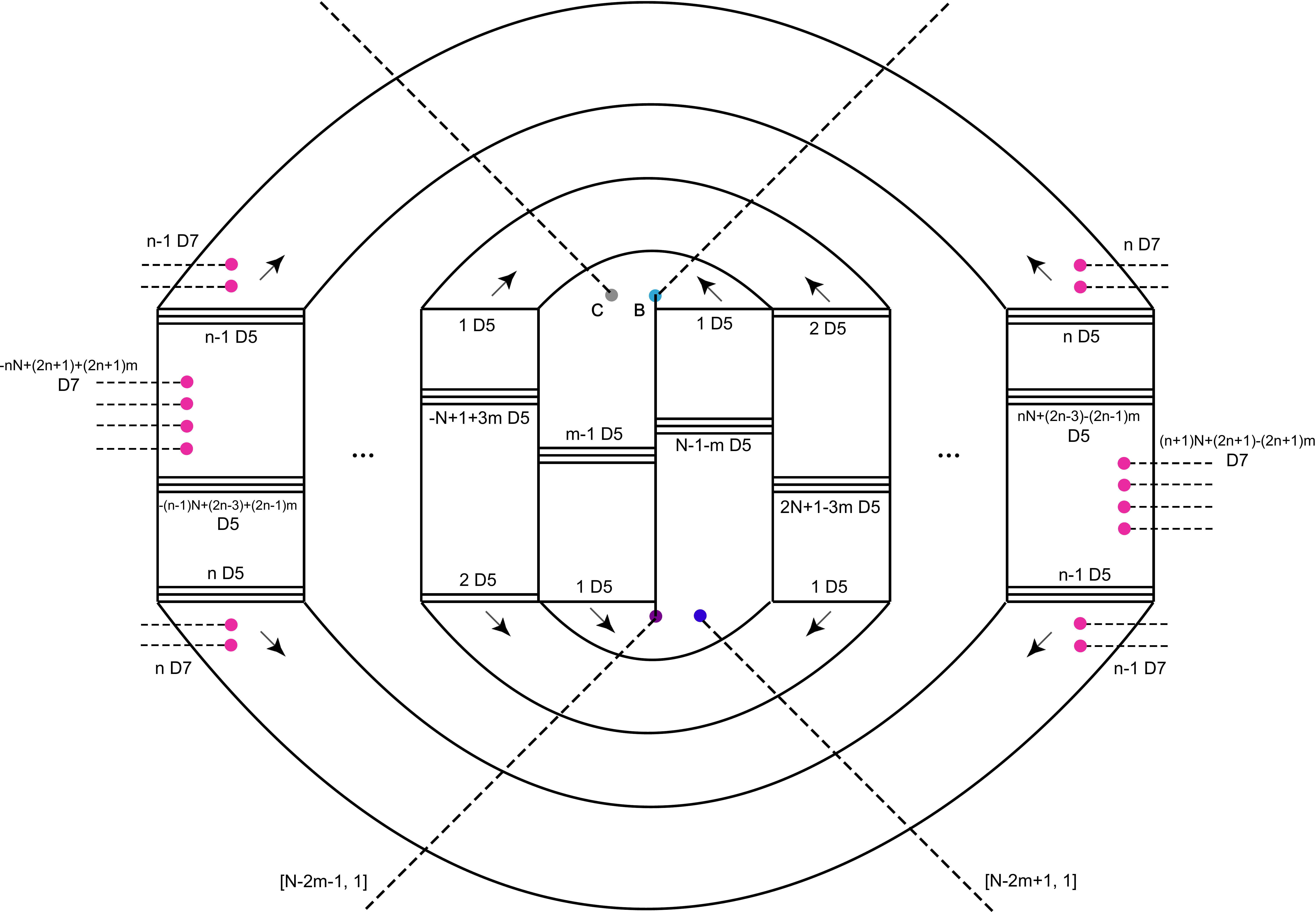}
\end{center}
\caption{The diagram after resolving two $O7^-$ planes in Figure \ref{Fig:Brane6dcanonicalA}.
The branch cuts of ${\bf B}, {\bf C}$ and ${\bf X}_{[N+2m-1, 1]}\; {\bf X}_{[N+2m+1,1]}$
are moved in such a way that the D5-branes and the D7-branes indicated by the arrows go across the cut.} 
\label{Fig: 5loopA}
\end{figure}

Next, we consider the resolution of the two $O7^-$-planes attached to 5-branes.
The charges of the split 7-branes are determined by the non-negative integer $m$ introduced above
due to the condition that one of such 7-branes should be attached to the 5-brane
to which originally $O7^-$-plane was attached.
If we fix the splitting type of the upper $O7^-$-plane to be 
{\bf B}, {\bf C}.
Then, the splitting type of the lower $O7^-$-plane is determined to be ${\bf X}_{[N-2m-1,1]}$, ${\bf X}_{[N-2m+1,1]}$. 
The resolution of the $O7^-$-planes create $n$ 5-brane loops as in Figure \ref{Fig: 5loopA}.
Then, we move the branch cut of these four 7-branes in such a way that 
some of the D5-branes and D7-branes go across the cut as dictated in the arrow \ref{Fig: 5loopA}.
Contrary to the case in Figure \ref{Fig:5braneloop2}, 
the number of D5-branes at upper left and lower right is one more than the ones at upper right and lower left.
Apart from this small asymmetry, the procedure is quite parallel to section \ref{subsubsec:5dsuquiver1}.

\begin{figure}
\begin{center}
\includegraphics[width=12cm]{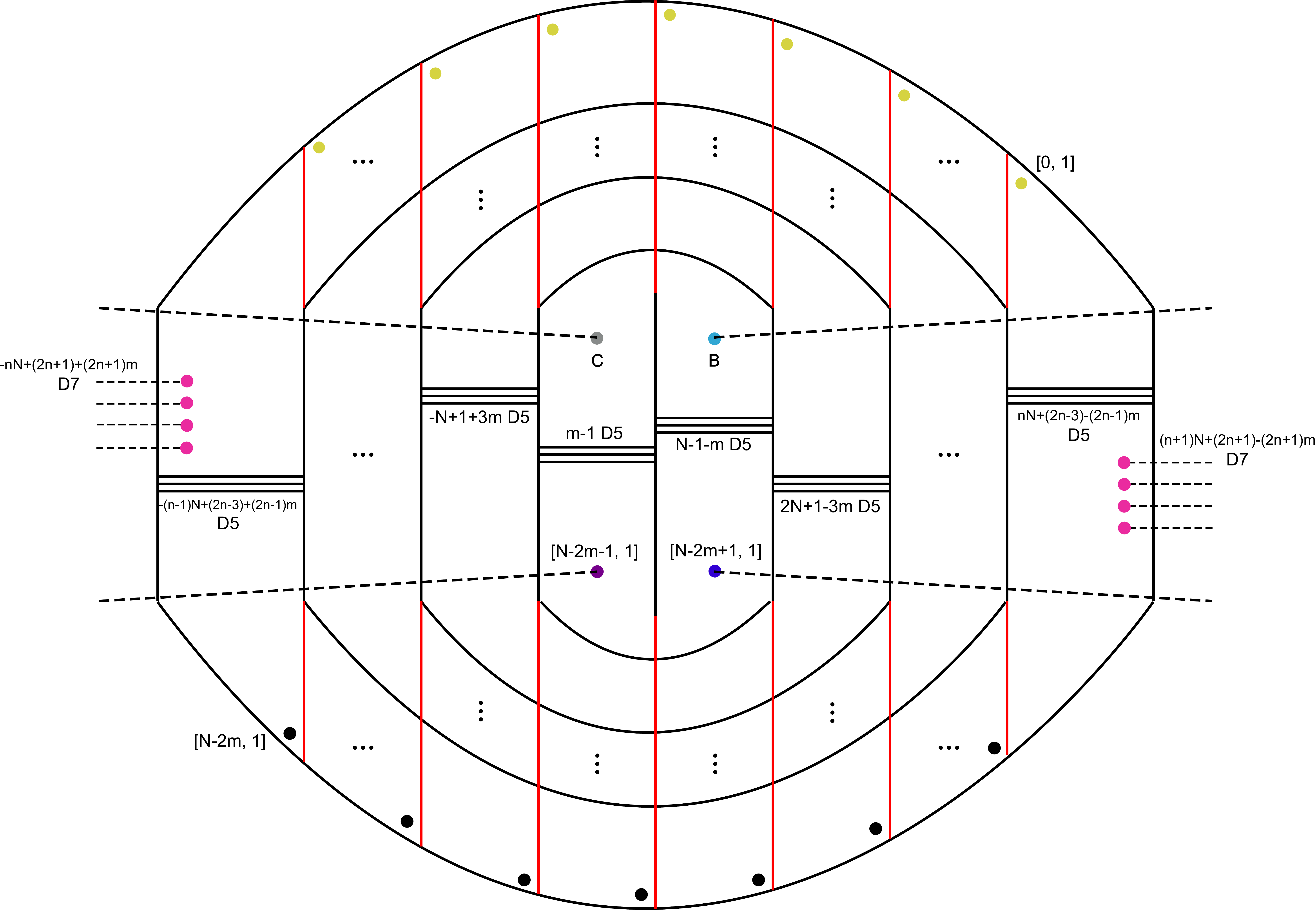}
\end{center}
\caption{Diagram after the motion in Figure \ref{Fig: 5loopA}.} 
\label{Fig:5loopA2}
\end{figure}

After this motion, we obtain the diagram in Figure \ref{Fig:5loopA2}.
The 5-branes depicted as red lines are the ones 
coming from the D5-branes which went across the 7-brane monodromy cut.
These 5-branes becomes part of the walls splitting each gauge node.
Moreover, it turns out that
the newly generated 5-brane loop gives
extra color D5-branes for each column
due to the tuned distribution parametrized by $m$ mentioned previously.
Then, from Figure \ref{Fig:5loopA2}, we see that 
$2n$ color branes are added to the two columns at the center
since all the $n$ 5-branes loops contribute as color D5-branes to these.
When we move to the next gauge node, the extra color D5-branes reduce by two
and there are only two additional color D5 branes at the columns at the both edges.

Therefore, after moving all the 7-branes outside,
we interpret this diagram as the following 5d quiver gauge theory:
\begin{eqnarray}
[N_0] - SU(N_1) - SU(N_2) - \cdots - SU(N_{2n}) - [N_{2n+1}]
\label{eq:gen5dquivA}
\end{eqnarray}
where
\begin{eqnarray}
N_0 &=& -nN+(2n+1)(m+1), \nonumber \\
N_\ell &=& 2n-1 + (-n+\ell) N + (2n-2\ell+1) m, \qquad \ell=1,\cdots, 2n, \\
N_{2n+1} &=& (n+1) N- (2n+1)(m-1). \nonumber 
\end{eqnarray}
The parameters are constrained such that the rank of each gauge group or the number of the flavors should be positive. 
In summary, we see that the 6d quiver theory (\ref{6dcanonicalA}) on $S^1$ is described by 
the 5d quiver gauge theory (\ref{eq:gen5dquivA}).


\subsubsection{5d \texorpdfstring{$SU$}{SU} quivers with an antisymmetric hypermultiplet}
\label{subsubsec:5dSUAgen}

\begin{figure}
\begin{center}
\includegraphics[width=12cm]{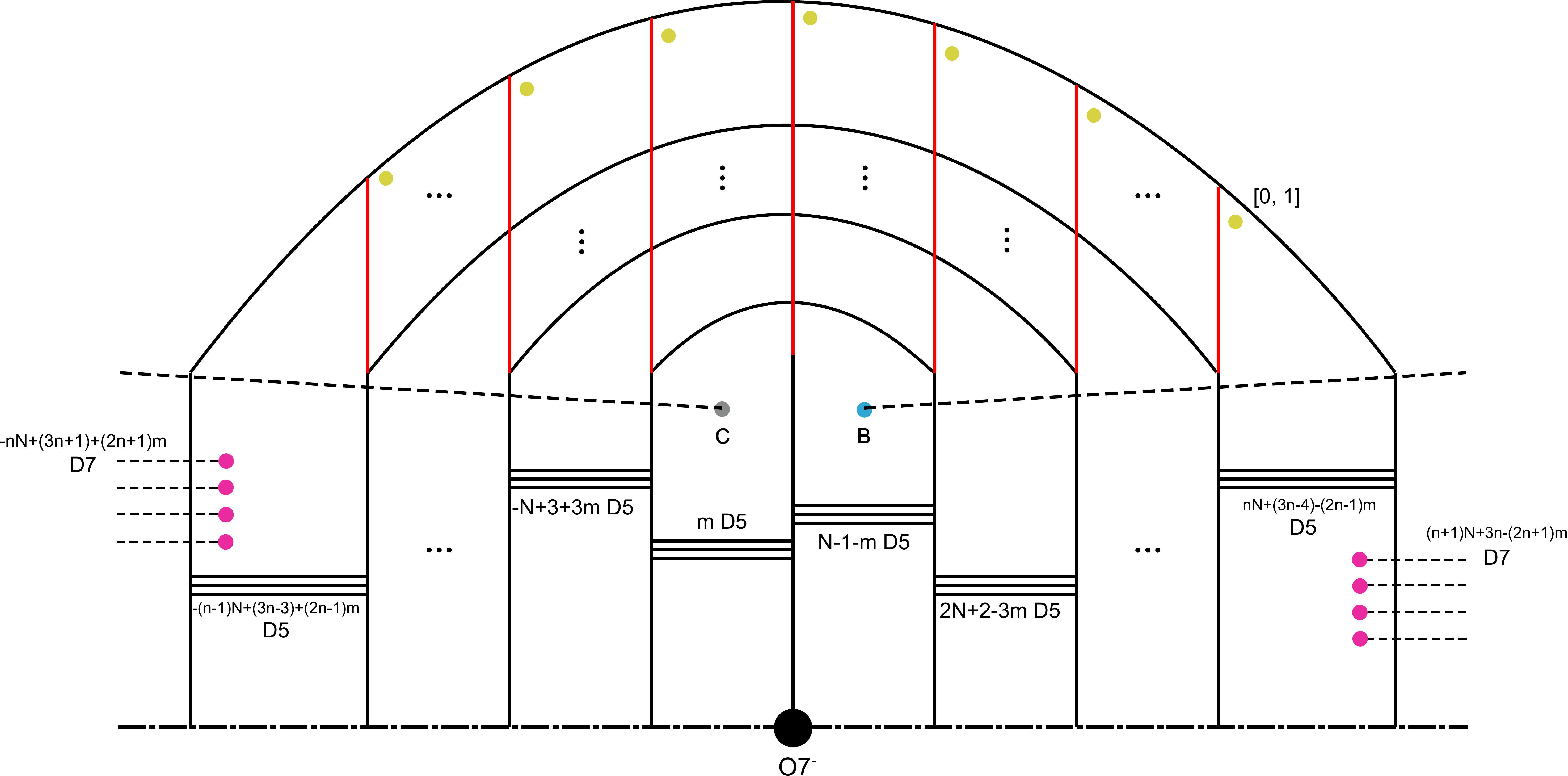}
\end{center}
\caption{The diagram for the case where only one of the two $O7^-$-planes is resolved.}  
\label{Fig:5AoneO7}
\end{figure}

Here, we consider the case of resolving only one of the two $O7^-$-planes in the diagram in Figure \ref{Fig:Brane6dcanonicalA}.  After we resolve the upper $O7^-$-plane into {\bf B} and {\bf C} 7-branes, 
we consider the same procedure, namely we move the branch cuts of the {\bf B} and {\bf C} 7-branes so that some of the D7-branes and D5-branes cross them. The explicit motion of the branch cuts or equivalently the motion of D7-branes and D5-branes is exactly the same one that we performed for the upper part of the diagram in Figure \ref{Fig: 5loopA}, 
while we keep the lower half part as it is.
After moving the branch cuts, it is straightforward to see that we obtain the diagram depicted in Figure \ref{Fig:5AoneO7}.
In this case, the distribution ambiguity does not matter as the $i$-th left column from the center is actually the same column as the $i$-th right column from the center by the orientifold action. In other words, we have $n$ $SU$ gauge nodes.

After moving all the 7-branes outside, we obtain a 5-brane web diagram. It is possible to can read off from the diagram that this theory is 
the following 5d quiver gauge theory:
\begin{eqnarray}
[1]_A - SU(N+2n-1) - SU(N+2n+3) - \cdots - SU(N+6n-5) - [N+6n+1]. \nonumber \\
\label{eq:gen5dA}
\end{eqnarray}
Note that we have an anti-symmetric hypermultiplet at the left end node as a 5-brane is attached to the lower unresolved $O7^-$-plane \cite{Bergman:2015dpa}.

\subsubsection{5d dualities}

In section \ref{subsubsec:5dSUquivgen} and in section \ref{subsubsec:5dSUAgen},
 we have discussed two different types of 5d description for the 6d theory (\ref{6dcanonicalA}),
 which are 5d quivers (\ref{eq:gen5dquivA}) and (\ref{eq:gen5dA}).
Since their type IIB diagrams come from the identical type IIA diagram for the 6d theory,
we claim that these two types of 5d theory are dual to each other,
which means that they have the identical 6d UV fixed point.

When we combine mass deformation and S-duality as discussed in section \ref{subsec:chain},
we will be able to obtain various dual 5d theories.
Furthermore, by considering the flavor decoupling limit,
all these dualities can be also reduced to the dualities for another set of theories 
which have an identical 5d UV fixed point.
Classifying all these dual 5d theories would be interesting future problem.

\subsubsection{Higgsed cases}
Now, we go on to the 6d theory (\ref{6dHiggsA}), which is obtained by Higgsing (\ref{6dcanonicalA}).
Diagrammatic derivation of the 5d description is quite parallel to what we did in section \ref{subsubsec:5dSUquivgen}
and in section \ref{subsubsec:5dSUAgen}.
The different point is that some of flavor D7-branes exist at internal columns.
The effect of such D7-branes are again quite parallel to what we discussed in section \ref{subsubsec:genHiggsSp}.
Here, we summarize the resulting 5d description.

When we resolve two $O7^-$ planes, we obtain the 5d quivers 
\begin{eqnarray}
{\overset{\overset{\text{\large$[L]$}}{\textstyle\vert}}{SU(N_1) }} 
- {\overset{\overset{\text{\large$[i_2]$}}{\textstyle\vert}}{SU(N_2) }} 
- \cdots 
- {\overset{\overset{\text{\large$[i_n]$}}{\textstyle\vert}}{SU(N_n) }} 
- {\overset{\overset{\text{\large$[j_n]$}}{\textstyle\vert}}{SU(M_{n}) }} 
- \cdots 
- {\overset{\overset{\text{\large$[j_2]$}}{\textstyle\vert}}{SU(M_{2}) }} 
- {\overset{\overset{\text{\large$[R]$}}{\textstyle\vert}}{SU(M_{1}) }} 
\label{eq:gen5dquivA2}
\end{eqnarray}
where
\begin{eqnarray}
L &=& -nN+(2n+1)(m+1) - \sum_{l=2}^{n}li_l ,
 \nonumber \\
N_p &=& 2n-1 + (-n+p) N + (2n-2p+1) m - \sum_{l=p+1}^n (l-p) i_l , \qquad p=1,\cdots, n, 
 \nonumber \\
M_p &=& 2n-1 + (n-p+1) N - (2n-2p+1) m - \sum_{l=p+1}^n (l-p) j_l, \qquad p=1,\cdots, n,
 \nonumber \\
R &=& (n+1) N- (2n+1)(m-1)  - \sum_{l=2}^{n}l j_l. \nonumber 
\end{eqnarray}
where $i_l + j_l = k_l$. 

When we resolve only one out of the two $O7^-$ planes, we obtain the 5d quivers 
\begin{eqnarray}
{\overset{\overset{\text{\large$[k_{n}, 1_A]$}}{\textstyle\vert}}{SU(N_1) }} 
- {\overset{\overset{\text{\large$[k_{n-1}]$}}{\textstyle\vert}}{SU(N_2)}} 
- {\overset{\overset{\text{\large$[k_{n-2}]$}}{\textstyle\vert}}{SU(N_3) }} 
 - \cdots 
- {\overset{\overset{\text{\large$[k_{2}]$}}{\textstyle\vert}}{SU(N_{n-1}) }} 
- {\overset{\overset{\text{\large$[k_1-2n+1]$}}{\textstyle\vert}}{SU(N_n) }} 
\end{eqnarray}
where
\begin{eqnarray}
N_s=N+2n-5 + 4s - \sum_{l=1}^{s-1}lk_{l+1+n-s}.
\end{eqnarray}

We claim that these two types of the theories have the same 6d UV fixed point.
Also, the 5d SU quiver theory (\ref{eq:gen5dquivA2}) with any possible value for $m$, $i_l$, and $j_l$ 
give the set of dual theories.
S-duality of these 5d theories will also give various dual theories.
Flavor decoupling limit will give other set of dual theories which have the same 5d UV fixed point.

%% file: section5.tex
\bigskip
\section{Special cases: the \texorpdfstring{6d quiver gauge theories of $O8^- + 8 D8$'s configuration}{O8+8D8}}\label{sec:speicalcases}

In this section, we consider the 6d SCFT configurations in tensor branch which are composed of an $O8^-$ plane together with eight D8 branes, leading to indefinite sequence of quiver type made of D6 branes stretched between two NS5 branes. These cases can be, in principle, obtained through the Higgsing of general 6d brane configuration discussed in the previous section. It is however non-trivial to get 5d description following the procedure described in the previous sections. It is partial because, to have Lagrangian description in 5d, many D7 branes are converted to other 7-branes resulting in lack of flavor 7-branes. On the other hand, we discuss below that one can still have an Lagrangian description in the S-dual frame or by implementing the $SL(2,\mathbb{Z})$ transformation on the $(p,q)$ plane. These special cases also give intriguing dual pictures depending the resolution of the $O7^-$ plane into a pair of 7-brane. 

\begin{figure}
\begin{center}
\includegraphics[width=7cm]{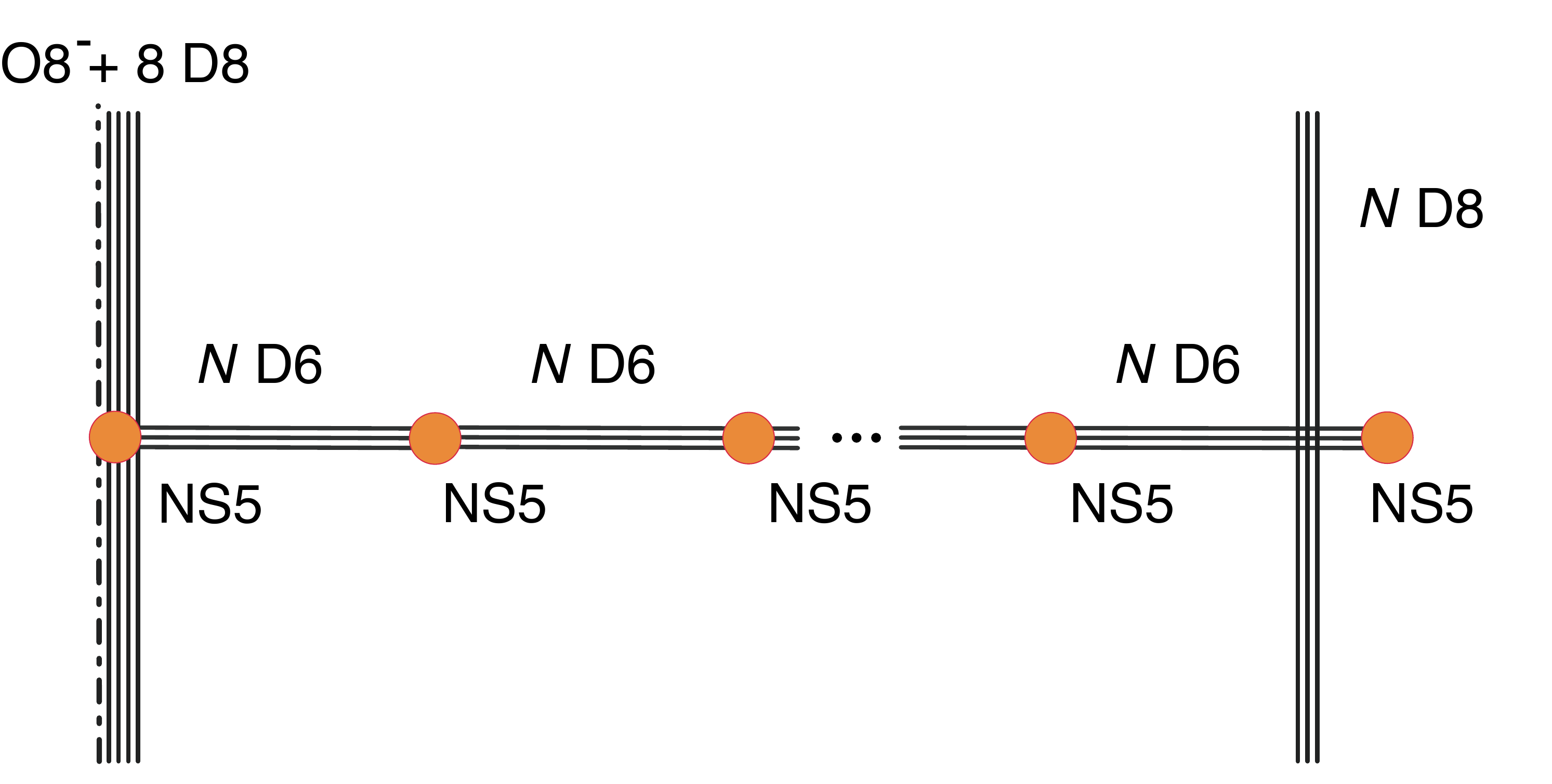}
\end{center}
\caption{6d brane configuration for the quiver $[1_A,8] - SU(N) - SU(N) - \cdots - SU(N) - [N]$ which has an antisymmetric hypermultiplet and eight fundamental flavors of the leftmost gauge group of the quiver also coupled to tensor multiplets.}
\label{Fig:A8SUSU}
\end{figure}
\subsection{\texorpdfstring{6d $[1_A,8]-SU(N)-SU(N)-\cdots-[N]$ quiver}{[A,8]-SU(N)-SU(N)-}}\label{subsec:18sususun}
We first consider the case where an $O8^-$ with eight D8 branes and a NS5 brane being on top of the $O8^-$ plane,  
\begin{align}
{\rm 6d} ~[1_A,8] - \underbrace{SU(N) - SU(N) - \cdots - SU(N)}_\text{$n$ nodes} - [N].
\end{align}
The corresponding 6d brane configuration is given in Figure \ref{Fig:A8SUSU}.

\paragraph{\texorpdfstring{$\mathbf{SL(2,\mathbb{Z})}$}{SL2Z} invariant 7-brane combinations.}
We will discuss various dualities along the reduction of the 6d theories to 5d theories by analyzing the 7-brane monodromies with or without the $O7^-$ planes. In order to see the dual picture, it is useful to consider combinations of 7-branes which are invariant under the $SL(2,\mathbb{Z})$ transformation on the $(p,q)$ 5-brane web plane. An $O7^-$ plane and four D7 branes would be an obvious example among many $SL(2,\mathbb{Z})$ invariant 7-brane combinations, as the total monodromy of them is proportional to minus identity matrix. 
It follows immediately that a pair of 7-branes as a resolution of the $O7^-$ plane, together with four D7 branes, is thus $SL(2,\mathbb{Z})$ invariant. For instance, $(\mathbf{B},\mathbf{C})$ or $(\mathbf{N},\mathbf{X})$ 7-branes with four D7 branes (four $\mathbf{A}$ 7-branes). We will show a few distinctive examples of such combinations involving 7-branes from different resolutions of the $O7^-$ plane. In particular, the cases where one of the 7-branes is attached to a 5-brane of the same charge will give rise to fruitful dualities among the resulting 5d quiver gauge theories. We consider such combinations frequently appearing through T-duality when reducing a 6d theory to 5d. 

\begin{figure}
\begin{center}
\includegraphics[width=9cm]{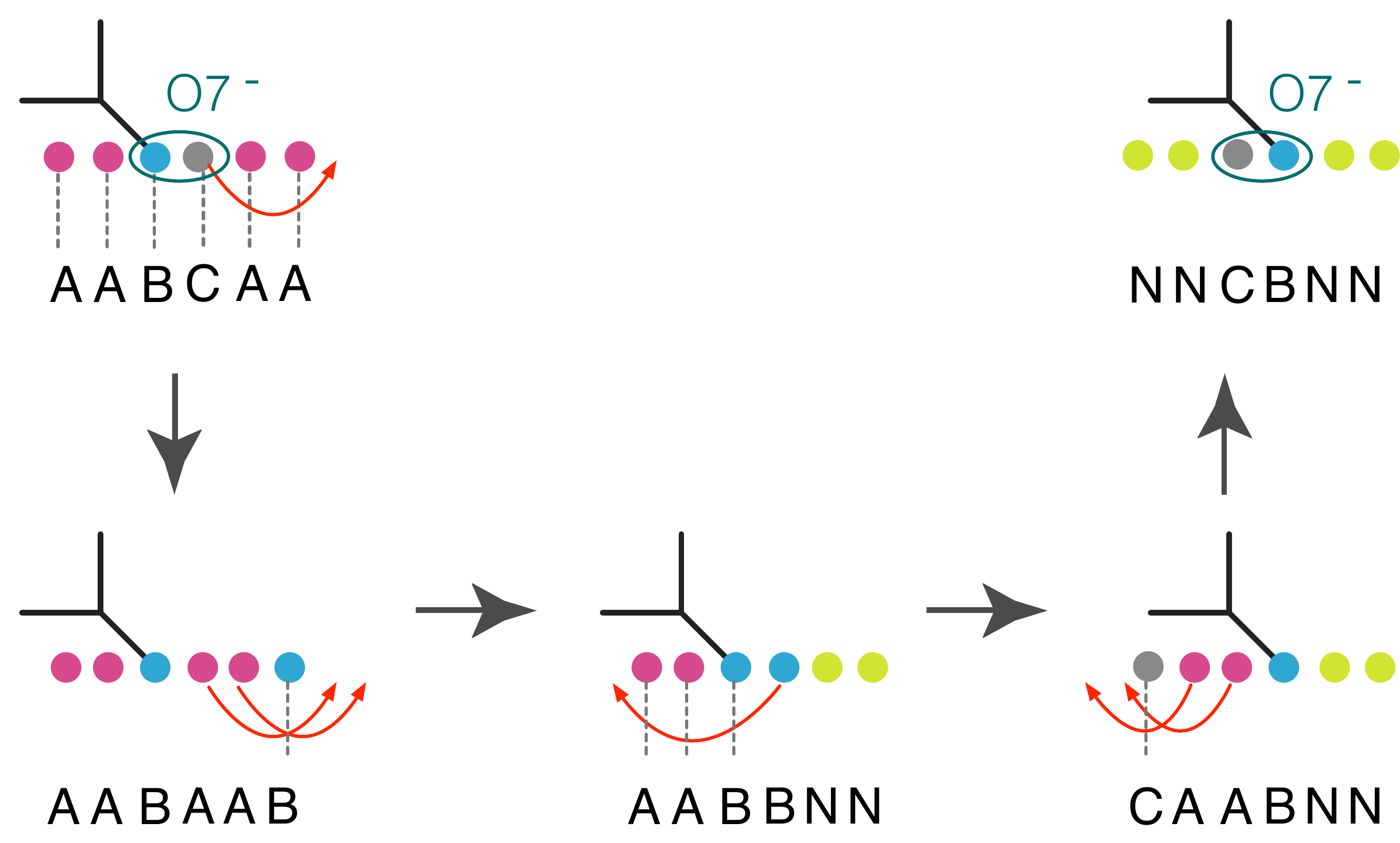}
\end{center}
\caption{(i) An $SL(2,\mathbb{Z})$ invariant 7-brane combination where $\bf B$ 7-brane is attached to 5-brane. A sequence of monodromy analysis makes $\bf A^2BCA^2$ into $\bf N^2CBN^2$ which will be converted into $\bf A^2BCA^2$ again via an S-dual action.} 
\label{Fig:SL2Zcom1}
\end{figure}
As discussed earlier, when going from 6d to 5d, one can distribute D7 branes and D5 branes with a suitable Wilson line, which enables us to allocate any number of D7 branes close to $O7^-$ planes. Regarding resolution of $O7^-$ planes, we restrict ourselves to two distinctive resolutions of $O7^-$ planes, which is either ($\bf B, C$) or ($\bf N, X_{[2,1]}$) 7-brane pair. 

(i) Consider the 7-brane configuration 
$\bf A^2BCA^2$ where $\bf B$ is attached to a $(1,-1)$ 5-brane as depicted in Figure \ref{Fig:SL2Zcom1}.  It is straightforward to show that it can be expressed as the $\bf N^2CBN^2$ 7-brane configuration where $\bf B$ is still attached to the 5-brane\footnote{Using the 7-brane monodromies, one finds that
\begin{align}
\bf CA = AN = NC, \quad NA = AB= BN, \quad {\rm or}\quad 
\bf N^2C= CA^2 = A^2B = BN^2, 
\end{align}
which yield
\begin{align}
{\bf A^2 \bar{B}C A^2} = {\bf A^2 \bar{B} A^2B} = {\bf A^2 \bar{B} BN^2 }= {\bf C A^2 \bar{B} N^2}=  {\bf  N^2 C \bar{B} N^2},
\end{align}
where the 7-brane $\bf \bar{B}$ is attached to $(1,-1)$ 5-brane. 
A pictorial version of this monodromies is given in Figure \ref{Fig:SL2Zcom1}.} 
(See Figure \ref{Fig:SL2Zcom1}).
By performing an S-duality transformation $(p,q)\to (-q,p)$, one sees that 
 ${\bf A^2 {B}C A^2} $ is an S-dual invariant combination
\begin{align}
{\bf A^2 \bar{B}C A^2} = {\bf  N^2 C \bar{B} N^2} \quad \overset{\rm S-dual}{\longrightarrow}\quad {\bf A^2 \bar{B}C A^2},
\end{align}
where we used a bar ``$\bar{~~~~}$'' to denote the 7-brane is attached to a 5-brane of the same charge.

\vskip 0.5cm

\begin{figure}
\begin{center}
\includegraphics[width=11cm]{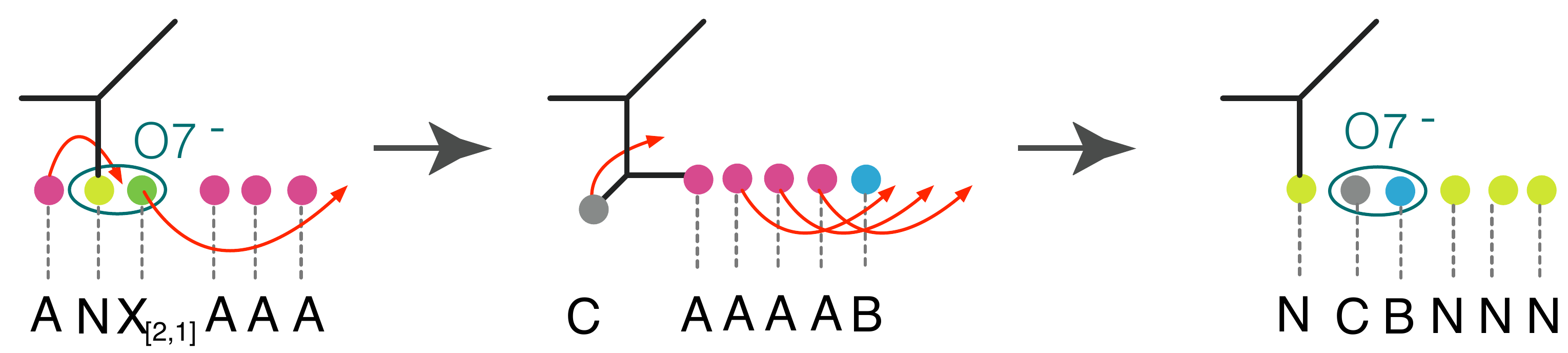}
\end{center}
\caption{(ii) An $SL(2,\mathbb{Z})$ invariant 7-brane combination where the 7-brane $\bf N$ is attached to (0,1) 5-brane. A sequence of monodromy analysis makes $\bf ANX_{[2,1]}A^3$ into $\bf NCBN^3$.}
\label{Fig:SL2Zcom2}
\end{figure}
\noindent (ii) Another example is ${\bf A {N}X_{[2,1]} A^3}$ with $\bf N$ 7-brane being attached to $(0,1)$ 5-brane. See Figure \ref{Fig:SL2Zcom2}. Again a little monodromy calculation 
\footnote{The relevant monodromy relation for this case is 
\begin{align}
\bf X_{[2,1]} A = AC , \quad CA = AN=NC, 
\end{align}
yielding
\begin{align}
{\bf A {N}X_{[2,1]} A^3} = {\bf CA^3 B} = {\bf NCBN^3} .
\end{align}
}
leads that ${\bf A {N}X_{[2,1]} A^3}$ is an $SL(2,\mathbb{Z})$ invariant 7-brane combination
\begin{align}
{\bf A \bar{N}X_{[2,1]} A^3} = {\bf \bar{N} CB N^3} \quad \overset{\rm S-dual}{\longrightarrow}\quad {\bf  \bar{A}BC A^3},
\end{align}
as $\bf NX_{[2,1]}$ and $\bf BC$ are related by a successive application $T$-action of the $SL(2,\mathbb{Z})$ transformation.  
Notice that before taking an S-duality, the $\bf N$ 7-brane is attached to $(0,1)$ 5-brane, but after the S-duality and a manipulation of monodromies as well as the Hanany-Witten transition, the $\bf A$ 7-brane is attached to a D5 brane. This procedure plays an important role in showing a dual description of 5d quiver theories which we discuss below.
\vskip 0.5cm

\begin{figure}
\begin{center}
\includegraphics[width=10cm]{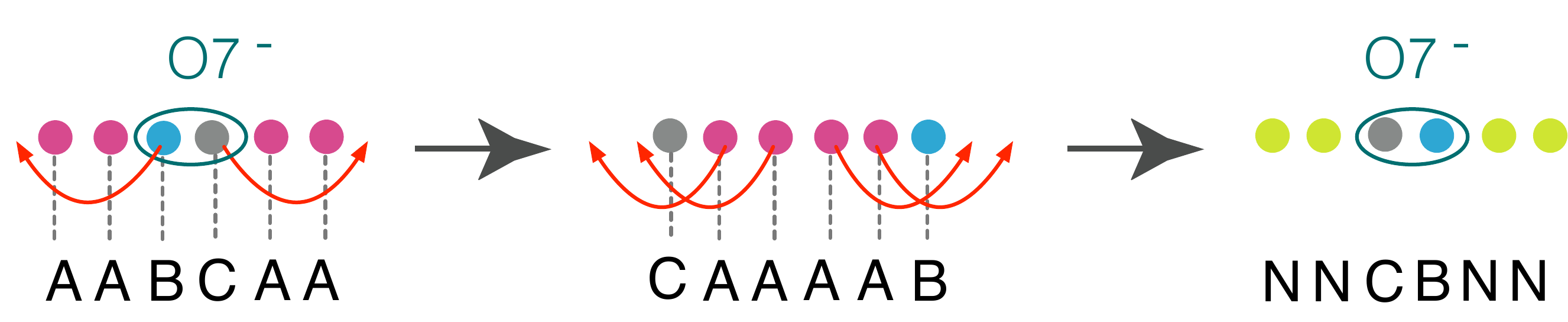}
\end{center}
\caption{(iii) An $SL(2,\mathbb{Z})$ invariant 7-brane combination where no 7-brane is attached to 5-brane. A sequence of monodromy analysis makes $\bf A^2BCA^2$ into $\bf N^2CBN^2$.
} 
\label{Fig:sl2zcom3}
\end{figure}
\noindent (iii) When no 5-brane is attached to 7-branes, it is easier for one to see $SL(2,\mathbb{Z})$ invariance. For instance, consider $\bf A^2 O7^- A^2$ or $\bf A^2 BC A^2$. (See also Figure \ref{Fig:sl2zcom3}.)
 Using the monodromy, one finds that ${\bf A^2 BC A^2}$ is S-dual invariant  
\begin{align}
{\bf A^2 BC A^2} = {\bf C A^2A^2B} = {\bf N^2 C BN^2} \quad \overset{\rm S-dual}{\longrightarrow}\quad {\bf A^2BC A^2}.
\label{a2bca2no5bra}
\end{align}

With all these $SL(2,\mathbb{Z})$ invariant 7-brane combinations, (i), (ii), and (iii), we study special brane configurations which yield various 5d quiver theories.

\bigskip
\paragraph{\texorpdfstring{6d $[1_A,8]-SU(1)-SU(1)-\cdots-[1]$ quiver and 5d $Sp$ theory with $N_f=8$ and $N_a=1$.}{[1_A,8]-SU(1)-SU(1)-}}
\begin{figure}
\begin{center}
\includegraphics[width=15cm]{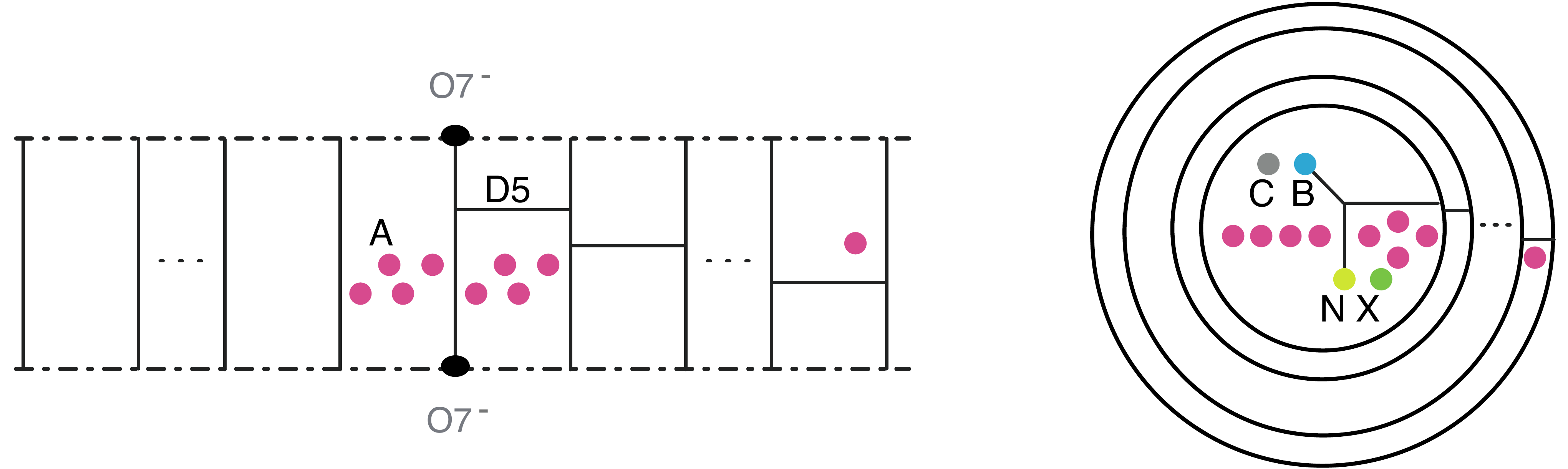}
\end{center}
\caption{Left: 5d brane configuration for the quiver $[1_A,8] - SU(1) - SU(1) - \cdots - SU(1) - [1]$. ~Right:
Splitting the $O7^-$ plane gives 5-brane loops which are denoted as circular loops. 
}
\label{Fig:A85dconf}
\end{figure}

\begin{figure}
\begin{center}
\includegraphics[width=9cm]{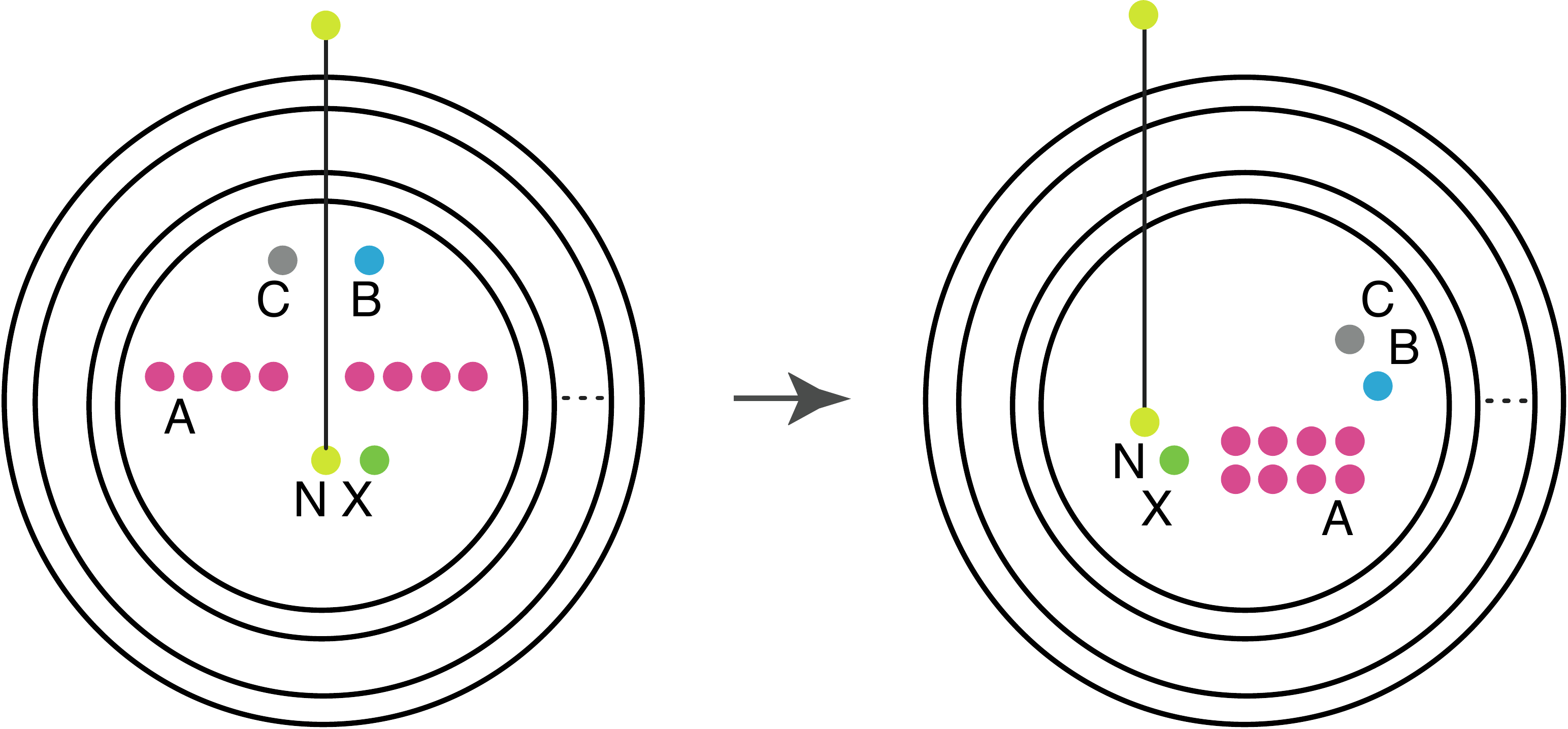}
\end{center}
\caption{6d $[1_A,8] - SU(1) - SU(1) - \cdots - SU(1) - [1]$ giving 5d $Sp(n)$ with $N_f=8$ and $N_a=1$. From the figure on the right of Figure \ref{Fig:A85dconf}, one moves $\bf B$ 7-brane along the direction of the charge $(1,-1)$ so that the $\bf B$ brane comes a freely floating 7-brane. This procedure gives a configuration on the left here.  Using the 7-brane monodromy, one can put all the 7-brane on one side like the figure on the right. 
}
\label{Fig:SpnwithA}
\end{figure}
For the $N=1$ case with $n$ quiver nodes, we claim that the 6d brane configuration gives rise to 5d $Sp(n)$ gauge theory with $N_f=8$ and $N_a=1$ hypermultiplets:
\begin{align}
{\rm 6d~} [1_A,8]-\underbrace{SU(1)-\cdots-SU(1)}_\text{$n$ nodes}-[1] ~\Rightarrow~  {\rm 5d}~ Sp(n)~{\rm theory~ with}~ N_f=8~ \& ~N_a=1.
\end{align}
A circle compactification followed by T-duality makes an $O8^-$ into two $O7^-$'s separated maximally along the T-dual circle. NS5 brane stuck on the $O8^-$ is now a brane junction of NS5 brane connecting two $O7^-$s as well as a D5 brane. Because of this junction, when resolving the $O7^-$ planes into two pairs of two 7-branes, each $O7^-$ is resolved differently, for instance, 
($[-1,1]$ and $[1,1]$) 7-branes for an $O7^-$ plane, and ($[0,1]$ and $[2,1]$) 7-branes for the other. See Figure \ref{Fig:A85dconf}. From 7-brane monodromies for this configuration, one finds that it leads to the configuration for 5d $Sp(n)$ gauge theory with $N_f=8$ and $N_a=1$ flavors, which describes a circle compactification of 6d higher rank E-string theory. We note that the web diagram for higher rank E-string theory of massless antisymmetric hypermultiplet was already discussed in \cite{Kim:2015jba} which is made out of eight D7 branes, and its flavor decoupling limit was also discussed in \cite{Kim:2014nqa} as a limit taking masses of D7 branes to infinite.
 Notice however that we  now have {\it nine} D7 branes in the original setup, and as a consequence of 7-brane monodromy, one of D7 branes is now converted to a $[0,1]$-brane. Hence the remaining D7 branes account eight flavors, but as we have the $[0,1]$ 7-brane which does not exit for the web configuration for the massless antisymmetric hypermultiplet, this 7-brane is thus associated with a non-zero mass of antisymmetric hypermultiplet. See Figure \ref{Fig:SpnwithA}. Taking the flavor decoupling limit for the fundamental flavors, one would find 5d web configuration for $Sp(n)$ gauge theory with $N_f\le 7$ and $N_a=1$ flavors, given in \cite{Bergman:2015dpa}. By pulling out all the 7-branes, it is not so difficult for one to find that it makes a Tao web diagram.  



\bigskip
\paragraph{\texorpdfstring{$N=2$ and $n$ nodes: duality between 5d  $SU(2n+1)$ with $N_a=2$ and  $Sp(n)\times Sp(n)$}{N=2}.}

Using an aforementioned $SL(2,\mathbb{Z})$ invariant combination with ($\mathbf{B}, \mathbf{C}$) 7-branes (a resolution of the  $O7^-$ plane) and four D7 brane, one finds that the $N=2$ case of $n$ quiver nodes 
\begin{align}
{\rm 6d~} [1_A,8]-\underbrace{SU(2)-\cdots-SU(2)}_\text{$n$ nodes}-[2]
\end{align}
gives rise to two 5d theories. Firstly, reducing it to 5d, we can take the S-duality in a $SL(2,\mathbb{Z})$ invariant way that the 7-branes are rearranging themselves to respect the original 7-brane structure. As this case can be understood as the case when $N$ is even, which we will discuss in what follows.  We state that the resulting 5d theories. This case leads to 5d $SU(2n+1)$ gauge theory with {\it two} antisymmetric hypermultiplets
\begin{align}
{\rm 5d~} [1_A,4] - SU(2n+1) -[1_A,4].
\end{align}
On the other hand, a different resolution of the $O7^-$ plane, e.g., $O7^-\to \bf N,\, X_{[2,1]}$, yields the quiver of $Sp(n)\times Sp(n)$ theory only with fundamental hypermultiplets
\begin{align}
{\rm 5d~}[4] - Sp(n) - Sp(n) - [4].
\end{align}
This suggests that these two 5d theories are dual to each other in the sense that they have the same UV fixed points.

\bigskip
\bigskip
\paragraph{\texorpdfstring{$N=$ even: Duality between $Sp$ quiver and $SU$ quiver.}{N=even}}
For $N=2m$ and $n$ quiver nodes, 6d theory is 
\begin{align}
{\rm 6d~} [1_A,8]-\underbrace{SU(2m)-\cdots-SU(2m)}_\text{$n$ nodes}-[2m].
\end{align}
Again, one uses $SL(2,\mathbb{Z})$ invariant 7-brane combinations introduced earlier and S-duality to obtain the following 5d theories:
\begin{align}
{\rm 5d~} & [1_A,4] - \underbrace{SU(2n+1) - SU(2n+1) - \cdots - SU(2n+1)}_\text{$m$ nodes} - [1_A,4]\\
{\rm 5d~}  & [3] - \underbrace{Sp(n) - {\overset{\overset{\text{\large$[1]$}}{\textstyle\vert}}{SU(2n+1)}} - SU(2n+1) - \cdots - SU(2n+1) -  {\overset{\overset{\text{\large$[1]$}}{\textstyle\vert}}{SU(2n+1)}}  - Sp(n)}_\text{$m+1$ nodes}  - [3].
\end{align}
\noindent The first case is realized when we split the $O7^-$ into $\mathbf{B}$ and $\mathbf{C}$ 7-branes. The second case is realized when we split the $O7^-$ into $\bf N$ and $\bf X_{[2,1]}$ 7-branes.
\begin{figure}
\begin{center}
\includegraphics[width=12cm]{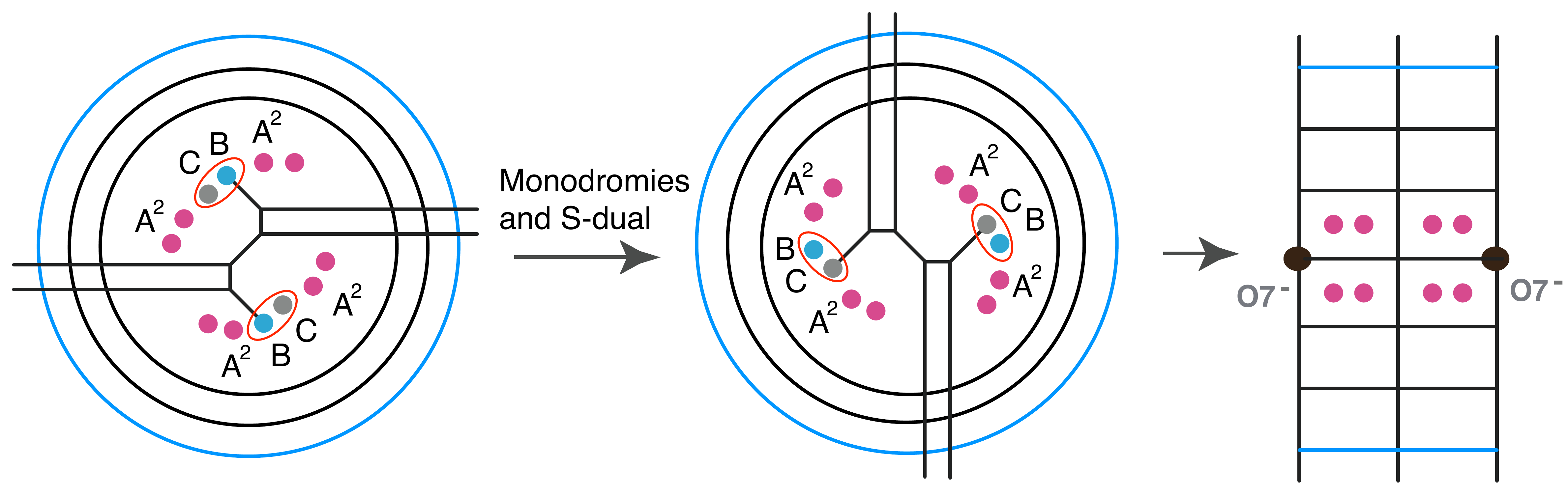}
\end{center}
\caption{A T-dual version of 6d $[1_A, 8]- SU(4)-SU(4)-SU(4)-[4]$ yielding 5d
$[1_A, 4]- SU(7)-SU(7)-[1_A,4]$ implementing S-dual invariant 7-brane combinations. 
}
\label{Fig:AABCAA}
\end{figure}

\begin{figure}
\begin{center}
\includegraphics[width=16cm]{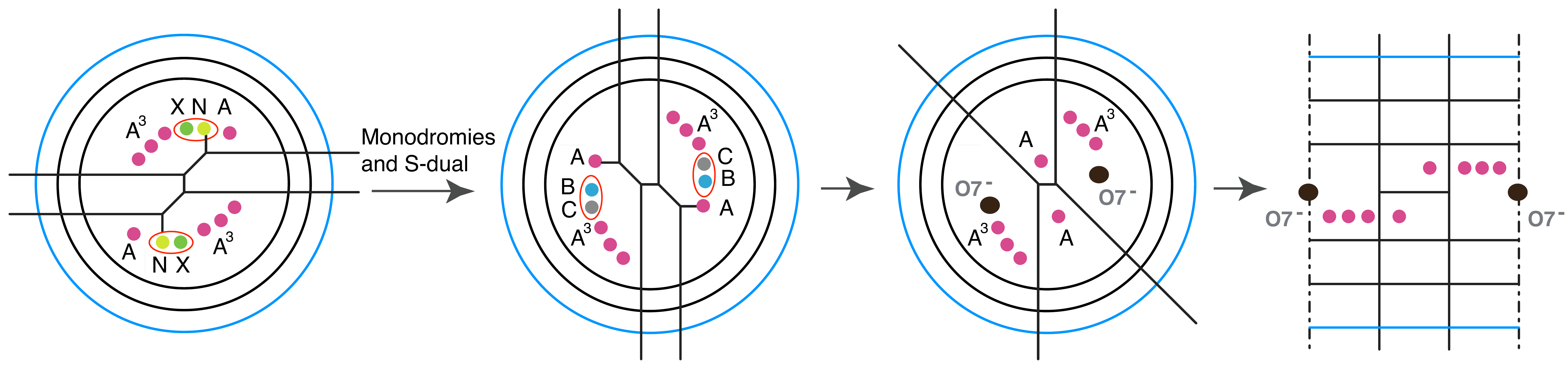}
\end{center}
\caption{A T-dual version of 6d $[1_A, 8]- SU(4)-SU(4)-SU(4)-[4]$ yielding 5d
$[3]- Sp(3)-( SU(7) - [2] )-Sp(3)-[3]$ implementing S-dual invariant 7-brane combinations. 
}
\label{Fig:ANXA3}
\end{figure}

As a representative example, the brane configuration for 6d $[1_A, 8]- SU(4)-SU(4)-SU(4)-[4]$ ($m=2$ and $n=3$) yielding 
\begin{align}
{\rm 5d~}[1_A, 4]- SU(7)-SU(7)-[1_A,4]
\end{align}
is depicted in Figure \ref{Fig:AABCAA}.  As there are eight D7 branes, one can relocate and arrange D7 branes with suitable Wilson lines, such that four D7 branes are located close to each the $O7^-$ plane so to make an $SL(2,\mathbb{Z})$ invariant 7-brane combination including the resolution of the $O7^-$ plane into the $\bf B, C$ 7-branes. Performing an S-duality and also taking a weak coupling limit to obtain the $O7^-$ plane out of two suitable 7-branes, one finds that the resulting web configuration gives rise to the quiver gauge theory with an antisymmetric hypermultiplet at the edge gauge node.  
In a different way of resolving the $O7^-$ planes, 6d brane configuration yields a seemingly different 5d quiver theory  
\begin{align}
{\rm 5d~} [3]- Sp(3)-{\overset{\overset{\text{\large$[2]$}}{\textstyle\vert}}{SU(7)}}-Sp(3)-[3]
\end{align}
is also depicted in Figure \ref{Fig:ANXA3}. Like the previous case, one has another $SL(2,\mathbf{Z})$ invariant 7-brane combination except for a different resolution of the $O7^-$ plane, $O7^-\to {\bf N, X_{[2,1]}}$. When performing an S-duality, the web configuration becomes completely different from the previous one as a D7 brane is being attached to a D5 brane which makes a floating 7-brane pair which can be converted into an $O7^-$ plane in the weak coupling limit. One can further move this D7 brane across 5-brane junction, and as the Hanany-Witten transition, one has freely floating D7 branes, which play the role of the fundamental flavors. The resulting configuration yields a quiver theory made out of $Sp$ and $SU$ gauge groups. See Figure \ref{Fig:ANXA3}. Therefore, we claim that two different quiver theories have the same 6d origin at UV and thus they are two dual description at IR. We note also that taking the flavor decoupling limit, this duality would still hold for less flavor cases.

\bigskip
\paragraph{\texorpdfstring{$N=$ odd: $Sp-SU$ quiver.}{N=odd}}

\begin{figure}
\begin{center}
\includegraphics[width=16cm]{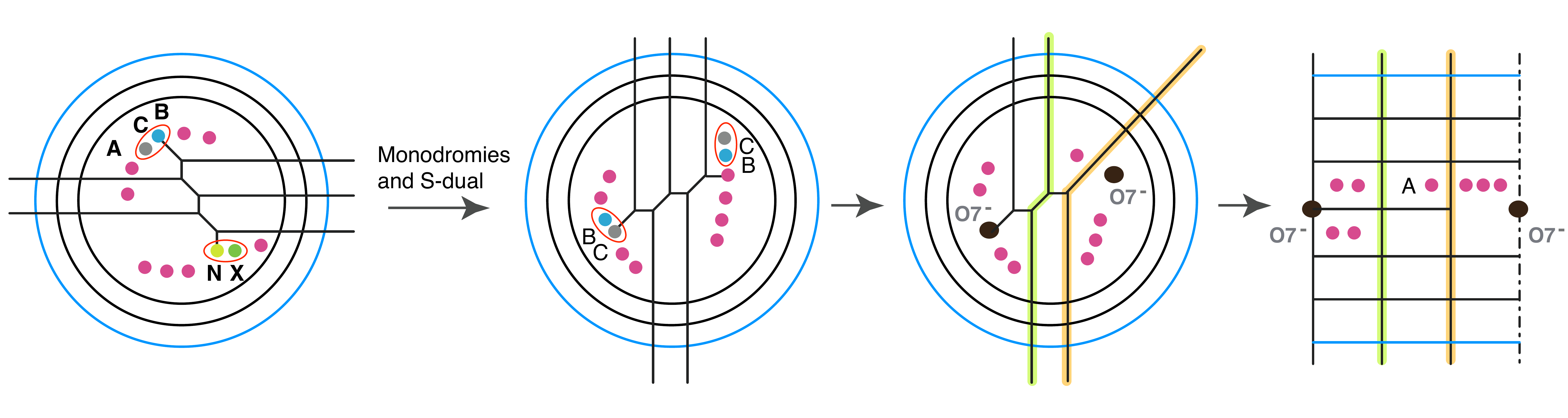}
\end{center}
\caption{A T-dual version of 6d $[1_A, 8]- SU(5)-SU(5)-SU(5)-[5]$ yielding 5d
$[1_A, 4]- SU(7)-( SU(7) - [1] )-Sp(3)-[3]$ implementing S-dual invariant 7-brane combinations. 
}
\label{Fig:Noddcase}
\end{figure}
In a similar fashion, one can obtain 5d theory for $N=2m+1$. In this case, the resolution of each $O7^-$ brane is different, one does get dual description, rather one finds the resulting 5d theory is a hybrid of the $N=2m$ cases: 
\begin{align}
{\rm 5d~}[3] -  \underbrace{Sp(n) - {\overset{\overset{\text{\large$[1]$}}{\textstyle\vert}}{SU(2n+1)}} 
- SU(2n+1) - \cdots - SU(2n+1)}_\text{$m+1$ nodes}  - [1_A,4].
\end{align}
As an example, 5d brane configuration for 6d $[1_A, 8]- SU(5)-SU(5)-SU(5)-[5]$ quiver theory is depicted in Figure \ref{Fig:Noddcase}, yielding
\begin{align}
{\rm 5d~}[1_A, 4]- SU(7)-{\overset{\overset{\text{\large$[2]$}}{\textstyle\vert}}{SU(7)}}-Sp(3)-[3].
\end{align}


\bigskip
\subsection{\texorpdfstring{6d $[8] - Sp(N) - SU(2N) - SU(2N) - \cdots - SU(2N) - [2N]$ quiver}{[8]-Sp-SU}}\label{sec:6d-8-Sp-SU}

\begin{figure}
\begin{center}
\includegraphics[width=7cm]{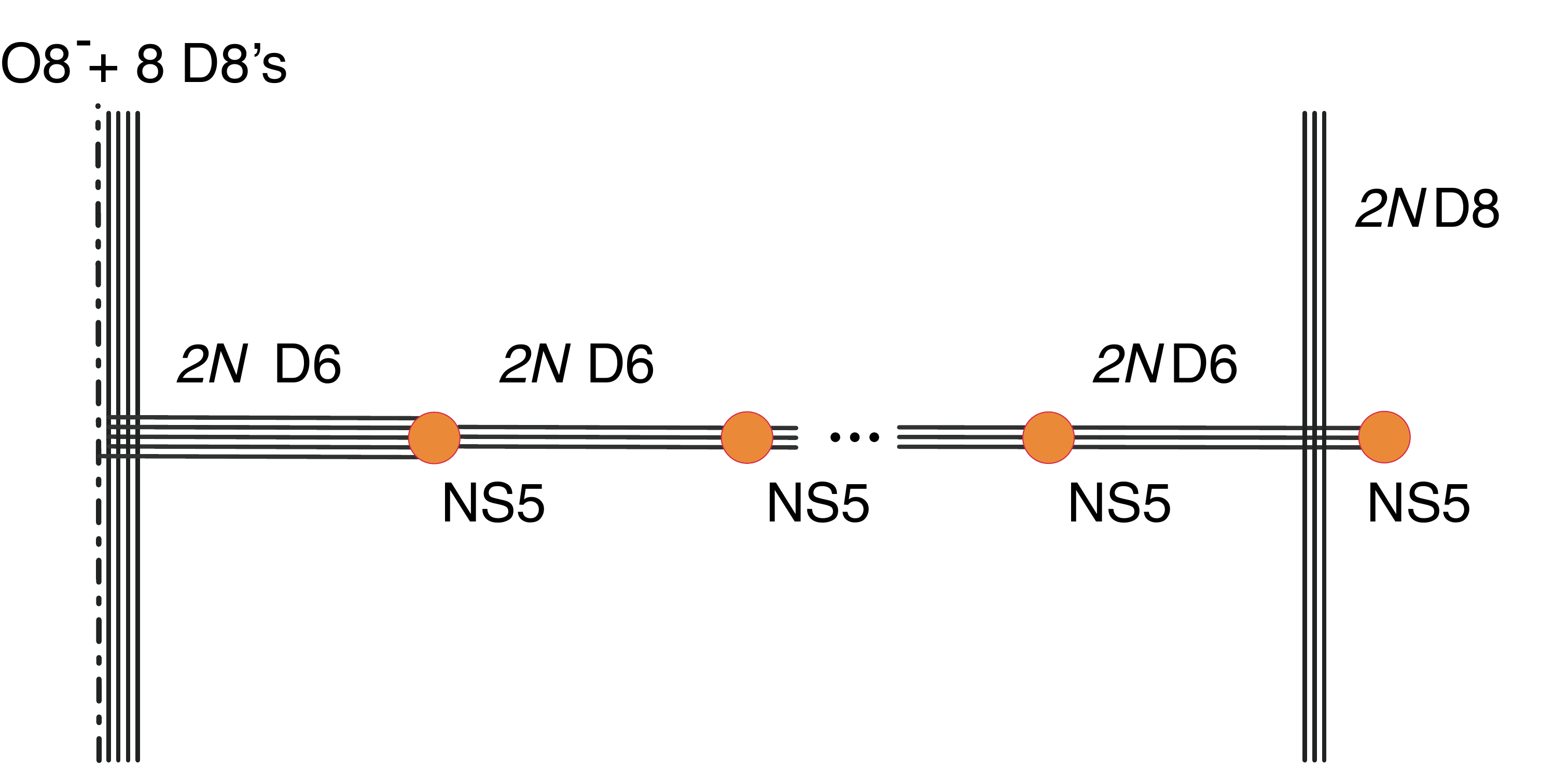}
\end{center}
\caption{6d brane configuration for the $[8] - Sp(N) - SU(2N) - SU(2N) - \cdots - SU(2N) - [2N]$ quiver theory.}
\label{Fig:A8SpSU}
\end{figure}

Next we consider another special case:
\begin{align}
{\rm 6d~}[8] - \underbrace{Sp(N) - SU(2N) - SU(2N) - \cdots - SU(2N)}_\text{$n$ nodes} - [2N],
\end{align}
whose brane configuration is given in Figure \ref{Fig:A8SpSU}, consisting of an $O8^-$ plane, and D8 branes, D6 branes and NS5 branes.  We claim that from a circle compactification and T-dual, one obtains the 5d theory given by 
\begin{align}
{\rm 5d~} [4] - \underbrace{Sp(n) - SU(2n) - \cdots - SU(2n) - Sp(n)}_\text{$N+1$ nodes} - [4].
\end{align}
\begin{figure}
\begin{center}
\includegraphics[width=12cm]{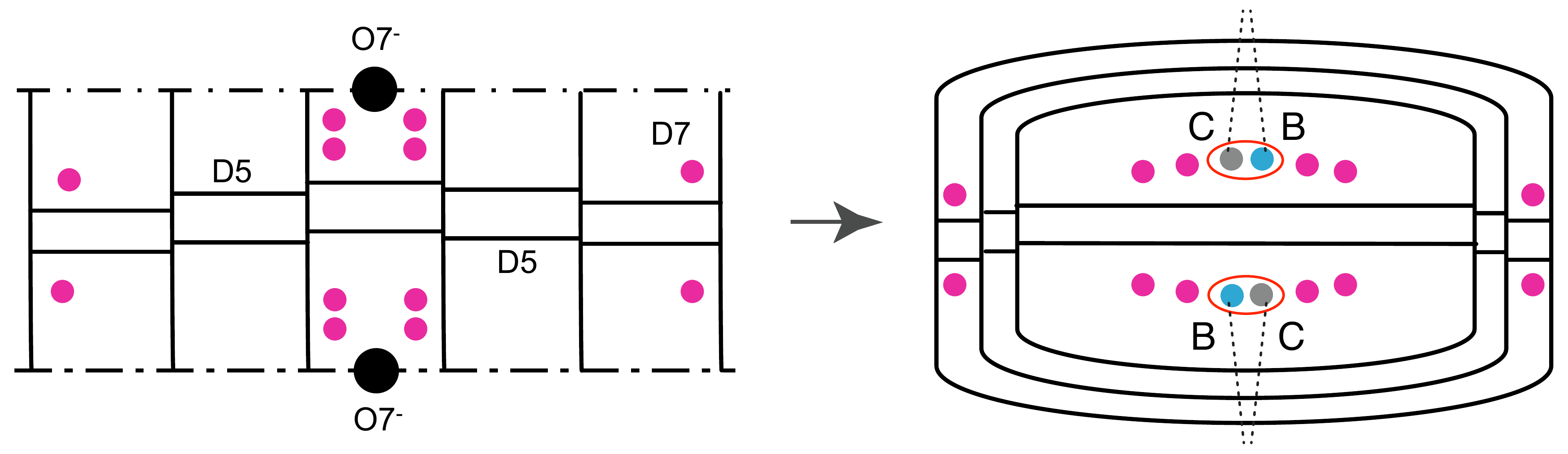}
\end{center}
\caption{T-dual version of 6d brane configuration for the $[8] -Sp(2) - SU(4) - SU(4)  -[4]$ quiver theory.}
\label{Fig:8sp2su4su4}
\end{figure}
As an example, consider 
\begin{align}
{\rm 6d~} [8] -Sp(2) - SU(4) - SU(4) -[4].
\end{align}
Reducing it to 5d, one gets a brane configuration of $O7^-$ planes, D7 branes and D5 branes given in the web configuration on the left of Figure \ref{Fig:8sp2su4su4}. By resolving two $O7^-$ planes, one obtains the configuration on the right of Figure \ref{Fig:8sp2su4su4}.
As discussed in the beginning of this section, using the $SL(2,\mathbb{Z})$ invariant 7-brane combination \eqref{a2bca2no5bra}, one sees that S-duality leads to 
\begin{align}
{\rm 5d~} [4] -Sp(3) - SU(6) - Sp(3) -[4].
\end{align}
The procedure is depicted in Figure \ref{Fig:4sp3su6}. 

\begin{figure}
\begin{center}
\includegraphics[width=11cm]{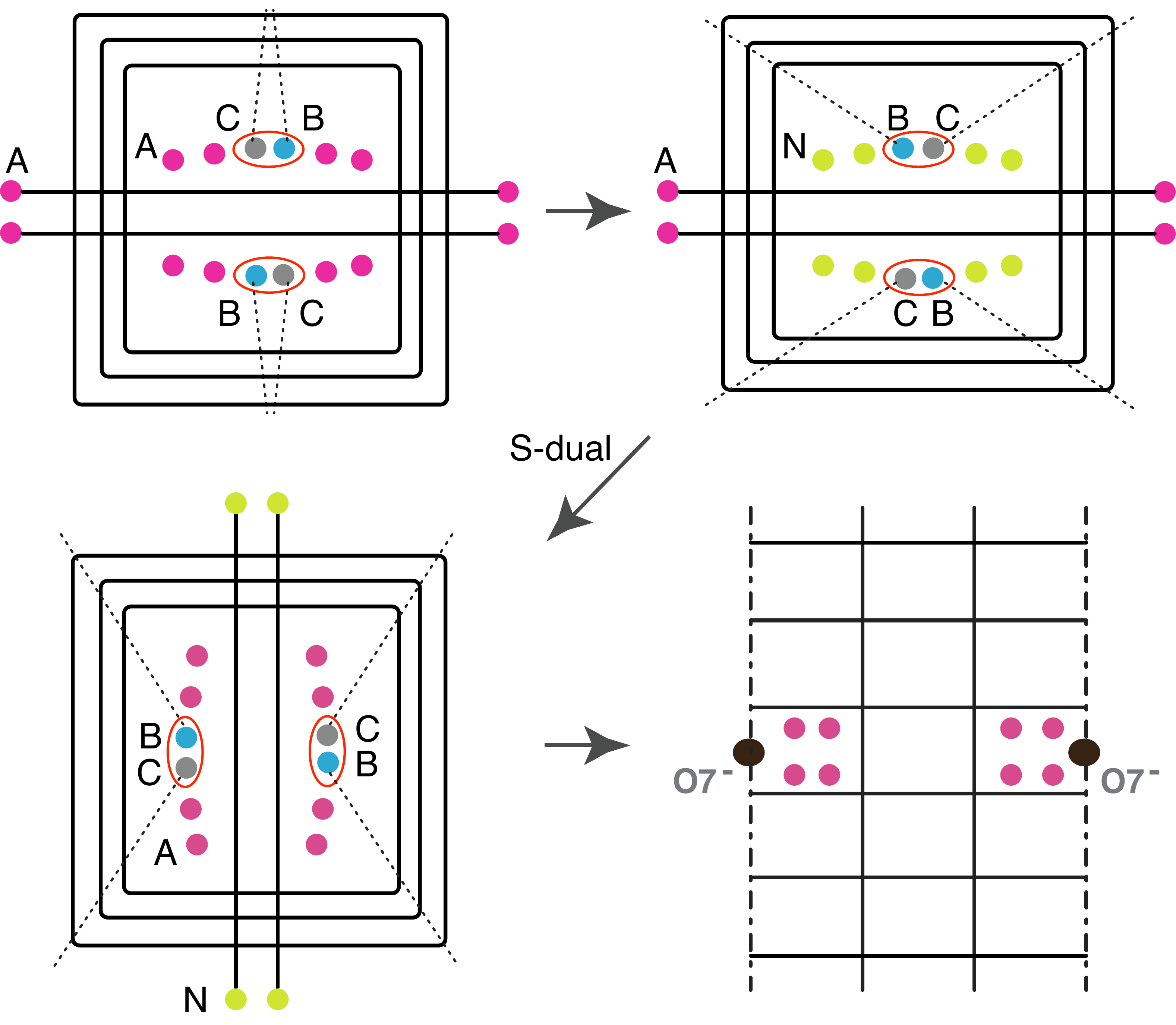}
\end{center}
\caption{The first figure is a simplified version of the right of Figure \ref{Fig:8sp2su4su4} for
6d $[8]-Sp(2) - SU(4) - SU(4) - SU(4) -[4]$ quiver.  The 7-branes are distributed and combined to form an S-dual invariant 7-brane combination. The second figure is the rearrangement of these 7-branes via monodromies explained in section \ref{subsec:18sususun}. The third one is the result of S-dual action of the second one. The last one is the corresponding brane configuration showing 
5d $[4]-Sp(3)-SU(6)-Sp(3)-[4]$.}
\label{Fig:4sp3su6}
\end{figure}

\bigskip
\subsection{\texorpdfstring{6d $[8+m] -  Sp(N)_{\rm T} - SU(2N-m)_{\rm T} - SU(2N-2m )_{\rm T} - SU(2N-3m)_{\rm T} - ...$ quiver}{[8+m]-Sp-SU-SU}}
As a generalization of Section \ref{sec:6d-8-Sp-SU}, we consider $m>0$ :
\begin{align}
	6d \quad [8+m] -  Sp(N)_{\rm T}- SU(2N-m)_{\rm T} - SU(2N-2m )_{\rm T}- SU(2N-3m)_{\rm T} - ...
\end{align}

%% file: section6.tex
\bigskip
\section{6d description of 5d \texorpdfstring{$T_N$}{TN} Tao theory}
\label{sec:TNTao}

\begin{figure}
\begin{center}
\includegraphics[width=8cm]{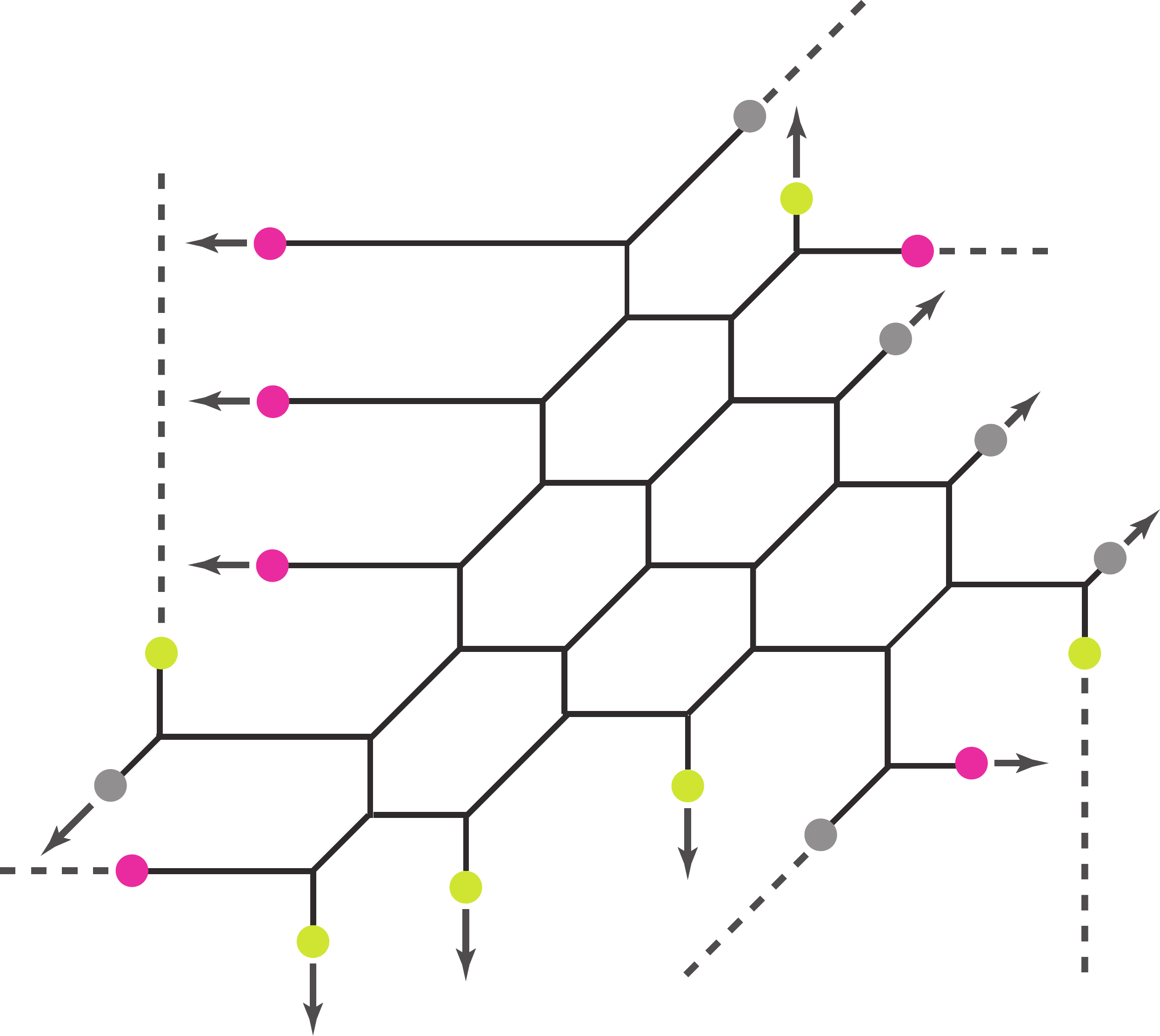}
\end{center}
\caption{The Tao web diagram which arises by adding flavors at the end nodes of the 5d quiver corresponding to the 5d $T_5$ theory.} 
\label{Fig:T5Tao}
\end{figure}

There is another important class of 5d theories obtained by Tao diagrams. It has been known that the 5d $T_N$ theory can be realized by a 5-brane web diagram \cite{Benini:2009gi}. By turning on certain mass deformations, the 5d $T_N$ theory flows to a 5d linear quiver theory $[N] - SU(N-1) - SU(N-2) - \cdots - SU(2) - SU(1)$ \cite{Hayashi:2013qwa, Aganagic:2014oia, Bergman:2014kza, Hayashi:2014hfa}. The last $SU(1)$ node can be understood as two flavors coupled to the $SU(2)$ gauge node, and hence the quiver is equivalently written as $[N] - SU(N-1) - SU(N-2) - \cdots - SU(2)- [2]$ \cite{Bergman:2014kza, Hayashi:2014hfa, Tachikawa:2015mha}. It is possible to construct a Tao web diagram by adding flavors to the two end nodes of the quiver \cite{Kim:2015jba}. An example of a Tao web diagram arising by adding flavors to the 5d $T_5$ theory is depicted in Figure \ref{Fig:T5Tao}. The 5d quiver theory realized by the Tao diagram is then 
\begin{equation}
[N+2] - SU(N-1) - SU(N-2) - \cdots - SU(3) - SU(2) - [3]. \label{quiver.TNTao}
\end{equation}
We will call the 5d quiver theory as 5d $T_N$ Tao theory. Since the 5d $T_N$ Tao theory is realized by the Tao diagram, the theory is expected to have a 6d UV completion. In this section, we propose a 6d description of the 5d $T_N$ Tao theory \eqref{quiver.TNTao} realized by the $T_N$ Tao diagram by examining the global symmetry as well as the number of the Coulomb branch moduli.

\subsection{Global symmetry from 7-branes}

The global symmetry of a 5d theory realized by a 5-brane web can be understood from the symmetry on 7-branes attached to external 5-branes in the web diagram \cite{DeWolfe:1999hj}. Namely, 7-branes can determine the global symmetry realized at the UV fixed point. In order to see the flavor symmetry, we pull all the 7-branes inside the 5-brane loops and try to put 7-branes on top of each other. The symmetry realized on 7-branes gives the non-Abelian part of the flavor symmetry of the 5d theory. Abelian part can be recovered by counting the number of parameters of the theory since the number of the parameters of the 5d theory should agree with the rank of the total global symmetry. The method was applied to the Tao diagram realizing the 5d $SU(n)$ gauge theory with $N_f = 2n+4$ flavors in \cite{Hayashi:2015fsa}, and the global symmetry was determined to be $SO(4n+8)$. We repeat the same procedure to determine the global symmetry of the 5d $T_N$ Tao theory. 

\begin{figure}
\begin{center}
\includegraphics[width=8cm]{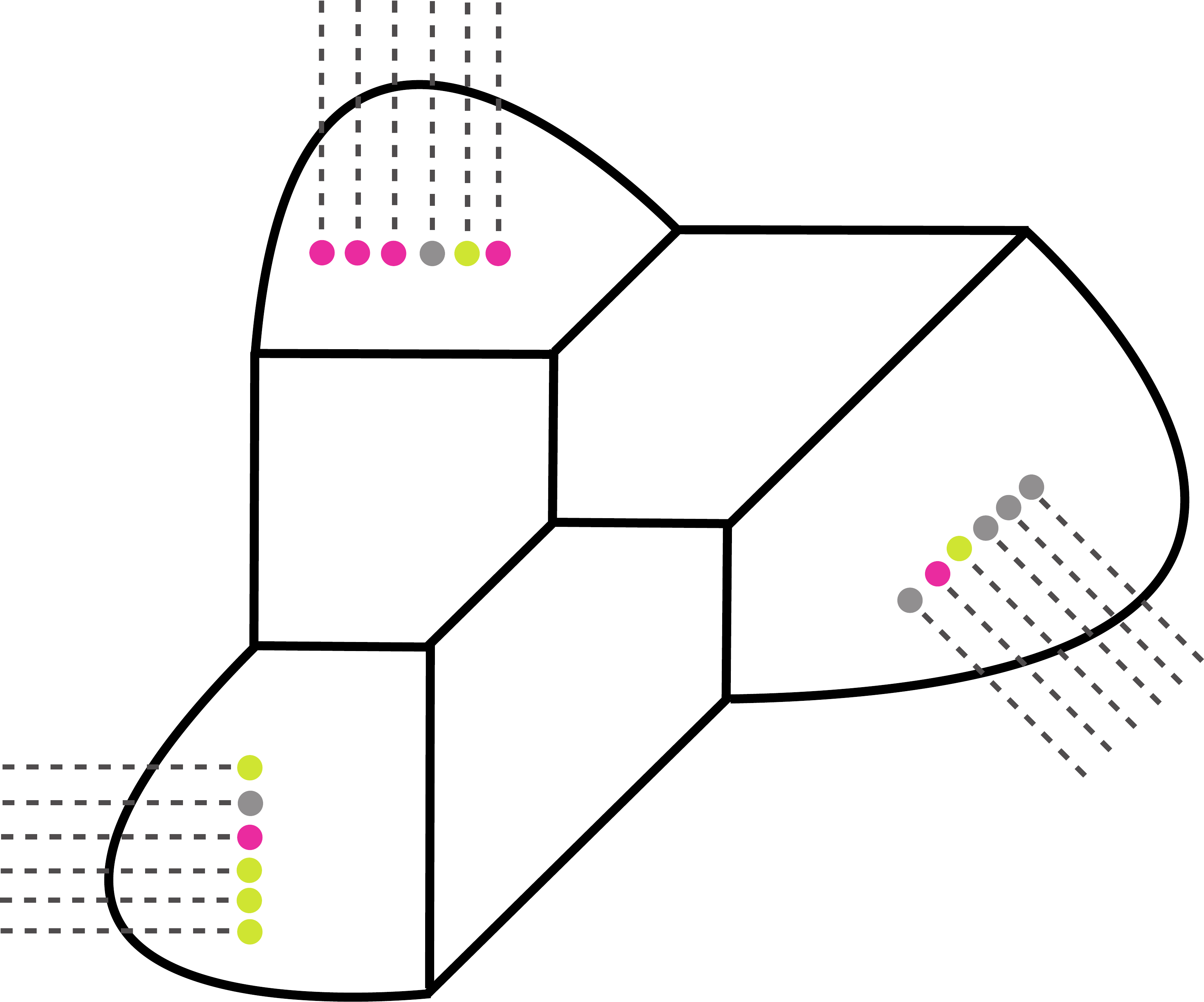}
\end{center}
\caption{The 5-branes web after pulling all the 7-branes inside the 5-brane loops. We write an example of the 5d $T_5$ Tao theory as an example.} 
\label{Fig:T5Tao2}
\end{figure}

We first pull all the 7-branes inside the 5-brane loops. Since $[p, q]$ 7-branes mutually commute with $(p, q)$ 5-branes, it is possible to collect the 7-branes into three chambers, the top-left one (1st chamber), the bottom-left one (2nd chamber) and the bottom-right one (3rd chamber) as in Figure \ref{Fig:T5Tao2}.  
The 7-branes in each chamber are schematically summarized as
\begin{equation}
{\bf ANCA}^{N-2}|| {\bf NCAN}^{N-2} || {\bf CANC}^{N-2} || \label{TN7branes}
\end{equation}
where ${\bf A}, {\bf C}$ and ${\bf N}$ represent $[1,0]$, $[1,1]$ and $[0,1]$ 7-branes respectively. We will also express $[1,-1]$ 7-brane by ${\bf B}$. In this expression of \eqref{TN7branes}, the branch cuts are assumed to extend in the lower direction, and the charge of a 7-brane change by the monodromy matrix
\begin{eqnarray}
\left(
\begin{array}{cc}
1-pq & p^2\\
-q^2 & 1+pq
\end{array}
\right)
\end{eqnarray}
when the 7-brane crosses the branch cut of a $[p, q]$ 7-brane counterclockwise.  Also, the first, the second and the third partition lines in \eqref{TN7branes} represent $(1,0)$ 5-branes, $(0,1)$ 5-branes and $(1,1)$ 5-branes respectively. Hence, 7-branes with the same charge as the 5-brane corresponding to the partition line can crosse its line. Therefore, we can move 7-branes in \eqref{TN7branes}, and obtain
\begin{equation}
{\bf C}^{N-2} {\bf ANC}|| {\bf A}^{N-2}{\bf NCAN}^{N-2} || {\bf CAN}|| \label{TN7branes1}
\end{equation}
By rearrangement of 7-branes inside each chamber, the 7-brane configuration becomes 
\begin{equation}
{\bf X}_{[2,1]}{\bf A}^N|| {\bf A}^{2N-2}{\bf B} || {\bf N}^2{\bf X}_{[2,1]}|| \label{TN7branes2}
\end{equation}
By moving 7-branes between the chambers with the rearrangemnt in the second chamber, we finally obtain
\begin{equation}
{\bf X}_{[2,1]}|| {\bf A}^{3N}{\bf B} || {\bf X}_{[2,1]}|| \label{TN7branes3}
\end{equation}
The presence of $3N$ A-branes on top of each other means that the non-Abelian part of the global symmetry of the 5d $T_N$ Tao theory is $SU(3N)$ symmetry. 

It is possible to count the number of the parameters of the 5d $T_N$ Tao theory. The number of mass parameters of the fundamental hypermultiplets is $N+5$. The number of mass parameters of the bi-fundamental multiplets is $N-3$. Also, the number of the gauge couplings or the mass parameters for the instantons is $N-2$. The total number is $3N$. Therefore, the full global symmetry of the 5d theory is $SU(3N) \times U(1)_I$. As expected, we have the $U(1)_I$ symmetry associated to the Kaluza-Klein modes from the $S^1$ compactification of a 6d theory. 

The dimension of the Coulomb branch moduli space can be determined easily by counting the rank of the gauge groups, and it is $\frac{(N-1)(N-2)}{2}$.

\subsection{5d \texorpdfstring{$T_4, T_5$}{T4, T5} and \texorpdfstring{$T_6$}{T6} Tao theories}

Let us move on to a candidate for a 6d description of the 5d $T_N$ Tao theory. We first focus on the 5d $T_4, T_5, T_6$ theories, which will be basic examples for the general 5d $T_N$ Tao theory. 

In fact, the 5d $T_4$ Tao theory is a very special case in the examples considered in section \ref{sec:6dSUNA}. The 6d description in the tensor branch is 
\begin{equation}
[1]_{A} - SU(3) - [11]. \label{T4}
\end{equation}
We also have one tensor multiplet corresponding to the number of the gauge nodes. The theory has one tensor multiplet and two vector multiplets in the Cartan subalgebra, and hence the sum of them reproduces the correct number of the 5d vector multiplets in the Cartan subalgebra of the 5d $T_4$ theory.

Let us then determine the flavor symmetry of the 6d theory \eqref{T4}. It is important to note that the anti-symmetric representation of the $SU(3)$ is equivalent to the anti-fundamental representation of the $SU(3)$. Although a 6d hypermultiplet contains a 6d Weyl spinor, the complex conjugation of the 6d Weyl spinor does not change its chirality. Therefore, the hypermultiplet in the anti-symmetric representation of $SU(3)$ is equivalent to a hypermultiplet in the fundamental representation of $SU(3)$. Namely, the 6d theory \eqref{T4} is equivalent to a 6d $SU(3)$ gauge theory with $12$ flavors with one tensor multiplet coupled. From this point of view, we have at least an $SU(12)$ flavor symmetry.

There can be a potential $U(1)$ flavor symmetry that may come from the overall part of $U(12)$. However, we argue that the $U(1)$ symmetry is broken by the anomaly $U(1)_{\text{global}} - SU(3)_{\text{gauge}}^3$, and does not appear as a global symmetry of the 6d theory. Since the third-order Casimir invariant for $SU(3)$ is non-trivial, it is impossible to cancel the anomaly if there is only one representation. Therefore, the 6d theory does not have an Abelian symmetry. 

After the circle compactification, the KK mode provides a $U(1)_I$ symmetry. Hence, the total flavor symmetry is $SU(12) \times U(1)_I$, which agrees with the flavor symmetry of the 5d $T_4$ Tao theory.

Let us then consider the 5d $T_5$ Tao theory. In this case, it is not included in the examples we have discussed so far. This means that it may not admit a brane configuration in Type IIA string theory. However, it has been conjectured that all the 6d SCFTs may be realized by F-theory compactifications \cite{Heckman:2013pva,  DelZotto:2014hpa, Heckman:2014qba, Heckman:2015bfa}. Then it is possible to find a candidate for its 6d description in the 6d SCFTs classified by the F-theory compactifications, by comparing the global symmetry as well as the dimension of the Coulomb branch moduli space. 

In order to realize a 6d SCFT from F-theory, we consider an F-theory compactification on a Calabi-Yau threefold which is given by an elliptically fibration over a  non-compact complex two dimensional K\"ahler manifold $B$. In the base $B$, there is a bunch of $\mathbb{P}^1$'s. Each size of the compact $\mathbb{P}^1$ corresponds to a vev of a scalar in a tensor multiplet. The elliptic fibration over the $\mathbb{P}^1$ can be singular, meaning that 7-branes wrap the $\mathbb{P}^1$ and yielding a gauge symmetry. When all the compact $\mathbb{P}^1$'s vanish simultaneously, the 6d theory is at the superconformal fixed point. This in fact constrains the shape of the sequence of the $\mathbb{P}^1$'s significantly \cite{Heckman:2013pva, Heckman:2015bfa}.

As for the 5d $T_5$ Tao theory, the dimension of the Coulomb branch moduli space is $6$. Therefore, the sum of the number of the tensor multiplets and the number of the vector multiplets in the Cartan subalgebra of the corresponding 6d theory must be $6$. Since the expected 6d flavor symmetry is $SU(15)$, we need one non-compact $\mathbb{P}^1$\footnote{We call non-compact $\mathbb{P}^1$ as a $\mathbb{P}^1$ whose size is taken to be infinity.} with a singular elliptic fiber giving the $SU(15)$ algebra on it. We denote the non-compact 2-cycle by $C_{SU(15)}$. Due to the strong constraint of the shape of the base geometry, only pairwise intersections are allowed. Furthermore, the non-trivial gauge algebra on the adjacent  $\mathbb{P}^1$ should be either $SU$ or $Sp$ \cite{Heckman:2015bfa}. 

Let us then consider the case where one compact $\mathbb{P}^1$ is attached to $C_{SU(15)}$. The theory has one tensor multiplet. Hence, the rank of the $SU$ or $Sp$ gauge group should be $5$. Then, the 6d anomaly cancellation condition with the $15$ flavors gives us a unique possibility 
\begin{equation}
\left[\frac{1}{2}\right]_{\Lambda^3} - SU(6) - [15], \label{T5}
\end{equation}
where $\left[\frac{1}{2}\right]_{\Lambda^3}$ stands for one half-hypermultiplet in the rank three anti-symmetric representation. This 6d theory on $S^1$ reproduces the correct number of the 5d vector multiplets in the Cartan subalgebra of the 5d $T_5$ Tao theory as well as the $SU(15)$ flavor symmetry. 
The potential $U(1)$ flavor symmetry which may come from the overall part of the $U(15)$ is again broken by the anomaly $U(1) - SU(6)^3$.

When one increases the number of compact $\mathbb{P}^1$'s attached to $C_{SU(15)}$, then one cannot find a candidate for a non-trivial gauge group on the compact $\mathbb{P}^1$'s that satisfies the 6d gauge anomaly cancellation condition with the $15$ flavors. If we increase the number of the compact $\mathbb{P}^1$'s, the total rank of the gauge groups should be less than or equal to $4$. Then, there is no possible choice for the gauge group on the compact $\mathbb{P}^1$ next to $C_{SU(12)}$ by the 6d anomaly cancellation condition. If there is no gauge symmetry on the compact $\mathbb{P}^1$ next to $C_{SU(15)}$, then the $15$ flavors on $C_{SU(15)}$ may give a $U(15)$ flavor symmetry at least\footnote{For example, the E-string theory on the tensor branch consists of $8$ hypermultiplets and one tensor multiplet, The flavor symmetry is $SO(16)$ which is further enhanced to $E_8 \supset SO(16)$.}. Therefore, it contradicts the flavor symmetry of the 5d $T_5$ Tao theory.

Hence, we propose that the 6d description of the 5d $T_5$ Tao theory is given by the 6d quiver of \eqref{T5}.

The last case is the 5d $T_6$ Tao theory. This is also a very special case in the examples considered in section \ref{subsec:6dSUquivgen}. The 6d description in the tensor branch is 
\begin{equation}
SU(1) - SU(9) - [17].
\end{equation}
Note that there is no hypermultiplet in the anti-symmetric representation for the $SU(1)$. Furthermore, the $SU(1)$ does not a dynamical vector multiplet. Therefore, the 6d theory is effectively described by the $SU(9)$ gauge theory with $18$ hypermultiplets in the fundamental representation with two tensor multiplets coupled. Therefore, after the circle compactification, it gives $2+8 = 10$ 5d vector multiplets in the Cartan subalgebras, which reproduces the correct number for the 5d $T_6$ Tao theory. Furthermore, the flavor symmetry is $SU(18) \times U(1)_I$, which also agrees with the flavor symmetry of the 5d $T_6$ Tao theory. The potential $U(1)$ symmetry of the overall part of $U(18)$ is again broken by the anomaly $U(1) - SU(9)^3$.

\subsection{Generalization}

We move on to the analysis for a 6d description of the 5d $T_N$ Tao theory. Given the 6d description of the 5d $T_4, T_5, T_6$ Tao theory, there is a very natural generalization to a general $N$. 

The 6d description of the 5d $T_4$ Tao theory is described by the 6d $SU(3)$ gauge theory with $12$ flavors, and it has the $SU(12)$ flavor symmetry. We then gauge the $SU(12)$ flavor symmetry. Due to the 6d anomaly cancellation condition, the $SU(12)$ gauge node needs to have $12$ more flavors. Namely, a 6d anomaly free quiver theory by gauging the $SU(12)$  flavor symmetry is 
\begin{equation}
SU(3) - SU(12) - [21]. \label{T7}
\end{equation}
It has two tensor multiplets and $13$  vector multiplets in the Cartan subalgebra. The total number is $15$, and this number in fact agrees with the number of the Coulomb branch moduli of the 5d $T_7$ theory. Furthermore, the flavor symmetry of the 6d theory \eqref{T7} after the $S^1$ compactification is $SU(21) \times U(1)_I$, which also agrees with the global symmetry of the 5d $T_7$ Tao theory. In this case, there can be potential $U(1) \times U(1)$ flavor symmetries. One may be associated to the one bi-fundamental hypermultiplet between $SU(3)$ and $SU(12)$, and the other may be associated to the $21$ fundamental hypermultiplets. However, any linear combination of the two $U(1)$'s has non-zero anomaly either from $U(1) - SU(3)^3$ or $U(1) - SU(12)^3$. Therefore, the 6d theory \eqref{T7} is a natural candidate for the 5d $T_7$ theory. 

It is easy to generalize this consideration by gauging the $SU(21)$ flavor symmetry. The general quiver theory by the successive gauging is 
\begin{equation}
SU(3) - SU(12) - SU(21) - \cdots - [9N+3].\label{T3N+1}
\end{equation}
It has $N$ tensor multiplets and $\frac{N(9N-5)}{2}$ vector multiplets in the Cartan subalgebra. The total number is $\frac{3N(3N-1)}{2}$, which exactly agrees with the number of the Coulomb branch moduli of the 5d $T_{3N+1}$ Tao theory. The flavor symmetry after the $S^1$ compactification is $SU(9M+3) \times U(1)_I$, which also agrees with that of the 5d $T_{3N+1}$ Tao theory. The potential $U(1)^{N}$ flavor symmetries are all broken by the [global - gauge$^3$] anomalies. Hence, the 6d quiver theory \eqref{T3N+1} is a candidate for the 6d description of the 5d $T_{3N+1}$ Tao theory with $N \geq 1$.  

It is interesting to note that the rank of the gauge group of the 6d quiver \eqref{T3N+1} increases by $9$. Hence, this 6d theory is not realized by a brane configuration with an $O8^-$-plane in Type IIA string theory. However, it may admit a F-theory realization, and a possible configuration may be
\begin{eqnarray}
\begin{array}{ccccc}
 \mathfrak{su}(3)&\mathfrak{su}(12) & \cdots & \mathfrak{su}(9N-6) & [\mathfrak{su}(9N+3)]\nonumber\\
 1 & 2 & \cdots & 2 & \nonumber
\end{array} 
\end{eqnarray}
Each column represents a $\mathbb{P}^1$ and the last one stands for a non-compact $\mathbb{P}^1$  in the base $B$. The upper row indicates the singular type of the elliptic fibration over the $\mathbb{P}^1$. The lower row gives the self-intersection number of each compact $\mathbb{P}^1$ inside $B$.  

In fact, the proposed 6d descriptions of the 5d $T_5$ and $T_6$ Tao theories also admit the same generalization. By the successive gauging of the flavor symmetries of \eqref{T5}, one arrives at 
\begin{equation}
\left[\frac{1}{2}\right]_{\Lambda^3} - SU(6) - SU(15) - \cdots - SU(9N-3) - [9N+6]. \label{T3N+2}
\end{equation}
The 6d quiver theory has $N$ tensor multiplets and $\frac{N(9N+1)}{2}$ vector multiplets in the Cartan subalgebra. The sum of the numbers give  $\frac{3N(3N+1)}{2}$, which is exactly the dimension of the Coulomb branch moduli space of the 5d $T_{3N+2}$ Tao theory with $N \geq 1$. The potential $U(1)^{N}$ global symmetries are all broken by the  [global - gauge$^3$] anomaly of some gauge group. Hence, the flavor symmetry after the circle compactification is $SU(9N+6) \times U(1)_I$. and it agrees with that of the 5d $T_{3N+2}$ Tao theory. Therefore, the 6d quiver theory \eqref{T3N+2} is a natural candidate of the 6d description of the 5d $T_{3N+2}$ Tao theory. 

Again, the 6d quiver theory \eqref{T3N+2} does not have a Type IIA bran realization, but it may have a F-theory realization. The possible configuration is 
\begin{eqnarray}
\begin{array}{ccccc}
 \mathfrak{su}(6)&\mathfrak{su}(15) & \cdots & \mathfrak{su}(9N-3) & [\mathfrak{su}(9N+6)]\nonumber\\
 1 & 2 & \cdots & 2 & \nonumber
\end{array} 
\end{eqnarray}
where the first $\mathbb{P}^1$ can support the half-hypermultiplet in the rank three anti-symmetric representation \cite{Heckman:2015bfa}. 

Let us finally turn to the generalization of the 5d $T_6$ theory by gauging the flavor symmetry. The 6d theory is a quiver theory described by 
\begin{equation}
SU(0) -  SU(9) - SU(18) - \cdots - SU(9N) - [9N+9]. \label{T3N+3}
\end{equation}
The first $SU(0)$ indicates an additional tensor multiplet without any gauge dynamics. The 6d theory has $N+1$ tensor multiplets and $\frac{N(9N+7)}{2}$. The sum of the numbers is $\frac{(3N+2)(3N+1)}{2}$, which agrees with the number of the Coulomb branch moduli of the 5d $T_{3N+3}$ theory. Due to the [global - gauge$^3$] anomalies, the potential $U(1)^N$ global symmetries are all broken. Hence, the flavor symmetry after the $S^1$ compactification is $SU(9N+9) \times U(1)_I$, and it correctly reproduces the global symmetry of the 5d $T_{3N+3}$ Tao theory. The possible F-theory realization of the 6d theory is 
\begin{eqnarray}
\begin{array}{ccccc}
 \emptyset &\mathfrak{su}(9) & \cdots & \mathfrak{su}(9N) & [\mathfrak{su}(9N+9)]\nonumber\\
 1 & 2 & \cdots & 2 & \nonumber
\end{array} 
\end{eqnarray}
where the first $\mathbb{P}^1$ does not support a gauge algebra. 

So far we have considered the 5d $T_{3N+3}$ Tao theory with $N \geq 1$. It is interesting to think of the case where $N=0$. This case admits a Type IIA brane configuration that was considered in section \ref{sec:speicalcases}. The 6d quiver description is 
\begin{equation}
SU(1) - [9].
\end{equation}
 In fact, this theory has been shown to be equivalent to the rank $1$ E-string theory in section \ref{sec:speicalcases}. This is perfect agreement with the formal substitution of $N=0$ into the 5d $T_{3N+3}$ Tao theory since the 5d $T_3$ Tao theory is nothing but the rank $1$ E-string theory \cite{Kim:2015jba}.

%% file: discussion.tex
\bigskip
\section{Discussions}
\label{sec:discussion}

We have started with 6d ${\cal N}=(1, 0)$ superconformal field theories whose Type IIA brane representations are made of NS5-branes, D6-branes, D8-branes and a single $O8^-$-plane. 
On their tensor branch, they are in general written as a linear quiver diagram with various matter hypermultiplets. 
After a circle compactification and T-duality, we have found a very rich group of 5d ${\cal N}=1$ gauge theories that are dual to each other. 
This diversity of the 5d theories for a given 6d theory originates from the choice in the resolution of two $O7^-$ planes and in the distribution of D5 and D7-branes as well as a part of the $SL(2,\mathbb{Z})$ duality. 
A decoupling of the same flavors from the dual 5d theories leads to another duality between 5d gauge theories which would have the identical 5d superconformal field theory.

While we provided the argument for the duality between 5d gauge theories by starting with the brane picture of their 6d mother theory, one can have different class of 6d superconformal field theories whose brane picture is not in the class considered here. There are also some 6d superconformal field theories with no obvious brane picture. 
It would be interesting to consider their 5d counterparts and the corresponding 5d dualities.

%% file: appendix.tex
\appendix
\section{A brief introduction of Tao diagram} \label{IntroTao}
A Tao diagram is a generalization of 5d $(p,q)$ web diagram with the critical number of flavors that gives the UV completion as 6d theory. A typical example is 5d $SU(2)$ superconformal theory of $N_f=8$ flavors, which is understood as a circle compactification of 6d E-string theory. Note that the theories of $N_f\le 7$ have the 5d UV fixed point \cite{Seiberg:1996bd} while the $N_f=8$ case has 6d UV fixed point. The corresponding $(p,q)$ web configuration has different features than the cases with $N_f\le7$. Salient features are that the web diagram form as periodic web configuration in a spiral whose period is identified as the instanton factor (squared) of the theory which is in turns identified as the KK mode of the circle compactified 6d theory. Infinite spirals correspond to an infinite KK spectrum coming from the compactification on a circle. The shape resembles a Taoism symbol which is why it is named to be Tao diagram. Suitable 7-brane monodromies give various yet equivalent diagrams, one of which can be made out of $(1,0), (0,1)$ and $(1,1)$ 5-brane charges, which enables one to compute the partition function via the topological vertex technique. As for $SU(2)$ gauge theory of $N_f=8$ flavors, an explicit computation \cite{Kim:2015jba} was done and reproduced the result of the E-string elliptic genus computation \cite{Kim:2014dza}. This confirms that the Tao diagram indeed describes a circle compactification of the 6d E-string theory. 
\begin{figure}
\begin{center}
\includegraphics[width=14cm]{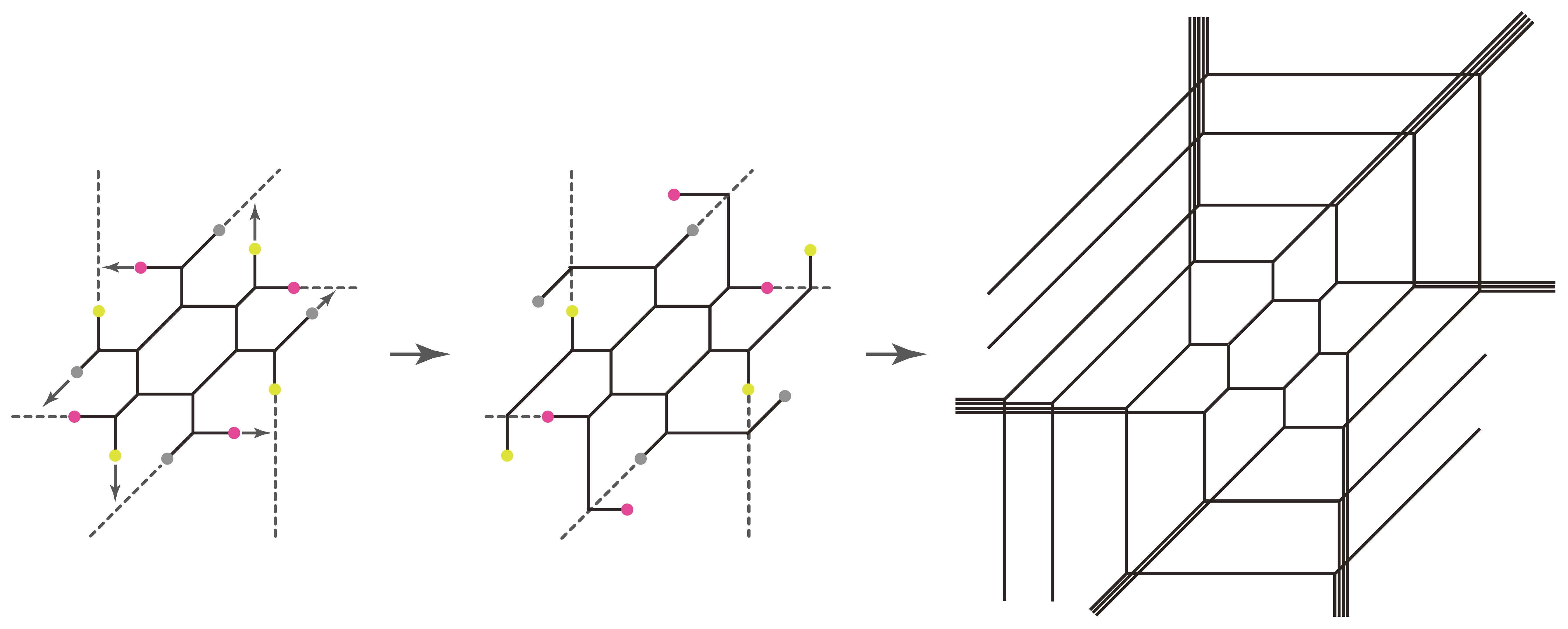}
\end{center}
\caption{An example of Tao diagram for 5d $SU(2)$ theory of $N_f=8$. There are twelve 7-branes of various charges which are the end point of 5-branes of the same charge. The colored dots denote 7-branes and different colors means different charges. Middle: Six branch cuts are chosen such that other remaining 7-branes are supposed to cross as being pulled to infinity.  Right: The charges of 7-branes are altered as they crossed the branch cut. A repeated process generates an infinite spiral web. Right: The last step is to pull out 7-branes associated with the branch cuts. In so doing, the Hanany-Witten transition creates 5-branes along the direction that the 7-branes are pulled out such that the newly created 5-branes are bound by one 7-brane. The resulting diagram makes a Tao diagram which possesses a constant period giving rise to KK mode, and infinite web diagram corresponding to infinite KK spectrum so that the Tao diagram describes a circle compactification of 6d SCFT.} 
\label{Fig:Taodiag}
\end{figure}

To see how infinite spiral shape arises, one first chooses some of the 7-brane branch cuts, and then pulls out other 7-branes along the geodesic direction given by the charge of 7-brane in the $(p,q)$ web plane. As 7-branes cross the branch cuts, the charges of the 7-branes change according to the monodromy associated with the branch cut. For instance, see Figure \ref{Fig:Taodiag}. When the number of flavors is critical, all the pulled-out 7-branes inevitably cross all the chosen branch cuts so that the charge of a pulled-out 7-brane becomes the same after one revolution making a spiral shape of a constant separation, or a constant period. The resulting diagram is a web diagram of infinite spiral. 

The constant period is expressed in terms of the instanton factor of the 5d theory. It is expected for a circle compactification of a 6d theory, and the constant period can be identified with the KK mode of the compactification. Hence, it possesses a compactified circle, more precisely, the radius of the T-dual circle.


For higher rank gauge theory or quiver theories, the configuration is basically same as shown in the Tao diagrams in the main text.

Tao web diagrams also give a computational tool using topological vertex methods.
